\documentclass[twocolumn]{aastex631}

\usepackage{upgreek}
\usepackage{verbatim}
\usepackage{hyperref}
\usepackage{url}            
\usepackage{booktabs}
\usepackage{amsfonts}
\usepackage{nicefrac}
\usepackage{microtype}
\usepackage{xcolor}
\usepackage{color}
\usepackage{bm}
\usepackage[utf8]{inputenc}
\usepackage{graphics}
\usepackage{graphicx}
\usepackage{amsfonts,amssymb}
\usepackage{amsmath}
\usepackage{subfigure}
\usepackage{mathrsfs}
\usepackage{multirow}
\usepackage{algorithm}
\usepackage{algorithmic}
\usepackage{CJK}

\shorttitle{Quasar Factor Analysis}
\shortauthors{Sun, Ting \& Cai}

\graphicspath{{./}{figures/}}

\begin{document}
\begin{CJK*}{UTF8}{gkai}
\title{Quasar Factor Analysis -- An Unsupervised and Probabilistic Quasar Continuum \\ Prediction Algorithm with Latent Factor Analysis}

\correspondingauthor{Zechang Sun}
\email{szc22@mails.tsinghua.edu.cn}

\author[0000-0002-8246-7792]{Zechang Sun (孙泽昌)}
\affiliation{Department of Astronomy, Tsinghua University, Beijing 100084, China}

\author[0000-0001-5082-9536]{Yuan-Sen Ting (丁源森)}
\affiliation{Research School of Astronomy \& Astrophysics, Australian National University, Cotter Rd., Weston, ACT 2611, Australia}
\affiliation{School of Computing, Australian National University, Acton, ACT 2601, Australia}

\author[0000-0001-8467-6478]{Zheng Cai (蔡峥)}
\affiliation{Department of Astronomy, Tsinghua University, Beijing 100084, China}
\affiliation{Department of Mathematics and Theories, Peng Cheng Laboratory, Nanshan, Shenzhen, China}

\begin{abstract}\label{sec:abstract}
Since their first discovery, quasars have been essential probes of the distant Universe. However, due to our limited knowledge of its nature, predicting the intrinsic quasar continua has bottlenecked their usage. Existing methods of quasar continuum recovery often rely on a limited number of high-quality quasar spectra, which might not capture the full diversity of the quasar population. In this study, we propose an unsupervised probabilistic model, \textit{Quasar Factor Analysis} (QFA), which combines factor analysis (FA) with physical priors of the intergalactic medium (IGM) to overcome these limitations. QFA captures the posterior distribution of quasar continua through generatively modeling quasar spectra. We demonstrate that QFA can achieve the state-of-the-art performance, $\sim 2\%$ relative error, for continuum prediction in the Ly$\alpha$ forest region compared to previous methods. We further fit 90,678 $2<\mathrm{z}<3.5$, SNR$>2$ quasar spectra from Sloan Digital Sky Survey Data Release 16 and found that for $\sim 30\%$ quasar spectra where the continua were ill-determined with previous methods, QFA yields visually more plausible continua. QFA also attains $\lesssim 1\%$ error in the 1D Ly$\alpha$ power spectrum measurements at $\mathrm{z}\sim 3$ and $\sim 4\%$ in $\mathrm{z}\sim 2.4$. In addition, QFA determines latent factors representing more physically motivated than PCA. We investigate the evolution of the latent factors and report no significant redshift or luminosity dependency except for the Baldwin effect. The generative nature of QFA also enables outlier detection robustly; we showed that QFA is effective in selecting outlying quasar spectra, including damped Ly$\alpha$ systems and potential Type II quasar spectra.
\end{abstract}


\section{Introduction}\label{sec:intro}
Powered by the accretion of matter into the supermassive black holes in the galactic nucleus, luminous quasars can shine across vast cosmic distances. As such, not only are they interesting astronomical objects which can reveal the enigmatic physics about the active galactic nucleus (AGN) \cite[e.g.,][]{SHEN2014}, they also serve as light beacons and shed light on topics that would otherwise not be accessible to us. The Sloan Digital Sky Survey \cite[e.g.,][]{SDSS2020} has been the singular powerhouse in the study of quasars. More than 750,000 quasars have been characterized by the Sloan Digital Sky Surveys to date with redshift up to $\sim 7.5$, spanning almost the entire cosmic history. Since quasars can reside in both the near and far Universe, the absorption features in quasar spectra are tell-tale witnesses of the evolution of the Universe.

The Ly$\alpha$ forest imprinted on the quasar spectra\cite[e.g.,][]{LYNDS1971, BAHCALL1971}, damped Ly$\alpha$ absorption systems \cite[e.g.,][]{DLAS2005} and Gunn-Peterson trough \cite[e.g.,][]{GP1965, GP2001} can all provide critical information about the physical state and matter distribution of the intergalactic medium (IGM). For instance, the measurement of baryonic acoustic oscillations (BAO) from the Ly$\alpha$ absorption forest \cite[e.g.,][]{BAO2013, BAO2014, PICCA2020} has been one of the mainstream methods to constrain cosmological parameters. The Ly$\alpha$ forest power spectrum can be used to infer the temperature distribution of IGM at high redshift \cite[e.g.,][]{1DPOWERSPECTRUM2013, 1DPOWERSPECTRUM2019} and reveal critical information about the epoch of reionization and cosmic dawn \cite[e.g.,][]{PAULO2020}. The Gunn-Peterson damping wing signatures in $\mathrm{z}\gtrsim 6$ quasar spectra is a direct probe to constrain the reionization history \cite[e.g.,][]{DAMPINGWING2018, BRAD2019, DAMPINGWING2020}. Moreover, cross-correlating the Ly$\alpha$ forest with the Ly$\alpha$ emission can be used to resolve the diffuse Ly$\alpha$ emission on cosmological scales which unravels the nature of galaxy evolution and outflow in the early Universe \cite[e.g.,][]{INTENSITYMAPPING2018, INTENSITYMAPPING2020, INTENSITYMAPPING2022}.
\begin{figure*}[t]
    \epsscale{1.2}
    \plotone{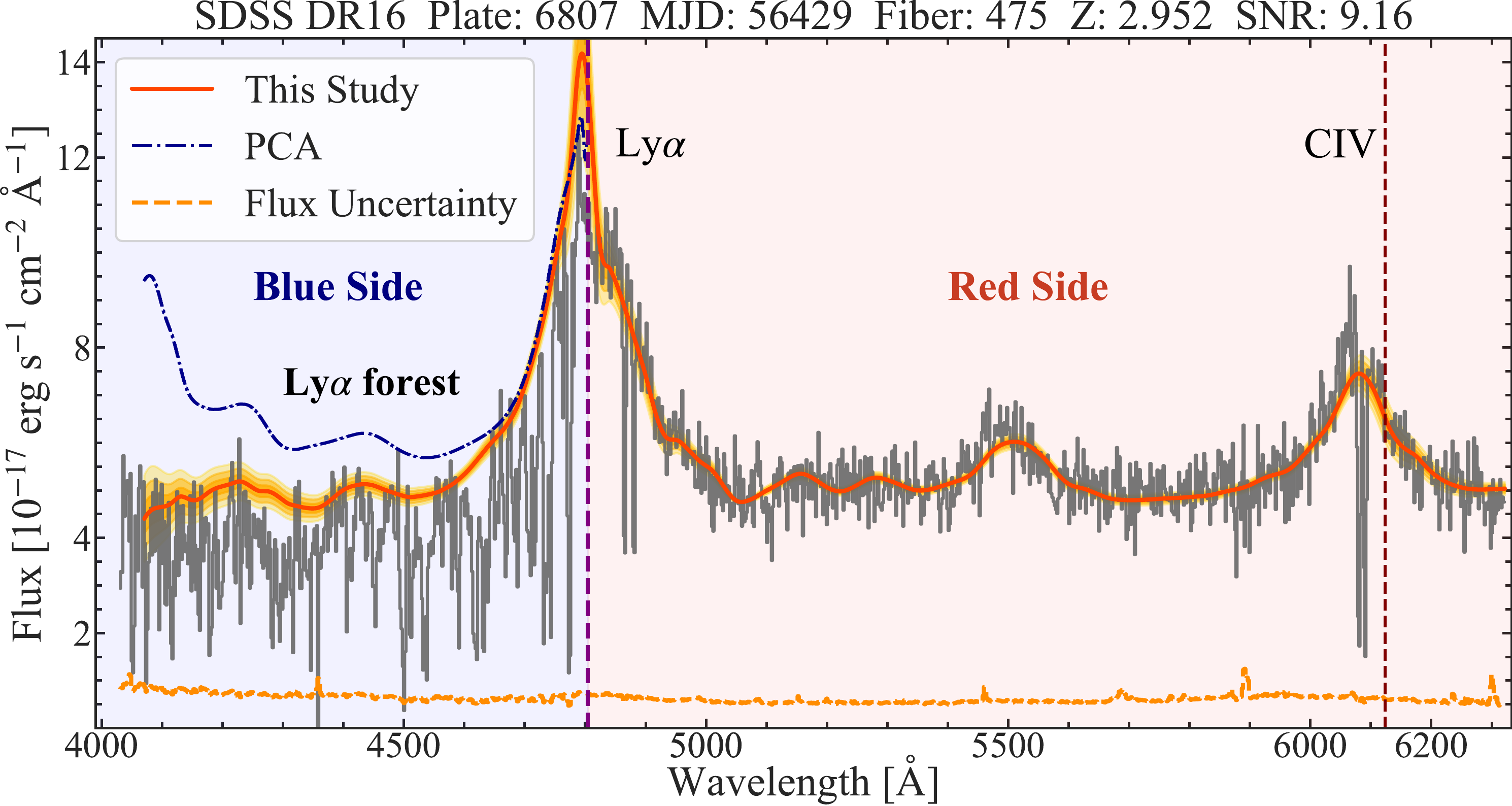}
    \caption{An example of quasar continuum prediction from the SDSS DR16 dataset. The regions with wavelengths longer than the Ly$\alpha$ emission line ($\lambda_\mathrm{rest}\geq 1215.67\,\mathrm{\AA}$), shaded in red, are free of the Ly$\alpha$ forest. The regions with wavelengths shorter than the quasar Ly$\alpha$ emission line, shaded in blue, ($\lambda_\mathrm{rest}\leq 1215.67\,\mathrm{\AA}$) are filled with the Ly$\alpha$ forest absorption. The unsupervised learning method proposed in this study, \textit{Quasar Factor Analysis}, is more resilient toward outlying quasar spectra and predicts a more robust continuum compared to the widely-adopted principal component analysis-based method (PCA, \citet{PARIS2011}). Regions shaded in orange denote the posterior uncertainty estimation from \textit{Quasar Factor Analysis}, showing $1\,\sigma$, $2\,\sigma$, and $3\,\sigma$ posterior confidence interval, respectively. \label{fig:example}}
\end{figure*}

However, how well we can use quasars to study the intervening gas between us and the quasars critically depends on how well we can infer the intrinsic background quasar continua. Despite its central role, deriving robust quasar continua, especially for noisy, low-resolution, and high-redshift quasar spectra, has proven to be non-trivial even for well-trained astronomers \cite[e.g.,][]{KIRKMAN2005, FG2008}. 

Our inability to accurately infer the continua has always been a significant source of uncertainties for using quasars as cosmological probes, especially for high-order statistics. For example, \citet{LEE2012} demonstrated that $\sim 2\%$ continuum prediction uncertainties could double the uncertainty in the study of the temperature-density relation \cite[][]{HUI1997} of the IGM. Furthermore, the current uncertainties in continuum determination can typically lead to $\sim 2-6\%$ bias in the measurements of Ly$\alpha$ power spectrum, \cite[e.g.,][]{1DPOWERSPECTRUM2013, 1DPOWERSPECTRUM2019}. On-going and upcoming large-scale spectroscopic sky surveys, e.g., the Dark Energy Spectroscopic Instrument \cite[DESI][]{DESIROADMAP2022}, typically acquire $\gtrsim 10^6$ quasar spectra or more \cite[e.g.,][]{DESITARGETSELECTION2021, DESITARGETSELECTION2022}. And the goal is to achieve cosmological measurements to a percent level of uncertainty \cite[e.g.,][]{DESIPRECISION2020}. As we further shrink the statistical errors via a larger sample, the high precision requirement hinges on our ability to infer the quasar continuum accurately.

Various methods have been proposed to infer the intrinsic quasar continua. A majority of the methods focuses on extracting information from wavelength redder than of the Ly$\alpha$ emission (hereafter, ``red-side"), which is largely devoid of the IGM absorption, to infer the continuum at wavelengths bluer than the Ly$\alpha$ emission line (hereafter, ``blue-side"). Among the most well-adopted methods, the power-law extrapolation methods \cite[e.g.,][]{FAN2006, JINYI2020} assume quasar continuum to be a power-law function and extrapolate the fitting from the red side to the blue side. The PCA-based methods \cite[e.g.,][]{PCA2005, PARIS2011, MFPCA2012, PCA2018, QSMOOTH2020} assume linear combinations of a few components can well represent quasar continua. The myriad of PCA-based continuum fitting methods mainly differs in the determination of the weight of each linear component, including projection \cite[][]{PCA2005, PARIS2011, PCA2018}, least-square fitting \cite[][]{MFPCA2012} and neural networks \cite[][]{QSMOOTH2020}. Apart from the mainstream PCA-based approaches, other continuum fitting methods include (1) \texttt{PICCA} continuum fitting \cite[][]{PICCA2020}, which fits for a polynomial correction to the mean quasar continuum to approximate various quasar continua in the blue side, and (2) deep learning based methods \cite[e.g., ][]{NF2020, IQNET2021}, which directly learn a high dimensional mapping between the red-side continua and the blue-side continua through neural networks, trained on mock datasets or a subsample of high-quality quasar spectra.

However, these existing methods come with several limitations. The supervised learning methods where we learn how to map the red-side continua to the blue-side continua -- including the PCA-based and the deep-learning-based methods -- rely on a small portion ($\lesssim 2\%$) of high signal-to-noise ratio (SNR) quasar spectra compared to the whole observations ($100\%$). The training sets typically comprise ad hoc continua derived from hand-fitting, \cite[e.g., ][]{PCA2005, PARIS2011}, or some automatic smoothing algorithms, \cite[e.g., ][]{PCA2018, NF2020, QSMOOTH2020, COMPARE2021}. These methods may be unable to generalize for the vast population of quasar spectra as the training sample is, intrinsically, biased toward brighter quasars. As for the existing unsupervised learning method, where the task is to generalize the red-blue connection through a parametric model -- including the power-law model and \texttt{PICCA} -- these parametric models might lack the flexibility which hinders their performance in real-life application. 

In light of these limitations, we propose a new unsupervised learning algorithm, \textit{Quasar Factor Analysis} (hereafter, QFA), to infer the posterior distribution of the intrinsic quasar continua using the entire observed spectrum instead of just harnessing the red-side information. At its core, QFA aims to model the full joint distribution of all quasar spectra by learning the distribution of quasar continua and the Ly$\alpha$ forest. As we will demonstrate, flexible unsupervised generative models with sufficient physical priors can capture the posterior distribution of quasar continua without ad hoc intervention. Since the generative task only maximizes the likelihood of the data, QFA can naturally harness information from all quasar spectra collected, taking into account their heteroskedastic noises.

The generative nature of QFA further leads to a few other advantages compared to existing methods. Firstly, QFA provides a robust posterior distribution of the quasar continua, which is handy for incorporating the continuum uncertainties into other downstream Bayesian inferences. Secondly, unlike PCA, latent factor analysis allows for a more flexible basis for the quasar continuum decomposition, leading to a more robust physical interpretation of the basis. Finally, as a fully probabilistic model of quasar spectra, missing pixels or outlying features (such as strong Ly$\alpha$ absorbers) in the quasar spectra can be rigorously dealt with through marginalizing over them

This paper is organized as follows: In Section~\ref{sec:method}, we will discuss the core idea and the probabilistic framework of our unsupervised learning algorithm. In Section~\ref{sec:data}, we present the quasar spectra studied in this paper, including both the SDSS DR16 quasar spectra and the mock spectra with realistic absorption fields. In Section~\ref{sec: result}, we will compare the performance of our model with existing methods and further investigate how continuum fitting affects the 1D Ly$\alpha$ forest power spectrum measurements. Besides, we will use QFA for out-of-distribution detection and study the quasar population's redshift evolution and luminosity dependency. Finally, we discuss the application of QFA for $\mathrm{z}>5$ quasars, its impact on Bayesian cosmology measurements, its limitations, and future directions in Section~\ref{sec:discussion}. We conclude in Section~\ref{sec:conclusion}.

\section{Method}\label{sec:method}

The basic premise of QFA is to build a statistical model to bridge the gap between the observables and the unobservables. Specifically, the quasar continua and the Ly$\alpha$ forest are not directly observable. Only the combination of them -- the quasar spectra are what we observe. QFA aims to independently model the distribution of both quasar continua and transmission fields and integrate them to obtain the marginal distribution of observed quasar spectra. By maximizing the marginal likelihood of quasar spectra, QFA will simultaneously learn the distribution of quasar continua and the transmission fields. The optimized model then allows us to infer the posterior distribution of the quasar continua, conditioning on the observed quasar spectra. 

\begin{figure*}[t]
    \epsscale{1.2}
    \plotone{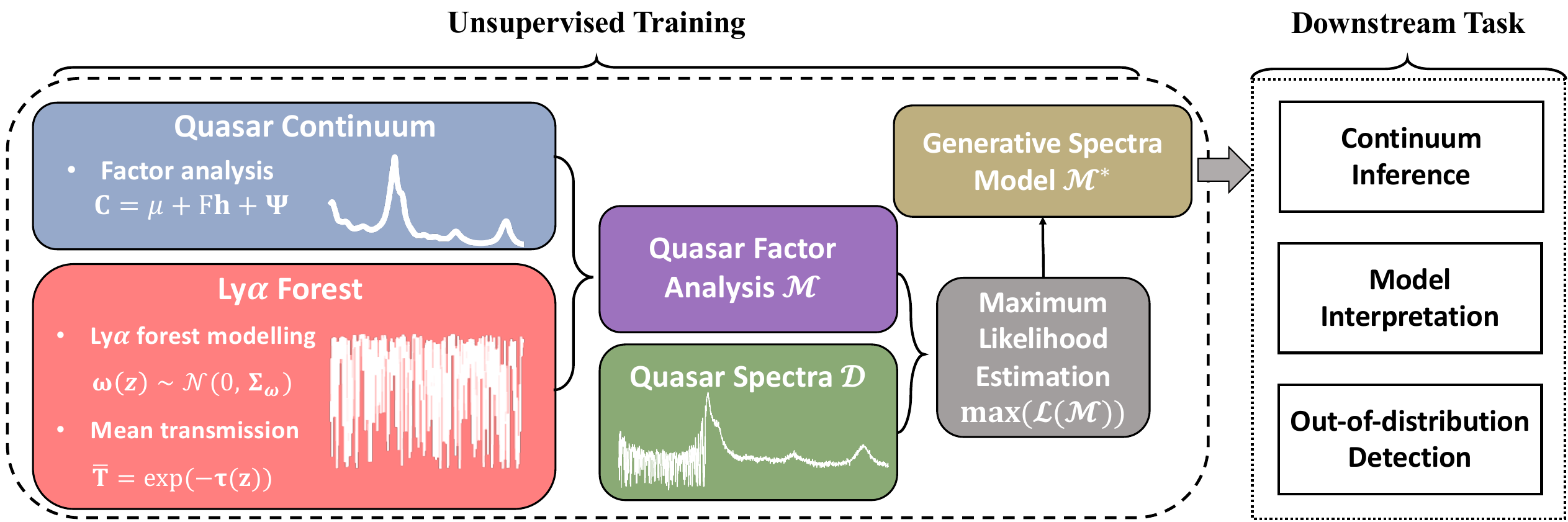}
    \caption{The schematic illustration of QFA. Quasar spectra are comprised of the intrinsic quasar continua and the Ly$\alpha$ forest. We assume the Ly$\alpha$ forest can be approximated as random Gaussian fluctuations and a factor model for the quasar continua (see Section~\ref{subsec: model}). We combine the two components to produce our unsupervised QFA model and train it via maximum likelihood estimation using the ensemble of observed spectra directly without any ad hoc training continuum (see Section~\ref{subsec:training}). The probabilistic nature of QFA leads to a statistically robust way of performing other downstream tasks, including a probabilistic inference of the continuum (see Section~\ref{subsec: pred} and Section~\ref{sec: result}).\label{fig:schematic}}
\end{figure*}

In the following, we will introduce QFA, how to train this model with maximal likelihood estimation and how to infer the posterior distribution of the quasar continua. Figure~\ref{fig:schematic} demonstrates a schematic summary of how QFA works. 

\subsection{Latent Factor Analysis}\label{subsec: fac}

QFA models the quasar continua based on latent factor analysis, hence the name. We will briefly summarize the basics of latent factor analysis to better orient the readers. For interested readers, we will refer to \citet{DAVID2011} and \citet{BEAUJEAN2017} for details. We denote all random variables in this paper in boldface and other deterministic variables otherwise. The model parameters we optimize through maximum likelihood are italicized.

Latent factor analysis \citep[][]{FA2011, FA2012, DAVID2011, BEAUJEAN2017} is a powerful statistical model which assumes that a high dimensional correlated data set (here, the quasar continua) can be expressed as linear combinations of a set of lower-dimensional latent factors. Formally, the observed data $\mathbf{X}$ is assumed to be

\begin{equation}
    \mathbf{X} = \mu + F\mathbf{h} + \mathbf{\Psi},
    \label{eq: FA}
\end{equation}

\noindent
where $\mu$ denotes the mean vector with size $\mathrm{N_s}$; $\mathbf{h}$ denotes the latent factors with size $\mathrm{N_h}$ ($\mathrm{N_h \ll N_s}$), $F$ denotes the factor loading matrix with size $\mathrm{N_s\times N_h}$ which signifies all the latent factors, and $\mathbf{\Psi}$ denotes the unaccounted ``error" term with size $\mathrm{N_s}$. 

Factor models assume the latent factor $\mathbf{h}$ and the $\boldsymbol{\Psi}$ follow Gaussian distributions. Since the factor $\mathbf{h}$ is only determined up to a rotation, as in any linear combination of $\mathbf{h}$ is interchangeable with the factor loading matrix $F$, $\mathbf{h}$ is preset to follow a multivariate normal distribution as $\mathbf{h}\sim\mathcal{N}\mathrm{(0, I)}$. The heteroscedastic error term, $ \boldsymbol{\Psi}$, accounts for the residual of the data set that the latent factors cannot explain. $ \boldsymbol{\Psi}$ is assumed to be distributed as $\boldsymbol{\Psi}\sim\mathcal{N}(0, \mathrm{diag}(\sigma^2_{\mathrm{\Psi}}))$, where $\sigma^2_{\Psi}$ is a free parameter vector to be optimized for. 

Generally, factor analysis can be viewed as a generalized version of PCA. The factor loading matrix $F$ plays the same role as the basis matrix in PCA, and the factor $\mathbf{h}$ works the same as the PCA coefficients. However, there are two key differences between a factor model and a PCA model. Firstly, PCA assumes an isotropic error term which is distributed as $\mathcal{N}(0, \mathrm{\sigma^2 I})$ while the error term $\mathbf{\Psi}$ in latent factor analysis has an anisotropic covariance matrix $\mathrm{diag}(\sigma^2_{\mathrm{\Psi}})$. Secondly, unlike PCA, which requires a set of orthogonal basis, factor analysis does not necessitate the linear components to be orthogonal. The anisotropic stochastic ``error" term and non-orthogonal linear components allow factor analysis more flexibility to adapt to real-life data and better interpretability. Furthermore, the statistical foundation of factor analysis, based on maximum likelihood estimation, makes it possible to construct a complex probabilistic spectrum model from this base model, including other physical priors, which we will explain next. The physical prior is critical in ensuring that the factor model focuses only on extracting the continuum, breaking the degeneracy between the transmission fields and the quasar continua.

\subsection{QFA -- A Generative Model of Quasar Spectra}\label{subsec: model}

Building upon the basic latent factor model, we leverage our physical priors on the Ly$\alpha$ forest to construct a full generative model of the quasar spectra. For ease of discussion, throughout this paper, we will split the quasar spectra into the red side and the blue side relative to the Ly$\mathrm{\alpha}$ emission line (rest-frame wavelength $\mathrm{\lambda_{RF}\approx 1215.67\AA}$) to distinguish various variables and their roles. We denote the red-side variables with a subscript $\mathrm{r}$ and the blue-side variables with a subscript $\mathrm{b}$.

We model the quasar continua $\mathbf{C}$ through a factor model:
\begin{equation}
    \mathbf{C} = \mu + F\mathbf{h} + \mathbf{\Psi}.
    \label{eq: cont_model}
\end{equation}

\noindent
The quasar spectra are further modified for the blue side by the Ly$\alpha$ forest. We assume the overall strength of the absorption can be captured by mean optical depth function $\tau_{\mathrm{eff}}$ following \citet{BECKER2013},
\begin{equation}
\tau_{\mathrm{eff}}(\mathrm{z}_{\mathrm{abs}})= 0.00958\times (1+\mathrm{z}_{\mathrm{abs}})^{2.90}-0.132,
\label{eq:optical_depth}
\end{equation}
\noindent
where $\mathrm{z}_{\mathrm{abs}}=\lambda/1215.67\mathrm{\AA} - 1$ is the redshift of the absorption systems. Apart from the mean optical depth, the series of stochastic Ly$\mathrm{\alpha}$ forest are modeled as random Gaussian fluctuations $\boldsymbol{\mathrm{\omega}}(\mathrm{z}_{\mathrm{abs}}) \sim \mathcal{N}(0, \mathrm{diag}(\sigma^2_{\omega}(\mathrm{z}_{\mathrm{abs}})))$. $\sigma_{\omega}(\mathrm{z}_{\mathrm{abs}})$ signifies the evolution of the absorption strength of the Ly$\mathrm{\alpha}$ forest. Similar to \citet[][]{GP2017, GP2020}, we adopt the function form of $\sigma^2_{\omega}(\mathrm{z}_{\mathrm{abs}})$ as
\begin{equation}
    \sigma^2_{\omega}(\mathrm{z}_{\mathrm{abs}}) = \omega_0\circ (1-\exp(-\tau_0 (1+\mathrm{z}_{\mathrm{abs}})^\beta )+c_0)^2,
\end{equation}
\noindent
where $\omega_0$, $\tau_0$, $\beta$, $c_0$ are free parameters, and ``$\circ$" denotes the element-wise product of two vectors.

Finally, the observational noise $\boldsymbol{\epsilon}$ is modeled as $\mathcal{N}(0, \mathrm{diag}(\mathrm{\sigma}_{\epsilon}^2))$, $\mathrm{\sigma}_{\epsilon}$ is the observed flux uncertainty. Putting them together, the observed quasar spectrum on the blue side can thus be written as
\begin{equation}
    \mathbf{S}_{\mathrm{b}} = \mathbf{C}_{\mathrm{ b}}\circ\exp(-\tau_{\mathrm{eff}}(\mathrm{z}_{\mathrm{abs}}))+\boldsymbol{\mathrm{\omega}}(\mathrm{z}_{\mathrm{abs}}) + \boldsymbol{\mathrm{\epsilon}}_{\mathrm{b}}.
    \label{eq: BlueMOdel}
\end{equation}

As for the red side, although there might a few minor absorption features, most of them are rejected in preprocessing, which leads to the simplified red-side model with only the continuum:
\begin{equation}
    \mathbf{S}_{\mathrm{r}} = \mathbf{C}_{\mathrm{r}} + \boldsymbol{\mathrm{\epsilon}}_{\mathrm{r}},
    \label{eq: redmodel}
\end{equation}

Finally, integrating the blue-side and red-side models leads to the final expression of the whole quasar spectrum
\begin{equation}
    \begin{split}
    \mathbf{S} &= \mathrm{A}\mathbf{C} + \boldsymbol{\mathrm{\Omega}} + \boldsymbol{\mathrm{\epsilon}}\\
    &=\mathrm{A}\mu + \mathrm{A}F\mathbf{h}+\mathrm{A}\boldsymbol{\mathrm{\Psi}}+\boldsymbol{\mathrm{\Omega}} + \boldsymbol{\mathrm{\epsilon}},
    \label{eq: obsMOdel}
    \end{split}
\end{equation}
where
\begin{equation}
    \mathrm{A} = \mathrm{diag}[\ \underbrace{\exp{(-\tau_{\mathrm{eff}}(z_{\mathrm{abs}}))}}_{\mathrm{N}_{\mathrm{b}}},\ \overbrace{1,\dots,1}^{\mathrm{N}_{\mathrm{r}}}\ ],
    \label{eq: Adef}
\end{equation}
and
\begin{equation}
    \boldsymbol{\mathrm{\Omega}} = [\ \underbrace{\boldsymbol{\mathrm{ \omega}}(z_{\mathrm{abs}})}_{\mathrm{N}_{\mathrm{b}}}, \overbrace{0, \dots, 0}^{\mathrm{N}_{\mathrm{r}}}\ ]^{\mathrm{T}}.
    \label{eq: Omega}
\end{equation}

We summarize all the learnable model parameters in QFA and their corresponding dimensions in Table~\ref{tab:param} and other variables in Table~\ref{tab:var-info}.

\begin{deluxetable}{lccccccc}
\tablecaption{Learnable model parameters in QFA
\label{tab:param}}
\tablehead{
\colhead{Symbol\tablenotemark{$\dagger$}} & \colhead{$\mu$}&\colhead{$\sigma^2_{\Psi}$} & \colhead{$\omega_0$} & \colhead{$\tau_0$}&\colhead{$\beta$}&\colhead{$c_0$}&\colhead{$F$}
}
\startdata
Size&$\mathrm{N}_{\mathrm{pix}}$&$\mathrm{N}_{\mathrm{pix}}$&$\mathrm{N}_{\mathrm{b}}$&$1$&$1$&$1$&$\mathrm{N}_{\mathrm{pix}}\times\mathrm{N}_{\mathrm{h}}$\\
\enddata
\tablenotetext{\dagger}{$\mathrm{N}_{\mathrm{b}}$ is the pixel number of the blue side, $\mathrm{N}_{\mathrm{r}}$ is the pixel number of of the red side, and $\mathrm{N}_{\mathrm{pix}}$ is the pixel number of the whole quasar spectrum. 
By definition, we have, $\mathrm{N}_{\mathrm{pix}}=\mathrm{N}_{\mathrm{b}}+\mathrm{N}_{\mathrm{r}}$. 
$\mathrm{N}_{\mathrm{h}}$ is the number of factors, which is a hyperparameter in our model, and
$\mathrm{N}_{\mathrm{h}} \ll \mathrm{N}_{\mathrm{pix}}$. In this work, $\mathrm{N}_{\mathrm{h}}$ is set to be $8$ and $\mathrm{N}_{\mathrm{pix}}$ is about $2\times 10^3$. In total, QFA model in this work has about $2\times 10^4$ learnable model parameters.}
\end{deluxetable}

\begin{deluxetable}{lll}[htbp!]
\tablecaption{Other variables in this study\label{tab:var-info}}
\tablehead{\colhead{Variable}&\colhead{Dimension}&\colhead{Definition\tablenotemark{$\dagger$}}}
\startdata
$\mathbf{S}_{\mathrm{b}}$&$\mathrm{N}_\mathrm{b}$&Blue-side observed flux spectrum\\
$\mathbf{C}_{\mathrm{b}}$&$\mathrm{N}_\mathrm{b}$& Blue-side quasar continuum\\
$\boldsymbol{\rm \epsilon}_{\mathrm{b}}$, $\mathrm{\rm \sigma}_{\mathrm{b}}$&$\mathrm{N}_\mathrm{b}$&Blue-side noise/flux uncertainty \\
$\mathbf{S}_{\mathrm{r}}$&$\mathrm{N}_\mathrm{r}$&Red-side observed flux spectrum\\
$\mathbf{C}_{\mathrm{r}}$&$\mathrm{N}_\mathrm{r}$&Red-side quasar continuum\\
$\boldsymbol{\rm \epsilon}_{\mathrm{r}}$, $\mathrm{\rm \sigma}_{\mathrm{r}}$ &$\mathrm{N}_\mathrm{r}$&Red-side noise/flux uncertainty \\
$\mathbf{\rm \lambda}$ & $\mathrm{N}_\mathrm{pix}$ & Rest-frame wavelength\\
$\mathbf{S}$ & $\mathrm{N}_\mathrm{pix}$ &  The entire observed flux spectrum \\
$\boldsymbol{\rm \epsilon}$, $\mathbf{\sigma_{\epsilon}}$ & $\mathrm{N}_\mathrm{pix}$ & The entire noise/flux uncertainty\\
$\mathbf{\Psi}$&$\mathrm{N}_\mathrm{pix}$&Unaccounted model error\\
$\mathbf{h}$&$\mathrm{N}_\mathrm{h}$&Latent factors\\
$\boldsymbol{\mathrm{\omega}}$&$\mathrm{N}_\mathrm{b}$& Ly$\alpha$ forest random fluctuations\\
$\mathrm{z}$&$1$&Quasar redshift \\
\enddata
\tablenotetext{\dagger}{``Blue side" denotes the part of the quasar spectrum that resides bluer than the Ly$\alpha$ emission line, and vice versa for the ``red side".}
\end{deluxetable}

\subsection{Model Training - Maximum Likelihood
Estimation}\label{subsec:training}

Thus far, we have developed an analytic description of quasar spectra. We note that most variables written are stochastic variables, and the analytic formulae described should be treated as the mathematical operation  (e.g., sum, product) on the random variables. The advantage of describing the quasar spectra as a stochastic process is that the probabilistic model defines the likelihood of any observed quasar spectrum. The best model would then be the one that maximizes the joint likelihood of all quasar spectrum observations. This section will derive the likelihood function and describe the model's training via the maximal likelihood estimation objective.

\subsubsection{Log-Likelihood}

A particular advantage of QFA is that it is specifically designed in a way where most random variable components can be described as a high-dimensional Gaussian distribution. This naturally stems from the fact that the Ly$\alpha$ forest is assumed to be a Gaussian distribution in QFA. Although, in reality, the Ly$\alpha$ forest is not strictly Gaussian distributed \cite[e.g.,][]{SDSSMOCK2015}. We will return to this point in Section~\ref{subsec:cal}. However, this trade-off allows us to describe the likelihood in a compact Gaussian, which facilitates many other operations, such as marginalizing over nuisance parameters or masked pixels. We will show that the Gaussian assumption only incurs minor systematics in the continuum inference and other downstream tasks (See Section~\ref{sec: result}).

Recall that, for multivariate Gaussian distributions, the two properties below, which will come in handy in some of our derivations, hold \cite[][]{JOHNSON2007}.
\begin{enumerate}
    \item If $\mathbf{X}$ distributed as $\mathcal{N}(\mu, \Sigma)$, then $B\boldsymbol{X}+d$ distributed as $\mathcal{N}(B\mu+d, B\Sigma B^\mathrm{T})$.
    \item For two independent multivariate Gaussian random variables $\mathbf{X}\sim\mathcal{N}(\mu_{X}, \Sigma_{X})$ and $\mathbf{Y}\sim\mathcal{N}(\mu_{Y}, \Sigma_{Y})$, their sum $\mathbf{X}+\mathbf{Y}$ is distributed as $\mathcal{N}(\mu_{X}+\mu_{Y}, \Sigma_{X}+\Sigma_{Y})$.
\end{enumerate}

Since $\mathbf{h}$ is distributed as $\mathcal{N}\mathrm{(0, I)}$ and $\boldsymbol{\mathrm{\Psi}}$ is distributed as $\mathcal{N}(0, \Sigma_{\Psi})$, it follows from properties above that the quasar continuum $\mathbf{C}$ is distributed as $\mathcal{N}(0, FF^\mathrm{T}+\Sigma_{\Psi})$. Furthermore, the intrinsic quasar continua, Ly$\alpha$ forest absorption, and observational noise are independent. It follows trivially from Equation~(\ref{eq: obsMOdel}) and the two properties above that, the distribution of the quasar spectrum $\mathbf{S}$ can be written as
\begin{equation}
    \mathbf{S}\sim\mathcal{N}(\mathrm{A}\mu, \mathrm{A}(FF^\mathrm{T}+\Sigma_{\Psi})\mathrm{A}^\mathrm{T}+\Sigma_{\Omega}+\Sigma_{\epsilon}),
    \label{eq: diStribution}
\end{equation}
where $\Sigma_{\Omega}=\mathrm{diag}\left[\omega(\mathrm{z}_{\mathrm{abs}}), 0,\dots,0\right]$,
and $\Sigma_{\epsilon}=\mathrm{diag}(\sigma^2_{\epsilon})$. 

The formula above characterizes the entire distribution of the quasar spectrum, given the model parameters $\mathcal{M} \equiv \left\{\mu, \sigma^2_{\Psi}, \omega_0, \tau_0, \beta, c_0, F\right\}$. The training process of QFA is to find the best model parameters $\mathcal{M}^*$, which maximizes the likelihood of all observed spectra. 

In particular, for any observed quasar spectrum $\mathrm{ (\lambda, S, z, \sigma_{\epsilon})}$, the log-likelihood of observing the spectrum is
\begin{equation}
\begin{split}
\mathcal{L}(\mathrm{S}|\mathrm{\lambda, z, \sigma_\epsilon}, \mathcal{M})
&= -\frac{1}{2}(\mathrm{N}_\mathrm{pix}\ln 2\pi + \ln\det\Sigma\\
 &\quad +(\mathrm{S}-\mathrm{A}\mu)^\mathrm{T}\Sigma^{-1}(\mathrm{S}-\mathrm{A}\mu)),
\end{split}
\label{eq:single_ll}
\end{equation}
where $\Sigma=\mathrm{A}(FF^\mathrm{T}+\Sigma_{\Psi})\mathrm{A}^\mathrm{T}+\Sigma_{\Omega}+\Sigma_{\epsilon}$. And for a data set $\mathcal{D}$ with $\mathrm{N}_{\mathrm{spec}}$ independent observed quasar spectra $\{ ( \mathrm{\lambda}^{(i)}, \mathrm{S}^{(i)}, \mathrm{z}^{(i)}, \mathrm{\sigma}^{(i)}_{\epsilon}
),i=1,2,\dots,{\mathrm{N}_\mathrm{spec}}\}$, their joint log-likelihood can thus be written as
\begin{equation}
    \mathcal{L}(\mathcal{D}|\mathcal{M})
    = \sum_{i=1}^{\mathrm{N}_\mathrm{spec}}\mathcal{L}(\mathrm{S}^{(i)}|\mathrm{\lambda}^{(i)}, \mathrm{z}^{(i)}, \mathrm{ \sigma}^{(i)}_\epsilon, \mathcal{M}).
\label{eq:total_ll}
\end{equation}

\subsubsection{Model Regularization}\label{subsec:regularization}

In practice, we found that maximizing the likelihood in Equation~(\ref{eq:total_ll}) often leads to non-physical local minima, i.e., continuum components with jagged features. Regularization tricks are necessary to facilitate the model convergence to local minima, better separating the distributions of the quasar continua and the Ly$\alpha$ forest.

We impose two regularization recipes. Firstly, previous PCA-based works demonstrated that the principal components of quasar continua always fluctuate around zero. The results suggested that despite the great diversity of quasar continua, their variations are small compared to the mean continuum. This motivates us to assume a prior for all model parameters to be close to zero. In particular, we assign each parameter a Gaussian prior centered around zero, equivalent to what is known as the ``L2 regularization" \citep{L22004}. With the regularization term, the loss function reads
\begin{equation}
    \mathrm{L}(\mathcal{M}) = -\frac{1}{{\mathrm{N}}_\mathrm{spec}}\mathcal{L}(\mathcal{D}|\mathcal{M}) + \alpha \mathrm{L}_2 (\mathcal{M}),
    \label{eq: loss}
\end{equation}
where $\alpha$ is a hyperparameter that controls the regularization strength. A larger $\alpha$ gives heavier penalties to the weight of model parameters, leading to smaller model parameters. $\mathrm{L}_2(\mathcal{M})$ denotes the L2 regularization, which is the square sum of all model parameters. We assume $\alpha=0.1$ in this study.

Secondly, we also enforce that quasar continua are smooth profiles. To implement this regularization, we apply a running median filter along the wavelength direction to smooth each component in the factor loading matrix $F$ every 20 optimization epochs. In this study, we set the filter width to 31 pixels ($\sim 7-11\,\mathrm{\AA}$ in the rest frame). Here, one optimization epoch is defined as the stochastic gradient-descent algorithm running over the entire dataset.

\subsubsection{Implementation}\label{subsec: opt}

Traditionally, a factor analysis model is optimized through singular value decomposition or expectation maximization (EM) algorithm \cite[][]{FA2012}. However, both methods are difficult to be implemented for QFA because of the heteroskedasticity nature of QFA. Also, the elaborate modeling of QFA with $\sim 2\times 10^4$ model parameters and the complex loss function (Equation~(\ref{eq: loss})) calls for a better optimization algorithm.

We adopt the \texttt{Adam} optimization algorithm \cite[][]{ADAM2014}, a robust optimization method based on gradient descent, which has been widely used in deep learning, to find the best model parameters that optimize the likelihood of the observed data. We implement QFA via PyTorch \cite[][]{PYTORCH2019} to speed up the matrix operations with GPU resources.
Due to the Gaussian nature of our model, the derivative of each parameter (see Appendix~\ref{appendix:derivative}) can be analytically derived, which we further harness to speed up the training process and optimize the GPU memory usage. For the details, we refer readers to Appendix~\ref{appendix:complexity}.

\subsection{Continuum Inference}\label{subsec: pred}

QFA depicts the joint distribution of the quasar spectra. The model breaks the degeneracy between the transmission fields and the continua by having two separate components for the continuum and the Ly$\alpha$ forest features. Consequently, given the best-fitted model $\mathcal{M}^*$, one can obtain the posterior distribution of the quasar continuum $\mathbf{C}$, conditioning on the observed quasar spectra, which we will elaborate on in this section.

QFA models the quasar continua $\mathbf{C}$ as $\mathbf{C} = \mu + F\mathbf{h} + \boldsymbol{\mathrm{\Psi}}$. We will neglect the ``error" term $\boldsymbol{\mathrm{\Psi}}$ in the quasar continuum inference. The term $\boldsymbol{\mathrm{\Psi}}$ is designed to account for the stochastic residuals of quasar continua that can not be explained by the linear model $\mu+F\mathbf{h}$. However, in practice, we found that $\boldsymbol{\mathrm{\Psi}}$ also incorporates other unaccounted absorptions and observation noises and can lead to quasar continuum inference with jagged features. Thus, we neglect $\boldsymbol{\mathrm{\Psi}}$ when inferring the posterior distribution of the continua. Such treatment may result in poor continuum fitting at emission peaks in a few cases, and may also lead to an underestimation of the derived continuum fitting uncertainty. In Appendix~\ref{appendix:uncertainty}, we provide a detailed explanation of the issue of underestimating the continuum fitting uncertainty and present a possible calibration process to counteract this shortcoming.

On top of that, since $F$ and $\mu$ are deterministic parameters, to derive the posterior of $\mathbf{C}$, it suffices to evaluate the posterior of $\mathbf{h}$. Recall that, given an observed spectrum and its properties $\mathrm{\left(\lambda, S, z, \sigma_{\epsilon}\right)}$, the posterior of $\mathbf{h}$ follows the Bayes rule:
\begin{equation}
    \mathrm{P}(\mathrm{h}|\mathrm{\lambda}, \mathrm{S}, \mathrm{z}, \sigma_{\epsilon}, \mathcal{M}) \propto \mathrm{P}(\mathrm{S}|\mathrm{\lambda, h, z, \sigma_{\epsilon}, \mathcal{M}}) \mathrm{P}(\mathrm{h}),
    \label{eq:pos_h_bayes}
\end{equation}
\noindent
where $\mathbf{h} \sim \mathcal{N}\mathrm{(0, I)}$. Recall that, in Section~\ref{subsec: model}, we derived the conditional distribution of the quasar spectra $\mathrm{S}$ can be written as 
\begin{equation}
    \mathbf{S}|_{\mathrm{\lambda, h, z, \sigma_{\epsilon}, \mathcal{M}}}\sim \mathcal{N}(\mathrm{A}F\mathrm{h}+\mathrm{A}\mu, (\mathrm{A}\Sigma_{\Psi}\mathrm{A}+\Sigma_{\omega}+\Sigma_{\epsilon})^{-1}).
    \label{eq:cond_s}
\end{equation}

\noindent
It follows from Equation~(\ref{eq:pos_h_bayes}) and Equation~(\ref{eq:cond_s}) that, $\mathrm{h}$ follows the distribution
\begin{equation}
\mathbf{h}|_\mathrm{\lambda, S, z, \sigma_{\epsilon}, \mathcal{M}}\sim\mathcal{N}(\Sigma_\mathrm{h} \tilde{F}^\mathrm{T}\Sigma_\mathrm{e}^{-1}\Delta, \Sigma_\mathrm{h}),
\label{eq:pos_h}    
\end{equation}

\noindent
where
\begin{equation}
\begin{split}
    \tilde{F} &= \mathrm{A}F\\
    \Delta &= \mathrm{S} - \mathrm{A}\mu\\
    \Sigma_\mathrm{e} &= \mathrm{A}\Sigma_{\Psi}\mathrm{A}+\Sigma_{\omega}+\Sigma_{\epsilon}\\
    \Sigma_\mathrm{h} &= (\mathrm{I}+\tilde{F}^\mathrm{T}\Sigma_\mathrm{e}^{-1}\tilde{F})^{-1}.\\
\end{split}
\end{equation}

With the posterior distribution of $\mathbf{h}$ at hand, the best-estimated continuum $\mathbf{C}$ given the observed spectrum can be evaluated as the continuum fitting with the maximum posterior probability density, or
\begin{equation}
\begin{split}
    \hat{\mathrm{C}} &= \mathbb{E}\left [ F\mathbf{h}+\mu |\mathrm{ \lambda,S, z}, \sigma_\epsilon, \mathcal{M}\right]\\
    &= F\mathbb{E}\left [ \mathbf{h}|\mathrm{\lambda,S, z}, \sigma_\epsilon, \mathcal{M}\right] + \mu\\
    &= F\Sigma_\mathrm{h} \tilde{F}^\mathrm{T}\Sigma_\mathrm{e}^{-1}\Delta + \mu,
\end{split}
\label{eq: pos_mu}
\end{equation}

\noindent
with the posterior variance
\begin{equation}
\mathrm{Var}\left[\mathrm{C}|\mathrm{\lambda,S, z},\sigma_{\epsilon}, \mathcal{M}\right] = \mathrm{diag}\left[F \Sigma_h F^\mathrm{T}\right].
\label{eq: pos_error}
\end{equation}

Modeling the full joint distribution of the quasar spectra comes with the advantage of taking the marginal posterior distribution by integrating over all masked pixels. Various factors, such as limited wavelength coverage of the spectrograph, can render part of the quasar spectra unavailable. Furthermore, some strong absorbers might violate the assumption of our models where the absorptions are assumed to be Gaussian distributed; including them as the conditional information would, therefore, bias our inference. 

Fortunately, since our posteriors are all Gaussian, taking the marginal distribution is trivial. More specifically, recall that for multivariate Gaussian distribution, the covariance matrix of the marginal distribution corresponds to the corresponding sub-matrices of the entire matrix, and the mean of the marginal distribution corresponds to the corresponding sub-mean of the mean vector.

\section{Data}\label{sec:data}

In this study, we apply QFA to observation and mock data to demonstrate its capability compared to other widely-adopted methods. We will apply QFA to observed quasar spectra from SDSS data release 16 (DR16) \citep[][]{SDSS2020}. Furthermore, we will also apply QFA for mock quasar spectra, of which we know their ground truth continua, and evaluate their performance compared to existing methods. 

\subsection{SDSS DR16}\label{subsec:sdss}

\begin{figure*}[t]
    \epsscale{1.2}
    \plotone{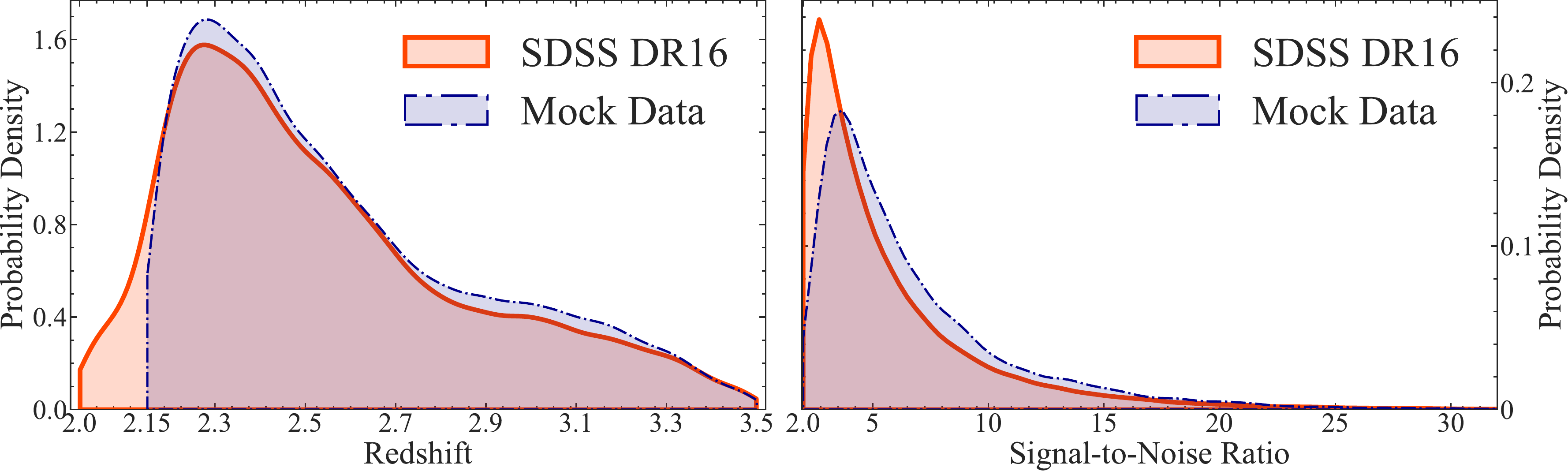}
    \caption{The SNR and redshift distribution of the quasar spectra adopted in this study. The left panel shows the redshifts, and the right panel the SNRs. The actual SDSS DR16 quasar spectra are shown in red, and the mock spectra are shown in blue. We specifically match the properties of the mock spectra to mimic the SDSS data to gauge the performance of QFA compared to other existing methods. Since the transmission field adopted to create the mock spectra truncates at $z=2.15$, this leads to a minor difference in redshift coverage between the mock dataset and the SDSS DR16 dataset.
    \label{fig:snrRedshift}}
\end{figure*}

The SDSS data release 16 quasar catalog\footnote{\hyperlink{https://data.sdss.org/sas/dr16/eboss/qso/DR16Q/}{https://data.sdss.org/sas/dr16/eboss/qso/DR16Q/}} \citep[DR16Q, ][]{SDSS2020} consists of a total of 750,414 quasar spectra conducted using the BOSS spectrographs on the $2.5\,\mathrm{m}$ wide-angle optical telescopes at Apache Point Observatory.  The spectrographs cover a wavelength range from $\mathrm{3,600\,\AA}$ to $\mathrm{10,400\,\AA}$ at a spectral resolution of $\lambda/\Delta \lambda \approx 2,000$. 

We exclude non-quasar spectra contaminants (e.g., star, blazar) by considering only spectra with confident visual ``quality flag", \texttt{CLASS\_PERSON=3}, as specified by the catalog. We compute the median SNR of each quasar spectrum at rest-frame wavelength $1280\pm 2\,\mathrm{\AA}$ and select only quasar spectra with median SNR greater than $2.0$ and redshifts $2 < \mathrm{z} < 3.5$ to ensure relatively high data quality and reliable mean optical depth measurements. Note that as various redshift estimates may differ in DR16Q \cite[][]{SDSS2020}, we use the \texttt{Z} column as a redshift reference. According to \citet{SDSS2020}, the redshifts from the \texttt{Z} column are expected to be the least biased (although, arguably with a higher variance). Since QFA can generally deal with redshift variance, the training of QFA thus benefits from the least biased estimator, which has led to our choice. We also discard those quasars flagged with broad absorption line systems (BAL). Although our algorithm can deal with masked pixels, e.g., the damped Ly$\alpha$ systems and missing data pixels (see Section~\ref{subsec: pred}), current techniques \cite[][]{BALCNN2019} cannot yet reliably mask the BAL regions. In total, 90,678 quasar spectra met our criteria. The SNR and redshift distribution of our final data set are shown in Figure~\ref{fig:snrRedshift}. Also, our data selection criteria and their corresponding number of spectra after these successive cuts are shown in Table~\ref{tab:data_selection}. We focus on the redshift interval $2<\mathrm{z}<3.5$, as it encapsulates the majority of quasars used for cosmological studies, for example, BAO measurements \cite[e.g., ][]{PICCA2020} and 1D Ly$\alpha$ forest power spectrum measurements \cite[e.g.,][]{1DPOWERSPECTRUM2019}.

\begin{deluxetable}{lc}[htbp!]
\tablecaption{SDSS DR16 Data Selection\tablenotemark{$\dagger$}\label{tab:data_selection}}
\tablehead{\colhead{Criteria}&\colhead{Number of spectra left}}
\startdata
All SDSS quasar spectra&750,414\\
SDSS quality flag&290,068\\
SNR$>2$&220,960\\
$2<\mathrm{z}<3.5$&125,156\\
Not BAL&90,678
\enddata
\tablenotetext{\dagger}{We impose data selection criteria from top to bottom, and the right column states the number of quasar spectra left after these criteria are upheld.}
\end{deluxetable}

\begin{figure}
    \centering
    \epsscale{1.3}
    \plotone{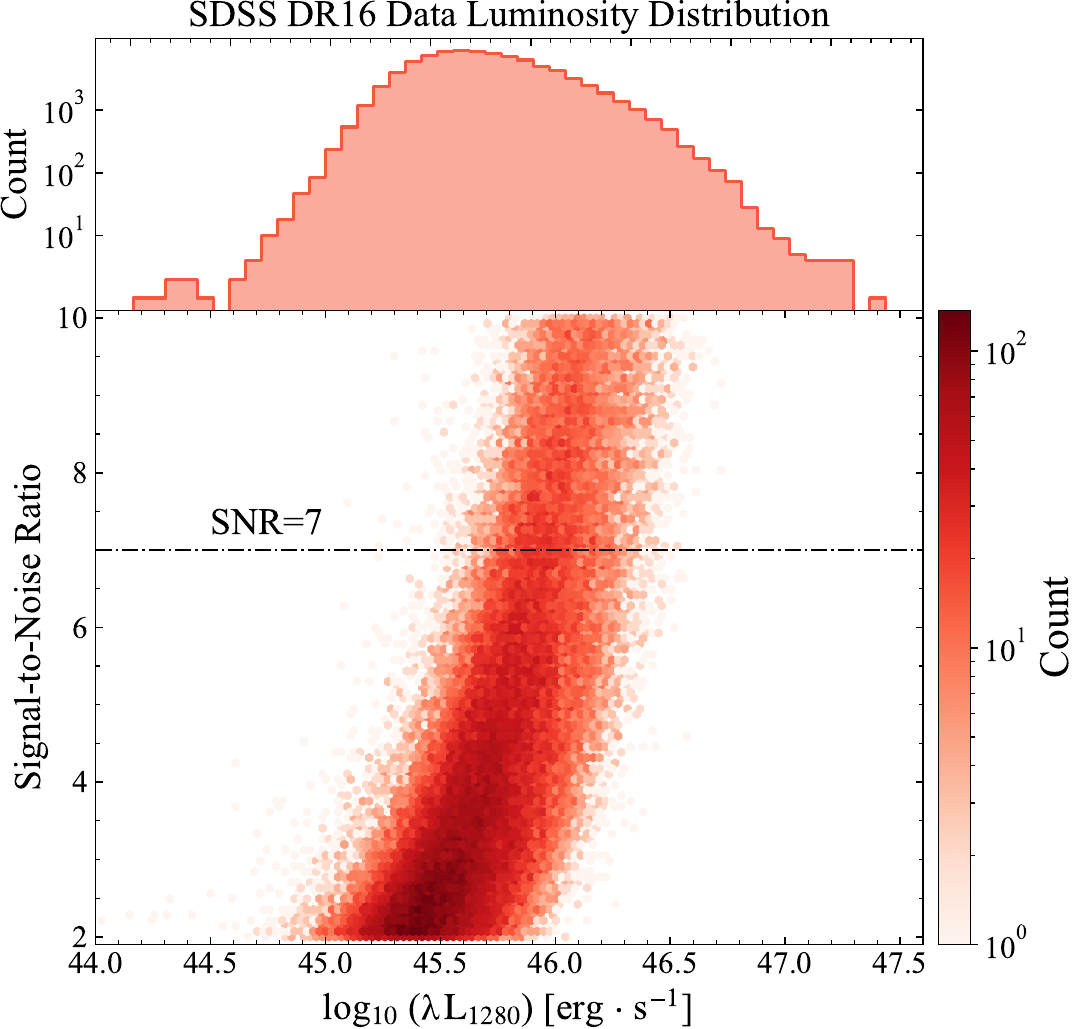}
    \caption{The luminosity distribution of the SDSS DR16 dataset. The bottom panel shows the dependency between the luminosity of quasars and the SNR, while the top panel gives the marginal distribution of quasar luminosity. the unsupervised nature of QFA enables us to incorporate low SNR quasar spectra into the training set, leading to a more expanded luminosity coverage ($10^{44}\,\mathrm{erg}\cdot \mathrm{s}^{-1}\sim 10^{47.5}\,\mathrm{erg}\cdot \mathrm{s}^{-1}$) in this study. The $\mathrm{SNR}>7$ selection criteria has been commonly adopted in previous work \cite[e.g., ][]{PCA2018}.}
    \label{fig:Lum}
\end{figure}

Unlike some of the existing methods, our unsupervised learning algorithm not only harnesses information from high SNR spectra (e.g., SNR$\gtrsim 10$, $\sim 10^2$ spectra in \citet[][]{PARIS2011}, and SNR$\geq 7$, $\sim 10^4$ spectra in \citet[][]{PCA2018}), but also low SNR quasar spectra (SNR$\geq 2$ in this study) without the need for continuum labels. As shown in Figure~\ref{fig:Lum}, taking into account of the low SNR quasar spectra enable us to cover a more extensive luminosity range\footnote{We calculate the monochromatic luminosity at rest-frame wavelength $1280\pm 2\,\mathrm{\AA}$.} than before (see figure~2 in \citet[][]{MFPCA2012}), and admit a much larger training dataset -- of the order of $10^5$ as opposed to $\sim 10^2-10^4$ in the supervised learning methods \cite[][]{PCA2005, PARIS2011, MFPCA2012, PCA2018, NF2020, IQNET2021}. 

We transform the observed quasar spectra to the rest frame and re-normalize them by dividing the whole spectra with the median value of the flux at the rest-frame wavelength $\mathrm{1280\,\AA}$. We perform piece-wise sigma-clipping ($3\,\sigma$) to reject erroneous absorption lines and pixels and interpolate the flux to conform each spectrum onto a uniform rest-frame wavelength grid spanning from $\mathrm{1030\,\AA}$ to $\mathrm{1600\,\AA}$ with a logarithmic uniform spacing. For piece-wise sigma-clipping, we adopt an adaptive filter window, adjusted based on the existence of the emission features: a 30-pixel window size ($\sim 7-11\,\mathrm{\AA}$ in rest frame) is used in the ``smooth" region devoid of emission peaks, and while a 10-pixel window size applies around emission peaks ($\sim 2-4\,\mathrm{\AA}$ in rest frame). The wavelength range covers the Ly$\alpha$ forest regions but not the Ly$\beta$ forest. As for quasar spectra whose rest-frame wavelength does not include the wavelength range of $\mathrm{1030\,\AA}$ to $\mathrm{1600\,\AA}$, we consider those regions to be masked.

\begin{figure*}[t]
    \epsscale{1.2}
    \plotone{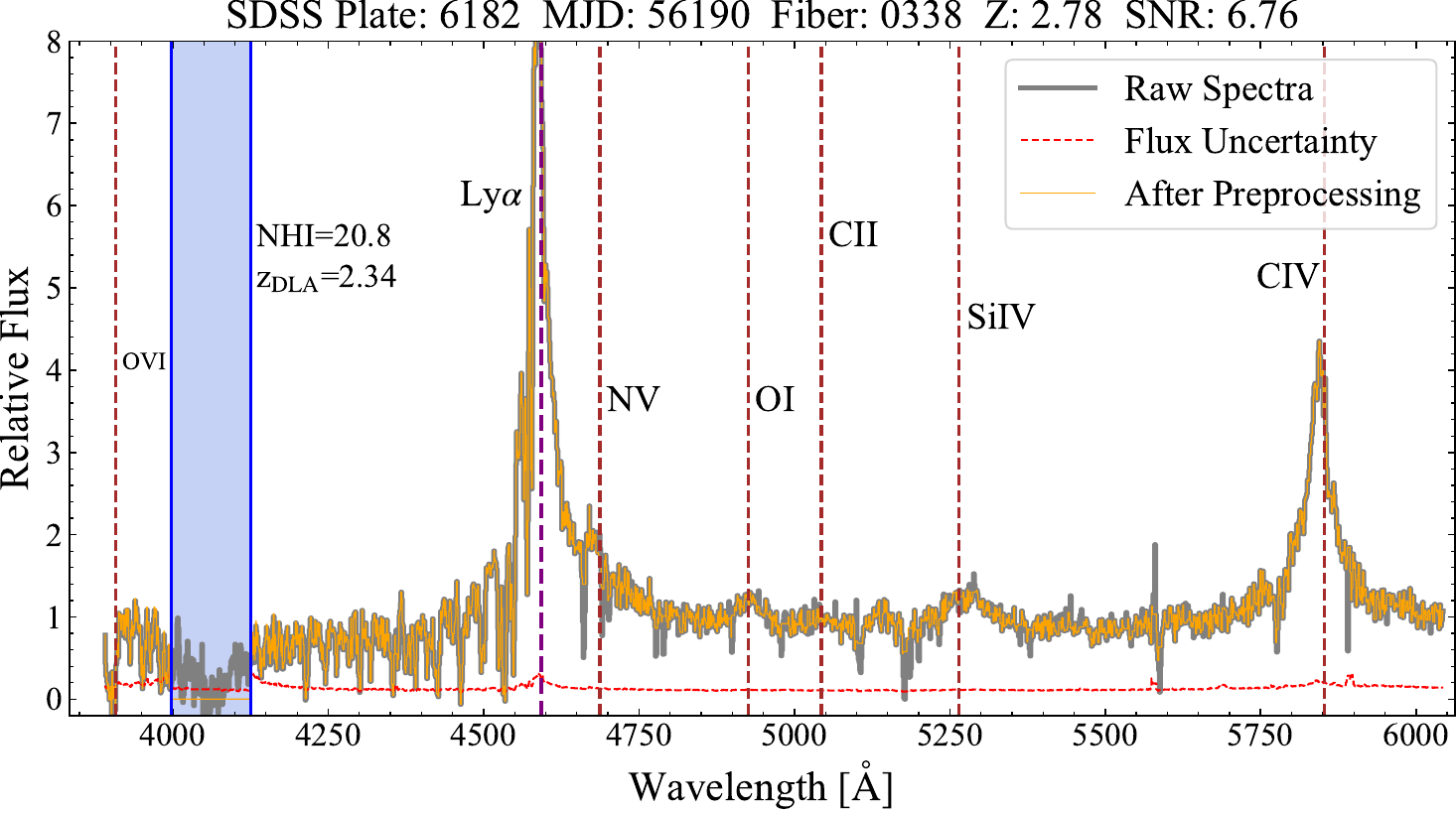}
    \caption{Data preprocessing for the SDSS DR16 spectra. The quasar spectrum after preprocessing is shown in orange, and the original quasar spectrum from SDSS is in grey. The raw and preprocessed quasar spectra are normalized at rest-frame wavelength $\mathrm{1280\,\AA}$. We reject absorption lines in the red side and noisy pixels by sigma clipping. The damped Ly$\alpha$ systems are masked, as shown in the blue shaded region, and the damping wing beyond two times the equivalent width is subsequently corrected following prescription in \citet{MFPCA2012}. \label{fig:preprocessing}}
\end{figure*}

Finally, as we only model the Ly$\alpha$ forest as in Section~\ref{sec:method}, the physical priors of our model do not include the damped Ly$\alpha$ systems (DLAs). DLAs \cite[][]{DLAS2005} are the population of strong absorbers with integrated neutral hydrogen (HI) column density $\mathrm{N}_{\mathrm{HI}}\geq 2\times10^{20}\,\mathrm{cm}^{-2}$, resulting in a broad absorption region ($\Delta v\sim 10^3$ $\mathrm{km/s}$). The broad absorption features bias our inferences and ought to be masked. Fortunately, various techniques have been developed to give a robust estimation of the DLA regions \cite[e.g.,][]{DAVID2018, GP2017, GP2020, WANG2022}. In this work, we adopt the DLA classification results in the SDSS DR16 quasar catalog \cite[][]{SDSS2020}, which include the central wavelength and column density information for each DLA. We mask the DLA regions within two times the equivalent width of each DLA \cite[][]{DRAINE2011} and further perform damping wing correction as depicted in \cite{MFPCA2012}. In Figure \ref{fig:preprocessing}, we show an example of our data preprocessing. The absorption systems on the red side of the quasar spectrum are filtered by sigma-clipping ($3\,\sigma$). We also mask the DLA region, shown in blue. 

\vspace{12pt}
\subsection{Mock Spectra}\label{subsec: mock}

Complementing our study of the SDSS data, we will also assume a set of mock quasar spectra of which the ground truth continua are known. For the mock continua, we assume the PCA templates from \citet{PARIS2011}. We adopt them as the basis quasar templates and fit our SDSS DR16 dataset. The fits yield a set of quasar continua given by PCA continuum fitting. We then draw continua from the fitted data set. 

As for the transmission fields, we adopt the mock transmission fields from the SDSS DR11 quasar-Ly$\alpha$ forest mock data sets \cite[][]{SDSSMOCK2015}\footnote{ \hyperlink{http://www.sdss.org/dr12/algorithms/lyman-alpha-mocks}{http://www.sdss.org/dr12/algorithms/lyman-alpha-mocks.}}. These mock transmission fields were well examined to mimic the actual non-Gaussian transmission fields and served as the baseline calibrator for the main BOSS Ly$\alpha$ Baryon Acoustic Oscillations (BAO) measurements \cite[][]{BAO2013, BAO2014, PICCA2020}. To test that our methods can robustly deal with high column density absorbers, we further inject to the mock spectra the high column density absorbers, including damped Ly$\alpha$ systems and Lyman limit systems in the SDSS DR11 quasar-Ly$\alpha$ forest mock datasets. 

\begin{figure*}
    \epsscale{1.2}
    \plotone{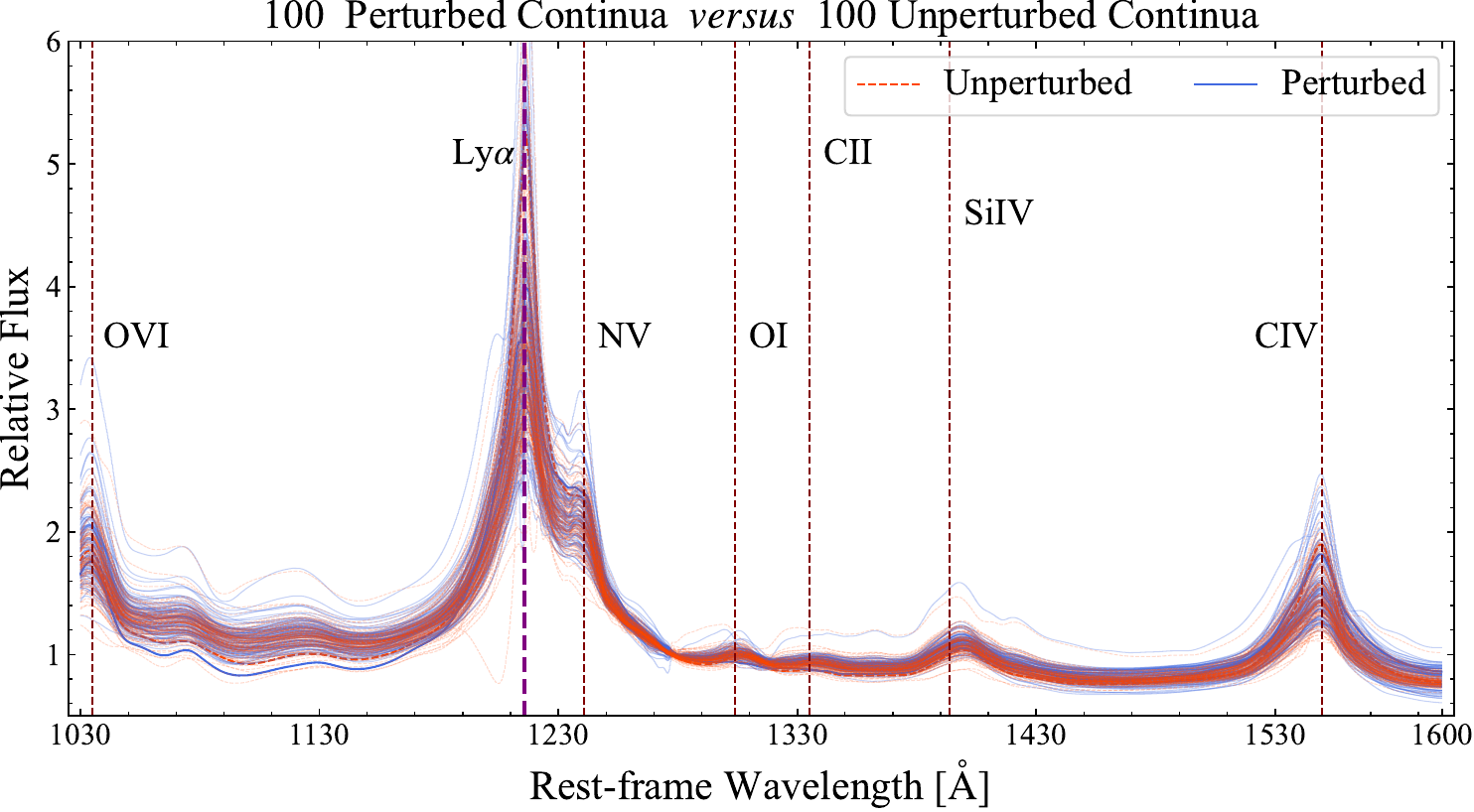}
    \caption{Mock quasar continua adopted in this study. The unperturbed continua are generated based on the PCA method \citep{PARIS2011}, fitted with the SDSS DR16 data. The perturbed continua further assume a perturbation of $10\%$ (see text for details). The perturbation allows for a broader diversity of quasar continua beyond the PCA basis. The perturbed mock data serves as a more stringent test for the continua inference algorithm, testing their ability to generalize beyond the high-quality SNR spectra from which the PCA basis was derived.\label{fig:perturb}}
\end{figure*}

As mentioned in Section~\ref{sec:intro}, the PCA basis from \citet{PARIS2011} is constructed based on a limited number of ($78$, SNR$\gtrsim 10$) quasar spectra. However, in practice, a PCA basis may poorly represent the real-world continuum distribution, potentially yielding extremely poor performance for spectra outside the PCA space. To quantitatively assess each method's robustness, we apply minor linear perturbations to the simulated quasar continua. More specifically, we assume a linear perturbation:
\begin{equation}
    \mathrm{C}_\mathrm{perturb}(\lambda) = \mathrm{C}_\mathrm{PCA}(\lambda) + \mathrm{M_0+M_1 \left(\frac{\lambda/\AA-1030}{1600-1030}\right)},
    \label{eq:perturb}
\end{equation}

\noindent
where $\mathrm{M}_\mathrm{0}$ is drawn from a uniform distribution from $-0.1$ to $0.1$, $\mathrm{M}_\mathrm{1}$ is drawn from a uniform distribution from $-0.2$ to $0.2$ and $\mathrm{C}_\mathrm{PCA}$ denotes the PCA constructed continua.
The perturbation given by Equation~(\ref{eq:perturb}) is no more than $10\%$ at rest-frame wavelength $1280\,\mathrm{\AA}$. We re-normalize each continuum after the perturbation to ensure that all spectra have the same relative flux. Such a perturbed spectrum can be viewed as a weighted combination of the linear perturbation and the original spectrum. As the random perturbation is constrained to be sufficiently minor, the original and perturbed distributions should appear visually similar. Figure~\ref{fig:perturb} demonstrates 100 unperturbed continua and 100 perturbed continua; perturbed continua have similar distribution as the unperturbed continua. While the linear perturbation term we introduce may seem arbitrarily simplistic and unrepresentative of the actual continuum distribution, we employ it solely as a proof-of-concept experiment to evaluate each method's robustness. As we show in Section~\ref{subsec:testOnMock}, even minor perturbations cause existing supervised techniques to fail in generalizing, unlike QFA. Moreover, Section~\ref{subsec:sdss} indicates real quasar spectra likely demonstrate more complex variations than our linear perturbation. In that event, the advantage of QFA becomes more prominent. 

Finally, we inject observational noise that mimics the one from SDSS DR16 spectra. We divide our SDSS DR16 dataset into five redshift bins from $\mathrm{z}=2$ to $3.5$ with a redshift interval of $0.3$. We randomly draw Gaussian random noise for each mock spectrum according to the SDSS DR16 observational uncertainty from the corresponding redshift bins. The SNR and redshift distribution of the mock spectra are shown in Figure~\ref{fig:snrRedshift}. The SNR and redshift distribution of our mock spectra follow the one from the SDSS DR16 dataset; testing on these realistic mock quasar spectra thus allows us to demonstrate the validity of our statistical assumption introduced in Section~\ref{sec:method} and yield a reliable quantification for the continuum recovery performance of different models. 

\section{Results}\label{sec: result}

In the following, we will demonstrate the performance of QFA for quasar continuum inference, compared to two representative existing continuum fitting methods: the PCA continuum fitting method from \citet{PARIS2011} and \texttt{PICCA} continuum fitting from \citet[][]{PICCA2020}. 
It should be noted that there exist various variants of PCA-based techniques that differ in the size of the training data and the mapping methods between the red side and the blue side. However, most are not publicly available, and assessing them is beyond the scope of this study. Furthermore, as demonstrated by \citet{COMPARE2021}, their performance is comparable ($\lesssim3\%$ difference) and they are subject to similar limitations as \citet{PARIS2011} when using a biased training set. Therefore, comparing against \citet{PARIS2011} shall not impact our conclusions significantly. We specifically compare these two representative methods (PCA and \texttt{PICCA}) because they best describe the state-of-the-art for the supervised learning methods (PCA) and the unsupervised learning methods (\texttt{PICCA}) and have been widely-demonstrated in practice. We leave the discussion about other existing methods to Section~\ref{subsec:other}. We note here that PCA only utilizes the red-side pixels to predict the blue-side pixels, whereas \texttt{PICCA} exclusively utilizes the blue-side pixels. In contrast, QFA encompasses both the blue-side and red-side pixels for prediction.

\subsection{A quantitative assessment of QFA compared to other existing methods}\label{subsec:testOnMock}

\begin{figure*}
\epsscale{1.3}
\centering
\gridline{
\fig{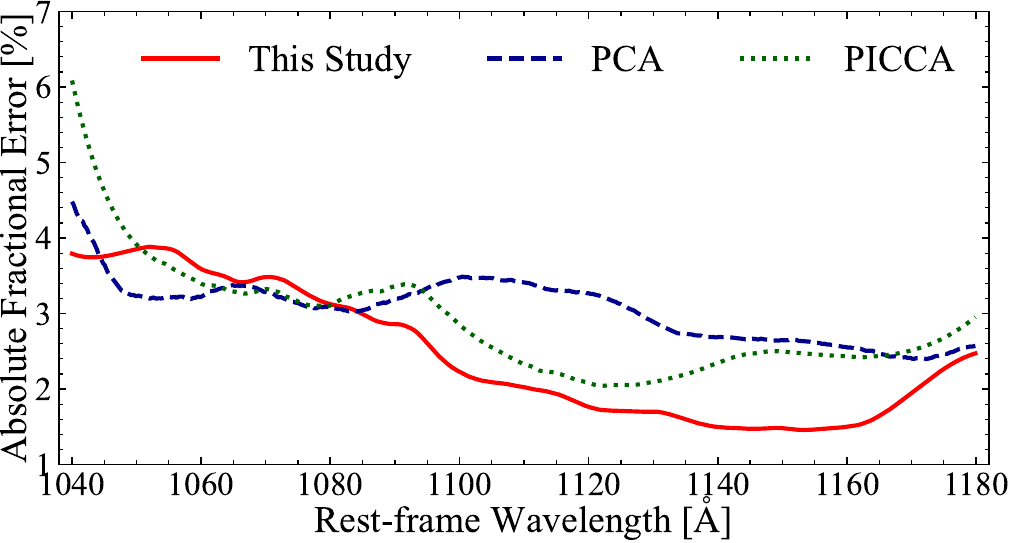}{0.5\textwidth}{(a) Continuum prediction error on the unperturbed dataset}
\fig{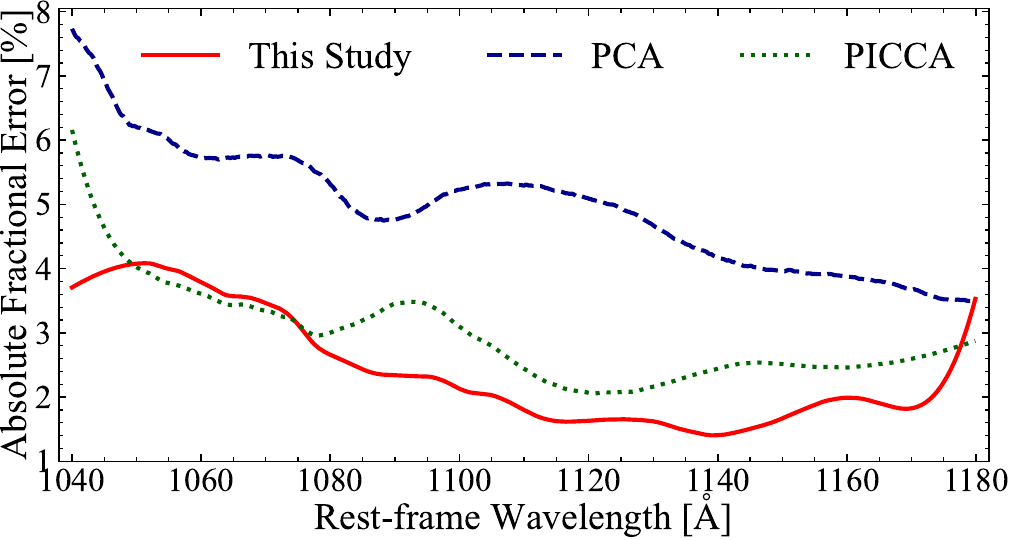}{0.5\textwidth}{(b) Continuum prediction error on the perturbed dataset}
}
\caption{Continuum prediction error at different wavelengths. Shown are the median errors at individual wavelength pixels. The left figure shows the model performance on the unperturbed mock spectra, and the right figure shows the perturbed mock spectra. QFA yields an AFFE of $\sim 2\%$ at $1040-1180\,\mathrm{\AA}$. PCA achieves similar prediction quality on the unperturbed spectra  $\sim 3\%$. But for the perturbed dataset, PCA performance degrades to $\sim 5\%$. \texttt{PICCA} is less susceptible to the perturbation, but its performance is worse than QFA, with an average AFFE of $\sim 3\%$ in both datasets. Both PCA and QFA incur larger errors towards the bluer parts, while \texttt{PICCA} performs worse on both the redder and bluer ends. 
\label{fig:model_performance_at_different_wavelength}}
\end{figure*}

First, we compare our model performance on the mock dataset from which the ground-truth continua are known. Throughout this paper, we will consider the following metrics to quantify the performance of QFA. We define the absolute fractional error of continuum prediction as
\begin{equation}
    \text{Absolute Fractional Error} = \left|\frac{\mathrm{C}_\text{pred}(\lambda)}{\mathrm{C}_\text{truth}(\lambda)}-1\right|\times 100 \%
\end{equation}

\noindent
to compare model performance at different wavelengths. We further quantify the overall performance by integrating over the wavelength. We define the absolute fractional flux error (AFFE) $\mathrm{\left|\delta C\right|}$ as in \citet{IQNET2021}:
\begin{equation}
    \left|\mathrm{\delta C}\right| = \int_{\lambda_1}^{\lambda_2}\left|\frac{\mathrm{C}_\mathrm{ pred}(\lambda)-\mathrm{C}_\mathrm{truth}(\lambda)}{\mathrm{C}_\mathrm{truth}(\lambda)}\right|\,\mathrm{d}\lambda /\int_{\lambda_1}^{\lambda_2}\,\mathrm{d}\lambda\times 100\%.
    \label{eq:AFFE}
\end{equation}
Briefly, AFFE measures, for individual spectrum, the average absolute error in the wavelength region of interest. Besides, we define the absolute fractional flux bias (AFFB) $\mathrm{\delta C}$ as:
\begin{equation}
    \mathrm{\delta C} = \int_{\lambda_1}^{\lambda_2}\frac{\mathrm{C}_\mathrm{pred}(\lambda)-\mathrm{C}_\mathrm{truth}(\lambda)}{\mathrm{C}_\mathrm{truth}(\lambda)}\,\mathrm{d}\lambda/\int_{\lambda_1}^{\lambda_2}\,\mathrm{d}\lambda\times 100\%
    \label{eq:AFFB}
\end{equation}
to better quantify the mean absolute bias of continuum fitting result. Compared with AFFE in Equation~\ref{eq:AFFE}, AFFB performs as an extra metric to evaluate the difference between the ground truth continua and fitted continua.

We also focus only on evaluating the performance at wavelength bluer than the Ly$\alpha$ emission (for simplicity, ``blue side"), i.e., from rest-frame wavelength $\mathrm{1040\,\AA}$ to $\mathrm{1180\,\AA}$, which covers most parts of the Ly$\alpha$ forest but exclude the effect of Ly$\beta$ forest \cite[e.g.,][]{JINYI2020} and proximity zone \cite[e.g.,][]{FAN2006, CARILLI2010}. This is a conservative assessment for QFA, as both PCA and \texttt{PICCA} are mainly designed for recovering quasar continua in the blue side. The red-side continuum prediction for PCA is not optimized in practice. In contrast, \texttt{PICCA} does not support red-side continuum prediction. Although not shown here, QFA outperforms PCA on the red side of the quasar spectra (see Appendix~\ref{appendix:red}).

\begin{table}[t]
\centering
\begin{tabular}{lrrr|rrr}
\hline\hline
&\multicolumn{3}{c}{Unperturbed}&\multicolumn{3}{c}{Perturbed}\\
 \hline
 AFFE [\%]& 5th& 50th & 95th  & 5th &  50th & 95th\\
 \hline
QFA {\it vs.} Truth &1.29&\textbf{2.47}&\textbf{4.82}&1.17&\textbf{2.52}&\textbf{5.02}\\
PCA {\it vs.} Truth&\textbf{0.67}&3.04&9.81&\textbf{0.99}&4.91&12.9\\
\texttt{PICCA} {\it vs.} Truth &1.26&3.13&6.65&1.32&3.20&6.72\\
\hline
\end{tabular}
\caption{Model performance on extracting quasar continua with different methodologies. We refer to Section~\ref{subsec: mock} for the definition of ``unperturbed" and ``perturbed" mock datasets. The results are evaluated over 10,000 mock spectra. The 5th, 50th, and 95th percentiles indicate the best, general, and worst performance of the different methods over the 10,000 mock spectra.
\label{tab:model_performance}}
\end{table}

\begin{table}[htbp!]
    \centering
    \begin{tabular}{lrrr|rrr}
    \hline\hline
&\multicolumn{3}{c}{Unperturbed}&\multicolumn{3}{c}{Perturbed}\\
 \hline
 AFFB [\%]& 25th& 50th & 75th  & 25th &  50th & 75th\\
 \hline
QFA {\it vs.} Truth &\textbf{-1.93}&1.56&\textbf{4.65}&\textbf{-2.15}&1.26&\textbf{4.67}\\
PCA {\it vs.} Truth&-8.25&\textbf{0.29}&7.35&-10.4&\textbf{1.06}&11.6\\
\texttt{PICCA} {\it vs.} Truth &-2.80&1.38&5.90&-3.05&1.38&5.95\\
\hline
    \end{tabular}
    \caption{Similar to Table~\ref{tab:model_performance}, we present the absolute fractional flux bias here. }
    \label{tab:AFFB}
\end{table}

Figure~\ref{fig:model_performance_at_different_wavelength} shows the absolute fractional error as a function of wavelength, evaluating the mock data set of $10,000$ spectra. QFA yields more accurate and robust continuum predictions than PCA and \texttt{PICCA} on the region of interest in the blue side regardless of the perturbation. The quality of the predictions from all three methods shows some dependency with respect to the wavelength, but for different reasons. For PCA, which learns the blue-side continua from the red-side information, the larger relative error is because the correlation between the bluer pixels to the red continua is less prominent. We also see the same limitation, as shown in Figure~\ref{fig:highz}. For \texttt{PICCA}, the fitted polynomial correction\footnote{A $k\,$th degree polynomial correction in \texttt{PICCA} is defined as $\sum_{i=0}^k a_k\Lambda^k$, in which $a_0,\dots,a_k$ are free parameters and $\Lambda = \log\lambda$ is the wavelength in log space. We adopt first order in this study as in \citet[][]{PICCA2020}.} can not fully describe the variations between the mean continuum and the continuum for individual spectrum, thus underfitting the continuum. The under-fitting affects not only the bluer end but also, the redder end.

For QFA, the reason for the more considerable uncertainty towards the bluer parts is two-fold. (a) The larger observational noise level in the bluer parts will inflate the prediction uncertainty, as shown in Figure~\ref{fig:compare}. (b) As we fixed the mean optical depth function as physical prior (Equation~(\ref{eq:optical_depth})), the inconsistency between the pre-defined and ground-truth mean optical depth assumption may cause QFA to perform worse toward the bluer end, a limitation of our current model. We will return to this in Section~\ref{subsec:cal}.

Table~\ref{tab:model_performance} summarizes the overall performance, integrating over the wavelength. As shown, the continuum predictions from QFA are more accurate and remain robust even with perturbations on the mock continua. QFA achieves an accuracy of $\sim 2\%$ in AFFE in both mock datasets, outperforming PCA and \texttt{PICCA}. PCA performs the best for the top five percentile of spectra. But we note that this is somewhat misleading because the mock continua are generated from the PCA template. Thus, it is unsurprising that PCA achieves the best performance in these limited cases when the PCA template perfectly matches the mock spectra. Nonetheless, even for the unperturbed case, QFA performs better generally. Importantly, as we perturb the continua, the recovery of the PCA method degrades from $\sim 3\%$ AFFE to $\sim 5\%$ AFFE, but QFA remains resilient to the perturbation. 

The unsupervised nature of \texttt{PICCA} makes it, similar to QFA, maintain similar performance on both datasets. Nonetheless, its predictions are worse than QFA because its rigid parametric assumption, based on a first-order polynomial correction. \texttt{PICCA} attains a $\sim 3\%$ in AFFE for both datasets. Among all three methods, QFA shows the smallest scatter, which is $\sim 1\%$ AFFE for the best cases and $\sim 5\%$ AFFE for the worst cases. PCA, in contrast, shows the largest scatter from $\lesssim 1\%$ AFFE to $\gtrsim 10\%$ AFFE. Although \texttt{PICCA} only performs worse than QFA at the $\lesssim 1\%$ level, its performance suffers from a larger scatter on a case-by-case basis, manifesting that the rigid polynomial correction performs subpar than QFA.

\begin{figure*}
\gridline{
\fig{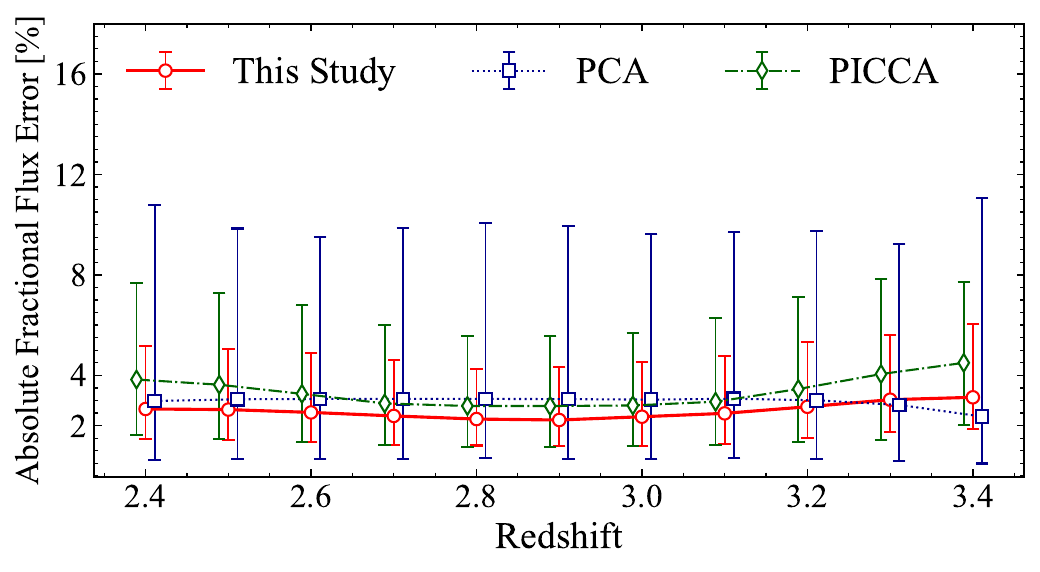}{0.50\textwidth}{(a) AFFE as a function of Redshift on the unperturbed dataset\label{fig:no-perturb-redshift}}
\fig{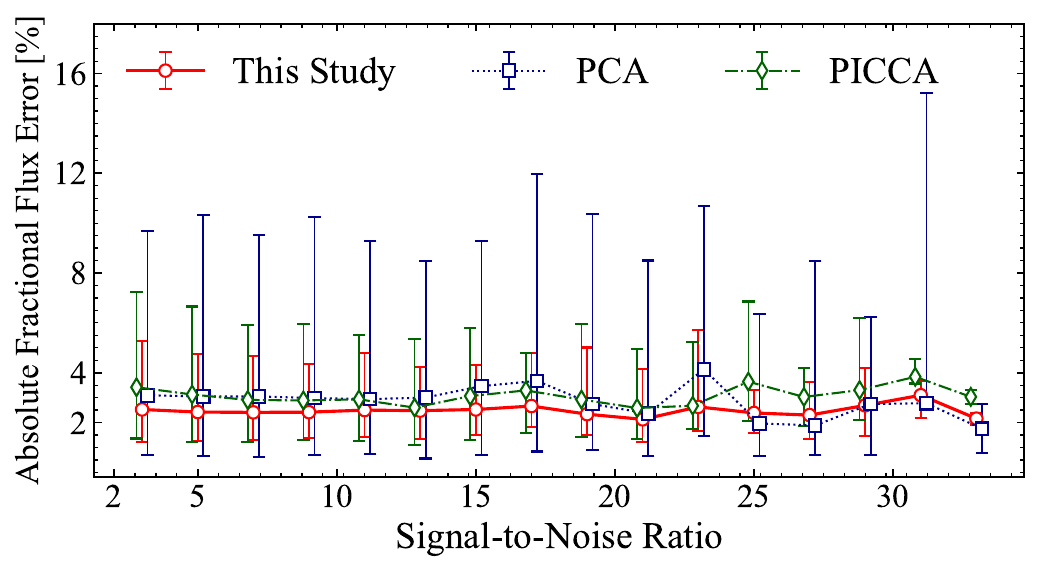}{0.50\textwidth}{(b) AFFE as a function of SNR on the unperturbed dataset\label{fig:no-perturb-SNR}}}
\gridline{
\fig{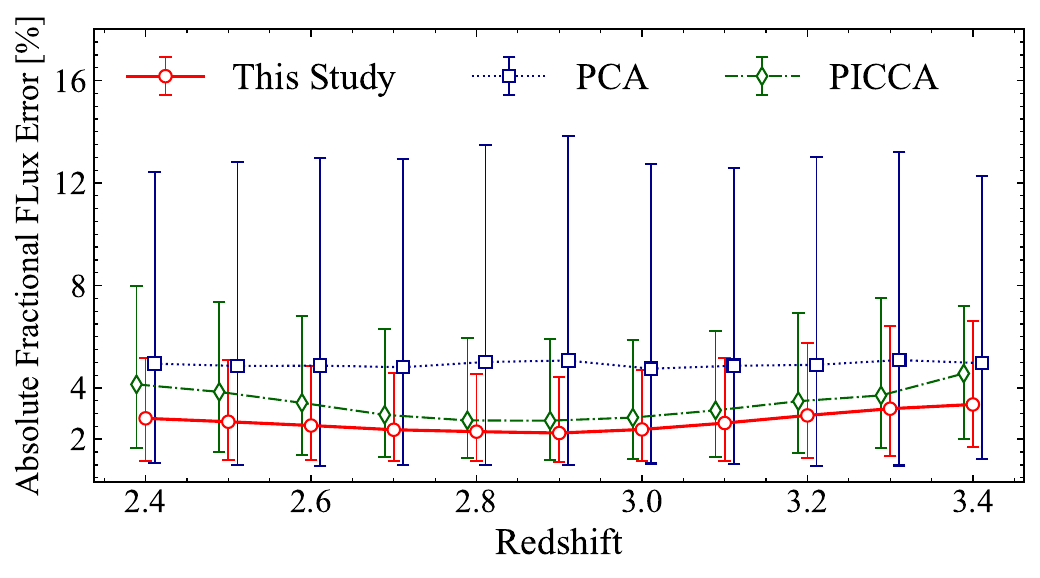}{0.50\textwidth}{(c) AFFE as a function of Redshift on the perturbed dataset\label{fig:with-perturb-redshift}}
\fig{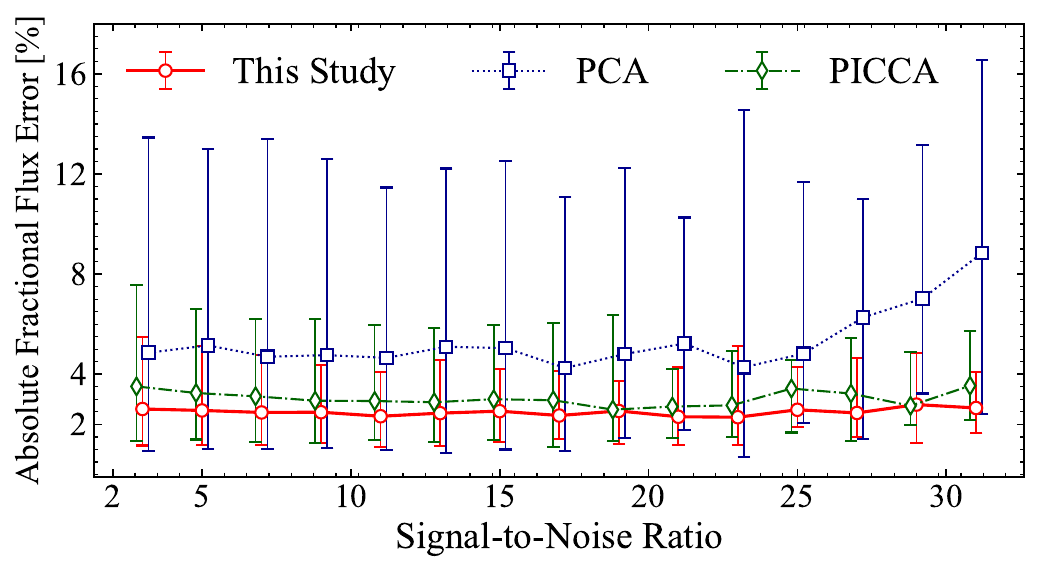}{0.50\textwidth}{(d) AFFE as a function of SNR on the perturbed dataset\label{fig:with-perturb-SNR}}}
\caption{Continuum prediction error as a function of redshift and SNR. Shown is the median value of AFFE for different redshifts and SNRs. The error bars denote the 95th and 5th percentile over the tested mock spectra. The upper panels show model performance on the unperturbed dataset and the lower panels on the perturbed dataset. The performance of all three methods -- QFA, PCA, and \texttt{PICCA} -- demonstrate little dependency on redshift and SNR. QFA outperforms PCA and \texttt{PICCA}; it has the lowest AFFE and the smallest scatter in all redshift and SNR bins, regardless of the perturbation.
 \label{fig:model_performance_evolution}}
\end{figure*}

Finally, a desired continuum fitting algorithm should yield robust predictions regardless of the redshift or SNR of the quasar spectra. In Figure~\ref{fig:model_performance_evolution}, we further evaluate the model performance as a function of redshift and SNR. Both PCA and QFA show little dependency on the redshift. \texttt{PICCA} shows a slightly stronger redshift dependence, which may be caused by the simultaneous fitting of both the mean optical depth function and quasar continua in \texttt{PICCA} as well as the redshift dependent large-scale variance introduced in \texttt{PICCA} (see equation~(4) in \citet[][]{PICCA2020}). Since all three methods take into account the observational noise, their performance shows little-to-no evolution with SNR. As before, compared to PCA and \texttt{PICCA}, QFA gives the smallest scatter compared to the existing methods.

\subsection{Application to SDSS quasar spectra}\label{subsec:visual}

\begin{figure*}[t]
\gridline{
\fig{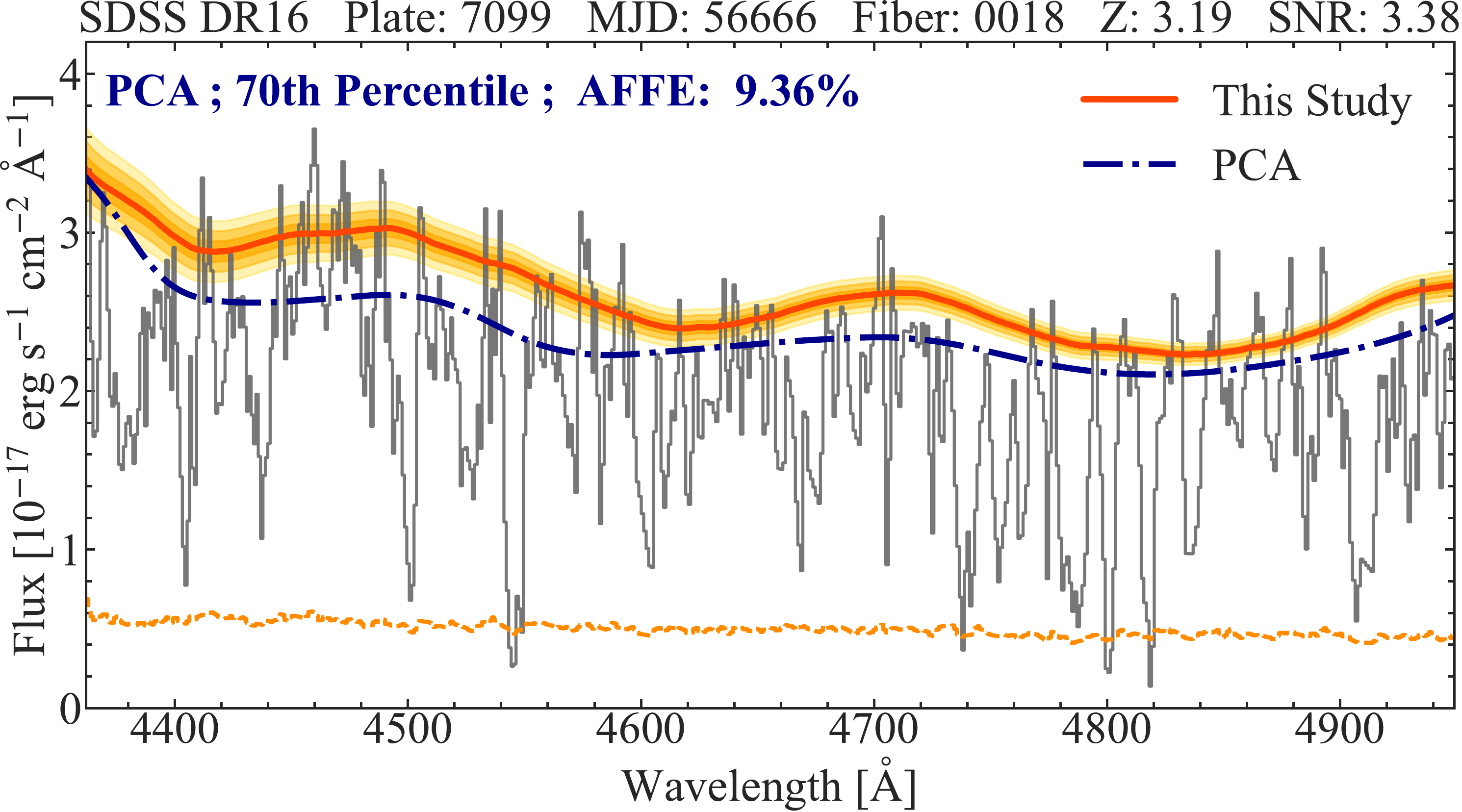}{0.5\textwidth}{(a) QFA {\it vs.} PCA 70th percentile ; AFFE: 9.36\%\label{fig:pca-70}}
\fig{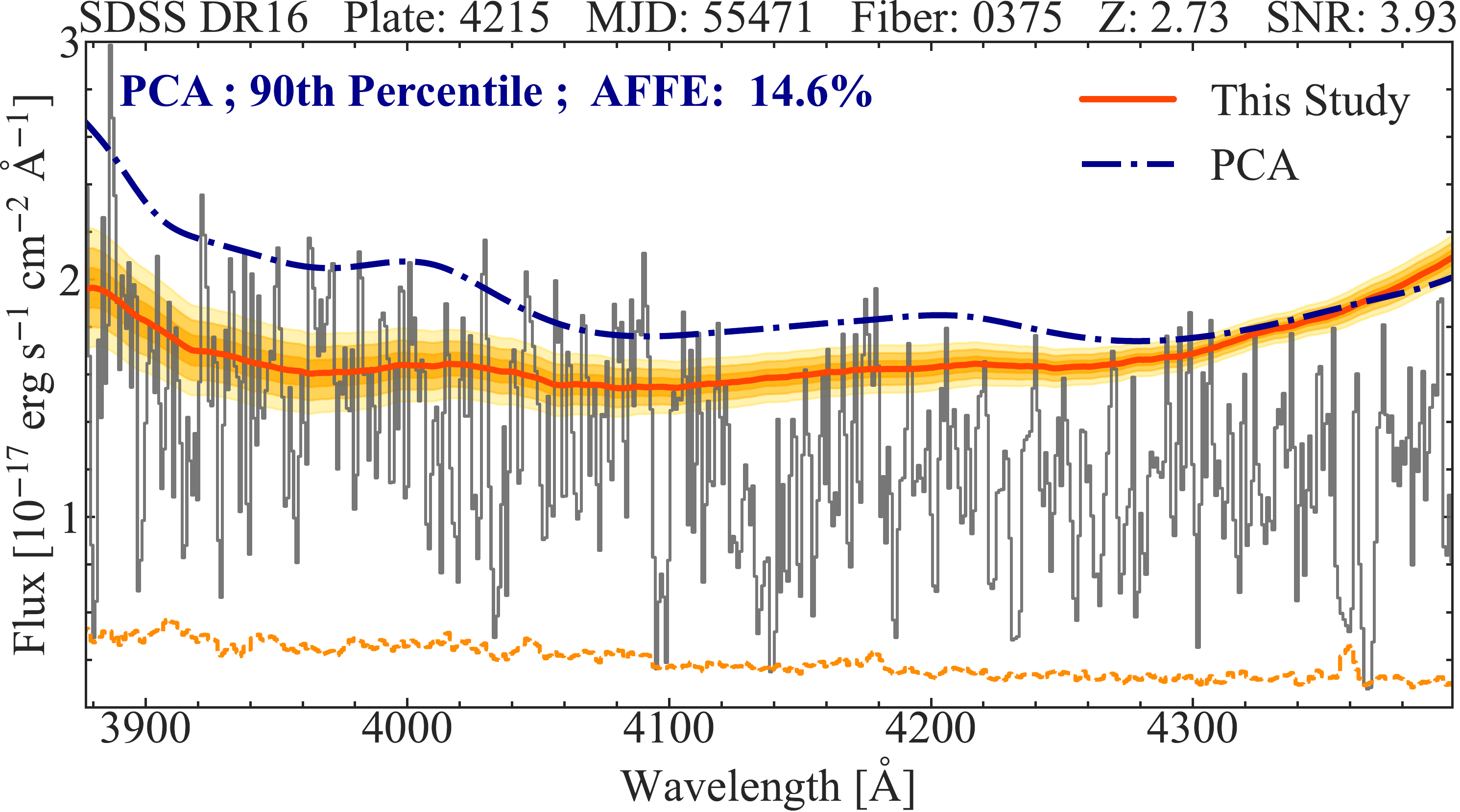}{0.5\textwidth}{(b) QFA {\it vs.} PCA 90th percentile ; AFFE: 14.6\%\label{fig:pca-90}}}
\gridline{
\fig{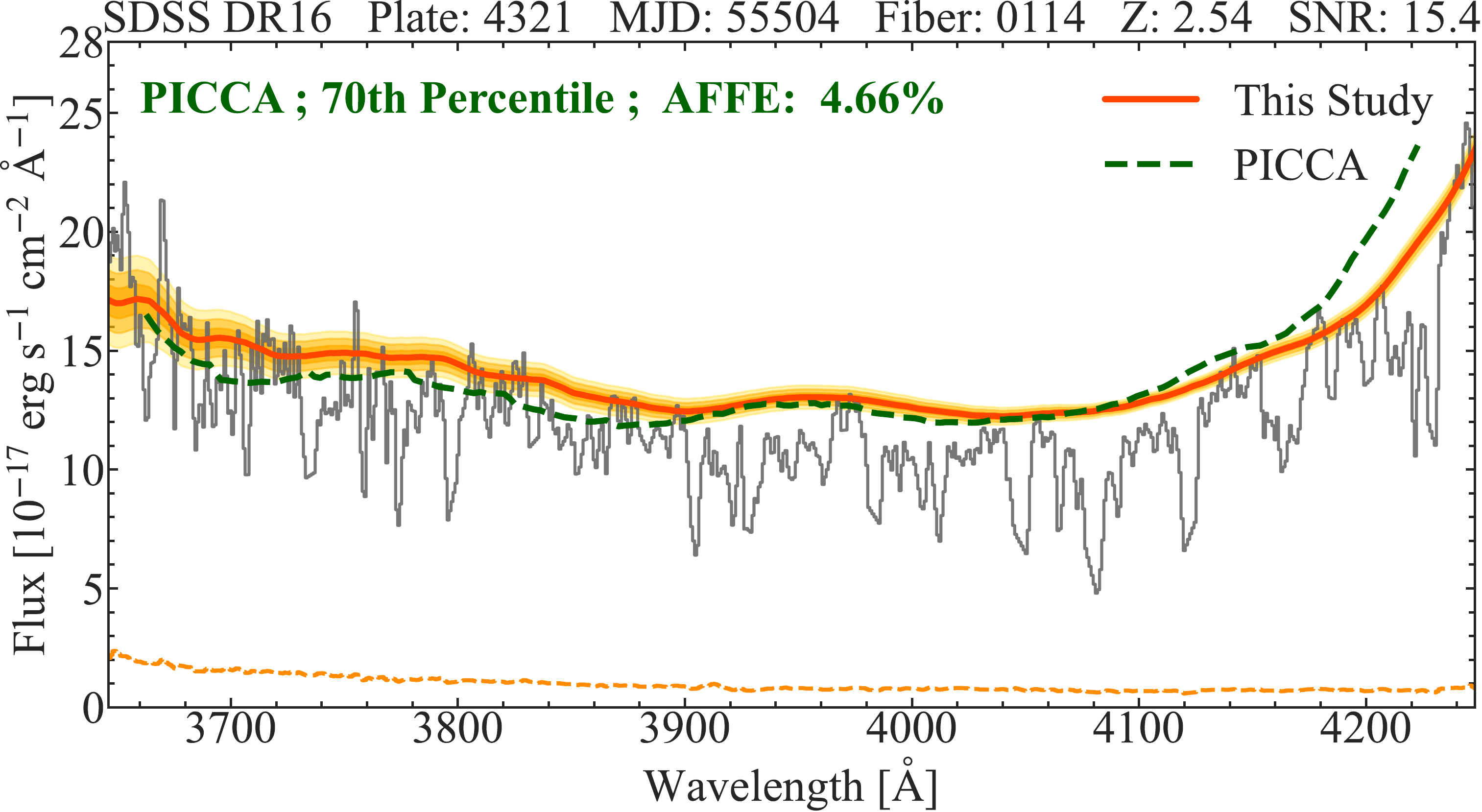}{0.5\textwidth}{(c) QFA {\it vs.} \texttt{PICCA} 70th percentile ;
AFFE: 4.66\%\label{fig:picca-70}}
\fig{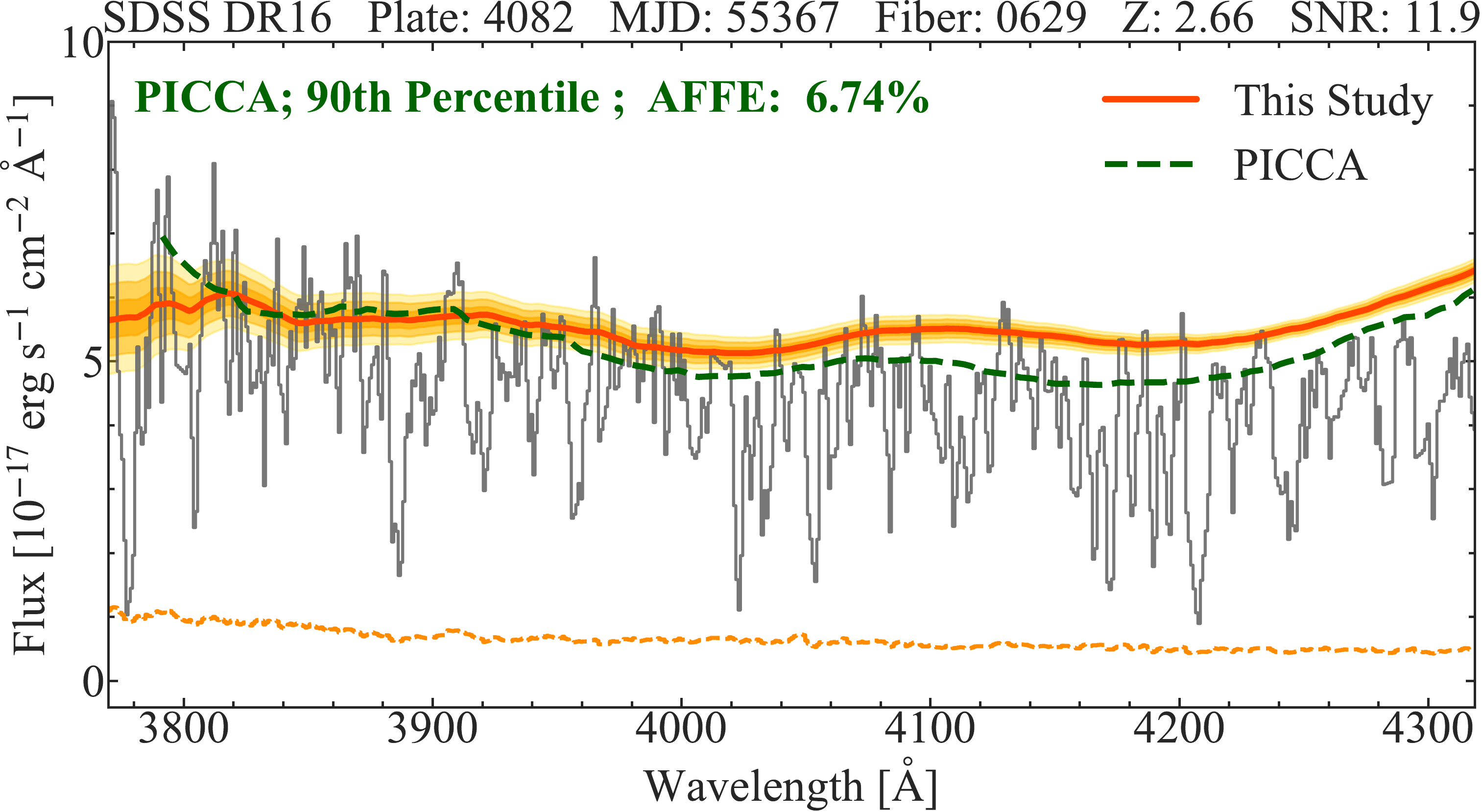}{0.5\textwidth}{(d) QFA {\it vs.} \texttt{PICCA} 90th percentile ; AFFE: 6.74\%\label{fig:picca-90}}}
\caption{Model performance on the SDSS DR16 dataset. We investigate the continuum predictions that show the most difference between PCA (top panels)/\texttt{PICCA} (right panels) and QFA. The left panels show the deviation at the 70th percentile (among the $37,548$ SDSS spectra), and the right panels at the 90th percentile. PCA underestimates or overestimates the quasar continuum in these outlying cases, and continuum predictions given by \texttt{PICCA} tends to diverge towards both ends.
\label{fig:compare}}
\end{figure*}

Besides mock spectra, we apply QFA to the SDSS DR16 data set. We train a separate QFA model. Since we do not know the ground truth continua for the SDSS spectra, we include 10,000 additional mock quasar spectra (without perturbation) in our training. These additional mock spectra, of which the ground truth continua are known, are included to gauge the convergence of our models. We found that the median AFFE estimated on the mock data set in the Ly$\alpha$ region attains a precision of $2.36\%$, a performance on par with the case when the training set only consists of mock spectra. It demonstrates that an extension of the quasar distribution with the actual spectra does not adversely affect QFA. Therefore, we expect our model to achieve about the same accuracy in the SDSS DR16 dataset as in the auxiliary mock data set in our training.

\begin{table}[htbp!]
\centering
\begin{tabular}{lrrr|rrr}
\hline\hline
&\multicolumn{3}{c}{2\textless{}SNR\textless{}5}&\multicolumn{3}{c}{SNR\textgreater{}5}\\
 \hline
AFFE [\%] & 25th & 50th & 75th & 25th & 50th & 75th\\
 \hline
PCA {\it vs.} QFA &3.76&6.53&10.4&3.31&5.93&9.51\\
\texttt{PICCA} {\it vs.} QFA &2.63&3.84&5.38&2.34&3.34&4.68\\
\hline
\end{tabular}
\caption{The difference in continua predictions for the SDSS DR16 quasar spectra, comparing PCA and \texttt{PICCA} with QFA. PCA and \texttt{PICCA} predict noticeably different continua than QFA.\label{tab:sdss}}
\end{table}

As for the actual SDSS spectra, since the ground truth is not directly accessible, we resort to evaluating the difference between the continuum estimates from QFA and the ones from PCA and \texttt{PICCA}. The differences are summarized in Table~\ref{tab:sdss}. As shown, there can be substantial deviations between the predictions from QFA and other methods. The predictions and QFA and PCA predictions can differ as much as $\gtrsim 6\%$ for $\sim 50\%$ quasar spectra, and $\gtrsim 10\%$ for $25\%$ quasar spectra, regardless of the SNR. Similarly, \texttt{PICCA} deviates $\gtrsim 4\%$ for $50\%$ quasar spectra and $\gtrsim 5\%$ for $25\%$ quasar spectra.  These non-negligible discrepancies beg the question of which methods infer the continua more accurately. 

In Figure~\ref{fig:compare}, we inspect the predictions from these different methods for SDSS spectra that show the most significant discrepancies. We show spectra that deviate at the 70th percentile level (of all the SDSS spectra, and on the right, at the 90th percentile level. As shown in the figure, while the ground truth continua are unknown, visual inspections suggest that the QFA estimates are more physically plausible. The inference from the PCA method tends to either overshoot or undershoot the observed spectra. \texttt{PICCA}, on the other hand, tends to diverge on the redder wavelength, which might be caused by the under-fitting of the polynomial correction in \texttt{PICCA}.

These non-negligible systematics of PCA and \texttt{PICCA} underline the importance of further advancing unsupervised continuum inference algorithms with QFA. As accurate continua are instrumental in constructing the Ly$\alpha$ forest, these systematic differences might lead to errors in cosmology estimations via the Ly$\alpha$ forest, which we will explore next.

\subsection{Ly$\alpha$ forest power spectrum measurements}\label{subsec:pk1d}

\begin{figure*}
    \gridline{
    \fig{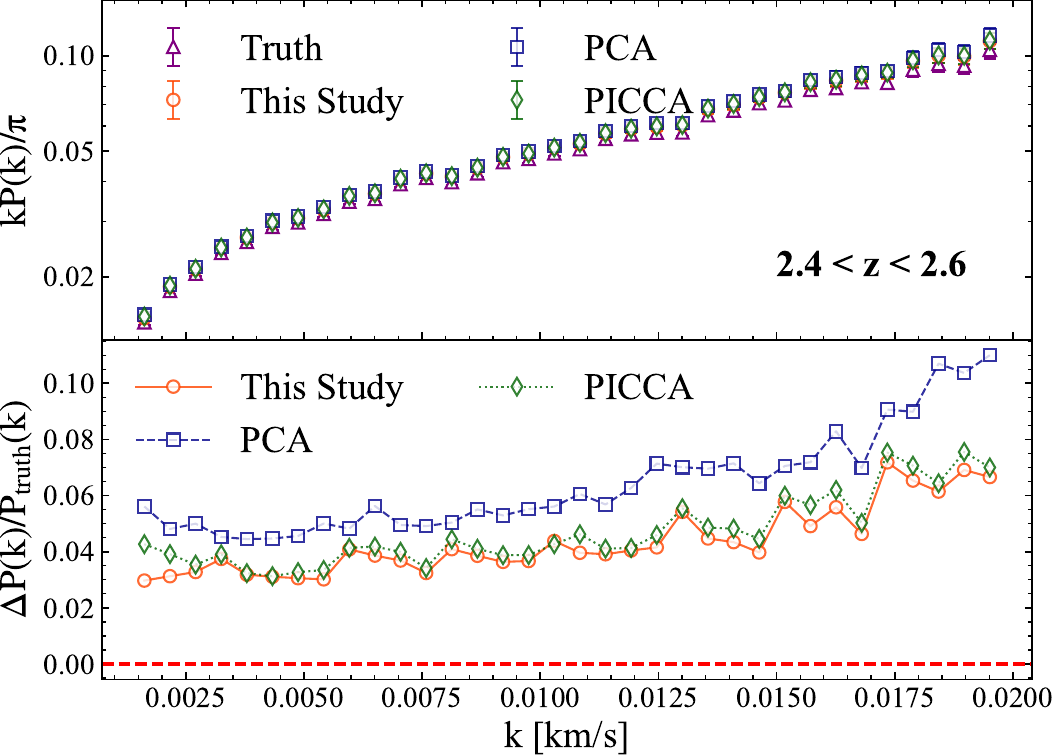}{0.5\textwidth}{(a) $2.4<\mathrm{z}<2.6$\label{fig:p1da}}
    \fig{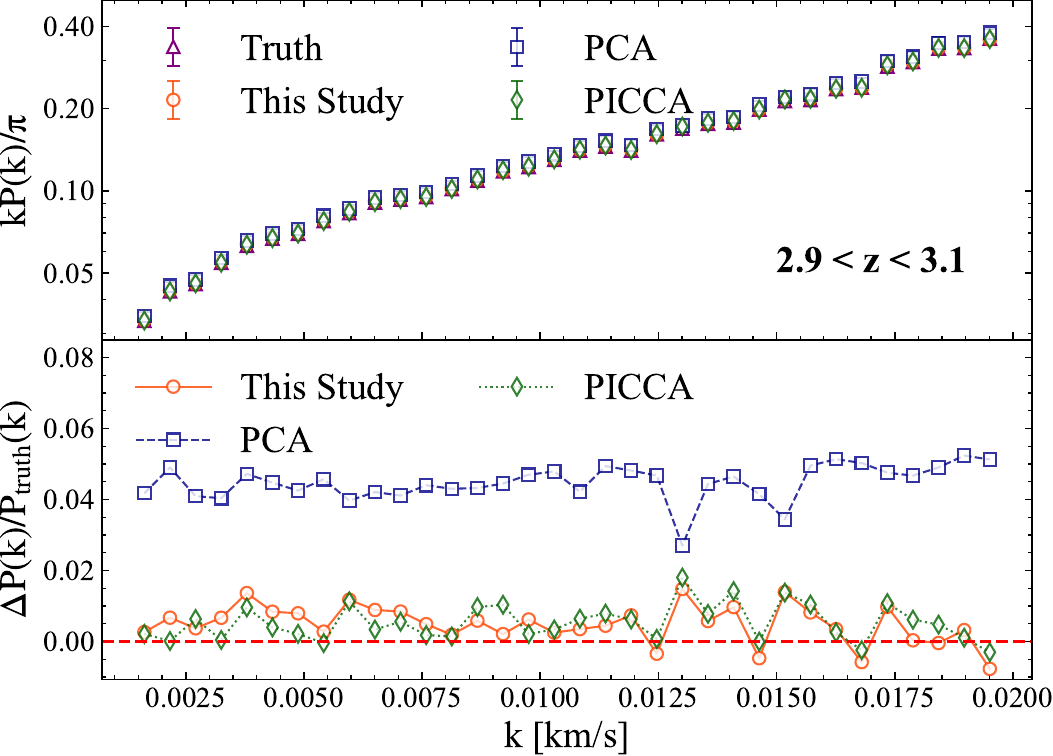}{0.5\textwidth}{(b) $2.9<\mathrm{z}<3.1$\label{fig:p1db}}
    }
    \caption{The fractional error in Ly$\alpha$ power spectrum measurements. The upper panels show the measured power spectrum, while the bottom panels display the relative deviation. The imperfect continuum reconstruction from the PCA leads to a larger systematic bias in the Ly$\alpha$ power spectrum compared to QFA. At redshift $2.3<\mathrm{z}<2.5$, QFA incurs a relative deviation of $\sim 4\%$, and PCA yields a poorer power spectrum estimate, with an error of $\sim 7\%$. At the redshift range of $2.9<\mathrm{z}<3.1$, QFA's estimate deviate at the level $\lesssim 1\%$, whereas PCA's estimate incurs an error of $\sim 5\%$. The recovery of the Ly$\alpha$ power spectrum correlates with the continuum prediction at different wavelengths (see Figure~\ref{fig:model_performance_at_different_wavelength} and the main text for details). The imperfect continuum recovery from QFA at the bluest wavelength causes a systematic bias in power spectrum measurement at the smaller redshift, albeit better than PCA. Compared to \texttt{PICCA}, QFA improves the 1D Ly$\alpha$ power spectrum measurements for $\sim 0.5\%$ at $2.4<\mathrm{z}<2.6$ and gives comparable result at $2.9<\mathrm{z}<3.1$.
    \label{fig:pk1d}}
\end{figure*}

The one-dimensional power spectrum $\mathrm{P_{1d}(k)}$ of Ly$\alpha$ forest \cite[e.g.,][]{1DPOWERSPECTRUM2013, 1DPOWERSPECTRUM2019, 1DPOWERSPECTRUM2020} is the tell-tale sign of the distribution of intergalactic medium in the distant universe. The accurate quantification of the Ly$\alpha$ forest power spectrum holds the key to many different sciences, including the thermal evolution of intergalactic medium \cite[e.g.,][]{GAIKWAD2021}, the neutrino masses in our universe \cite[e.g.,][]{YECHE2017, DARKRADIATION2017}, the dark radiation \cite[e.g.,][]{ROSSI2015} and the nature of dark matter \cite[e.g.,][]{VID2017, GARZILLI2019}. The power spectrum of Ly$\alpha$ forest has much-renewed interest thanks to the ongoing and upcoming large-scale spectroscopic surveys, such as DESI \cite[][]{DESIROADMAP2022}, 4MOST \cite[][]{4MOST2019} and WEAVE \cite[][]{WEAVE2016}. However, any continuum residual, as demonstrated in the previous section, can cause a non-negligible effect on the measurements of the power spectrum. 

In the following, we will evaluate how well QFA can extract the Ly$\alpha$ forest power spectrum as opposed to other existing methods. We will evaluate our performance, qualitatively on mock datasets described in Section~\ref{subsec: mock}. The transmission field of mock spectra translates into the ground truth power spectrum, which can then be compared with the recovered power spectrum from the different continuum extraction methods.

For a given spectrum with flux $\mathrm{S}$, continuum $\mathrm{C}$ and mean optical depth function $\tau_\mathrm{eff}(\mathrm{z}_\mathrm{abs})$, the flux-transmission field can be evaluated as 
\begin{equation}
    \delta = \frac{\mathrm{S}}{\mathrm{C}\circ\exp(-\tau_\mathrm{eff}(z_\mathrm{abs}))}-1.
    \label{eq:delta}
\end{equation}

\noindent
To calculate the Ly$\alpha$ forest power spectrum, we then plug the flux-transmission field into the \texttt{PICCA} pipeline \footnote{\hyperlink{https://github.com/igmhub/picca}{https://github.com/igmhub/picca}, Version: v4.2.0 } \citep{PICCA2020}. Briefly, the \texttt{PICCA} pipeline takes the flux-transmission fields as input and calculates the raw power spectrum of each transmission field through fast Fourier Transform (FFT). The resolution effect, metal absorbers, observational noise, and other systematic errors from the data pipeline are taken into account to recover the underlying Ly$\alpha$ forest power spectrum. The final Ly$\alpha$ forest power spectrum is the ensemble average over those forest spectra in the corresponding redshift bin. We refer interested readers to \citet[][]{1DPOWERSPECTRUM2019} for the details of the Ly$\alpha$ forest power spectrum calculation. In practice, we calculate the transmission fields and, subsequently, the estimated power spectra with both the ground truth continua and the estimated continua. The difference between the two tells us how much the imperfect continuum predictions from different algorithms imprint on the power spectrum estimate. In the following, we denote the ground truth power spectrum as $\mathrm{P_{truth}(k)}$, and the estimated one as $\mathrm{P(k)}$.

Figure~\ref{fig:pk1d} shows the measured Ly$\alpha$ forest power spectrum, comparing QFA with PCA and \texttt{PICCA}. We randomly select 5,000 mock spectra from the perturbed mock dataset and measure the 1D Ly$\alpha$ forest power spectrum over two redshift bins. At $2.3<\mathrm{z}_\mathrm{abs}<2.5$, QFA recovers the Ly$\alpha$ forest power spectrum to about $\sim 4\%$, while PCA incurs an error of $\sim 6-10\%$. At $2.9<\mathrm{z}_\mathrm{abs}<3.1$, QFA predicts a close-to-perfect Ly$\alpha$ forest power spectrum measurement, with an error of $\lesssim 1\%$, whereas PCA maintains an error of $\sim 5\%$. \texttt{PICCA}, as the state-of-the-art continuum fitting algorithm for the 1D Ly$\alpha$ power spectrum measurements \cite[e.g.,][]{1DPOWERSPECTRUM2019}, gives slightly worse performance ($\sim 4.72\%$ relative error) compared to QFA ($\sim 4.37\%$ relative error) at $2.4<\mathrm{z}<2.6$, and achieve comparable results ($\lesssim 1\%$ relative error) at $2.9<\mathrm{z}<3.1$. To thoroughly examine the performance of different models on the unperturbed dataset, we present the 1D Ly$\alpha$ forest power spectrum measurements obtained using various continuum fitting methods in Appendix~\ref{appendix:p1d}. It should be noted that the increasing error in the 1D Ly$\alpha$ forest power spectrum towards smaller scales in Figure~\ref{fig:pk1d} is likely due to the bias in that specific redshift range, which arises from the intrinsic degeneracy between the mean optical depth function and quasar continua. We reported this bias in Section~\ref{subsec:testOnMock} and will further discuss the dependency of our model on the mean optical depth in Section~\ref{subsec:cal}. Additionally, we present an ablation study of the mean optical depth in Appendix~\ref{ablation:opt}.

The difference in performance can be understood intuitively based on how well individual algorithms recover the quasar continua at different wavelengths. The error for the Ly$\alpha$ forest measurements is on par with the continuum fitting error. For instance, at the wavelength range of $1040\,\mathrm{\AA}$ to $1100\,\mathrm{\AA}$, in which most of the $2.4<\mathrm{z}_\mathrm{abs}<2.6$ Ly$\alpha$ forest absorbers reside, QFA recovers the continuum at the $4\%$ level, and PCA $7\%$ (see Figure~\ref{fig:compare}). This translates into the same error in terms of power spectrum measurements at the corresponding redshift of $2.3<\mathrm{z}_\mathrm{abs}<2.5$. Similarly, for redshift bin $2.9<\mathrm{z}_\mathrm{abs}<3.1$, the majority of Ly$\alpha$ forest absorbers resides in at $1100\,\mathrm{\AA}$ to $1180\,\mathrm{\AA}$, the Ly$\alpha$ forest power spectrum measurement error with QFA is $\lesssim 1\%$, consistent with the continuum fitting results from $1100\,\mathrm{\AA}$ to $1180\,\mathrm{\AA}$ (see Figure~\ref{fig:compare}). In contrast, PCA incurs an $\sim 5\%$ fractional error. As discussed in Section~\ref{subsec:testOnMock}, both the mean optical depth prior and the low SNR in the bluer regions might contribute to the QFA larger continuum prediction error at the bluer end. 

For \texttt{PICCA}, as its average performance only differs from QFA at the level of $1\%$ on the mock dataset, it achieves almost the same precision as QFA for 1D Ly$\alpha$ forest power spectrum measurements, which is $\sim 4\%$ at $2.4<\mathrm{z}<2.6$ and $\lesssim 1\%$ at $2.9<\mathrm{z}<3.1$. Our results are consistent with figure~7 in \citet[][]{1DPOWERSPECTRUM2019}, which found that \texttt{PICCA} introduces $\sim 2-4\%$ relative error in the Ly$\alpha$ forest power spectrum measurements on the mock spectra. However, we caution that mock datasets might have simplified the question at hand. As shown in Table~\ref{tab:sdss}, \texttt{PICCA} and QFA still give noticeably different continuum predictions at the level of $\gtrsim 3\%$ for $\sim 50\%$ SDSS quasar spectra, and even $\gtrsim 5\%$ for $\sim 25\%$ SDSS quasar spectra, far larger than in the mock datasets, which is about $\lesssim 3\%$ for $\sim 75\%$ mock quasar spectra (see Section~\ref{subsec:testOnMock}). As such, a detailed comparison between QFA and \texttt{PICCA} on real-life data is needed to resolve this issue. But this is beyond the scope of this paper, and we will leave it to future work.

\subsection{Quasar Outliers in SDSS}\label{subsec:outlier}

\begin{figure*}[t]
    \epsscale{1.2}
    \plotone{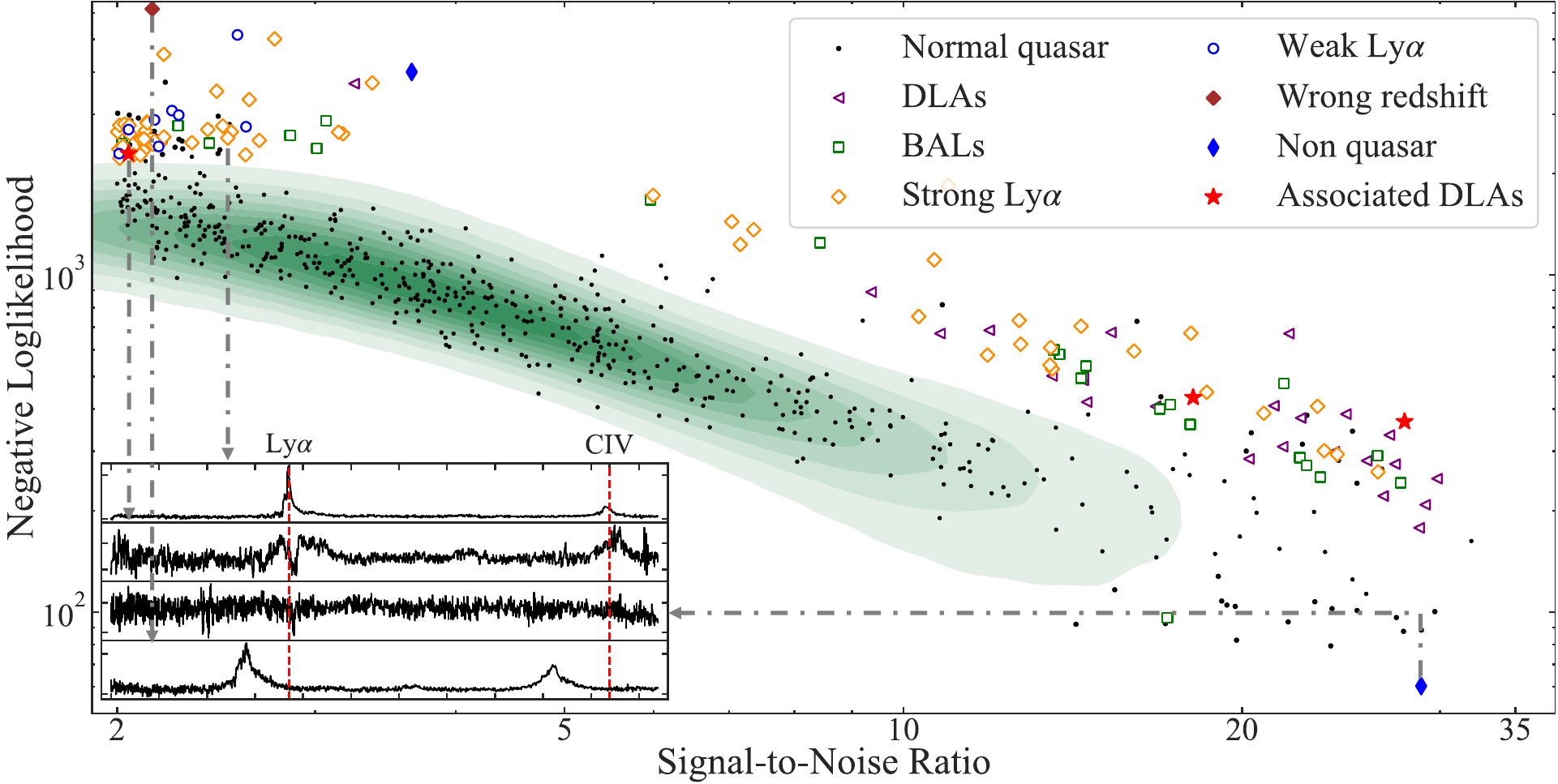}
    \caption{Outlier detection with QFA. The background contour shows the likelihood evaluated with QFA of the $37,548$ SDSS DR16 spectra. The colored symbols show outliers objects with low probability likelihood. We visually inspect these outliers and classify them into different classes, as shown in the legend. The inset displays some examples of these outliers (see also Appendix~\ref{appendix:outlier}). The black symbols within the contours are the values from 500 non-outlier SDSS quasars for illustration. \label{fig:outlier}}
\end{figure*}

Massive datasets from modern-day large-scale spectroscopic surveys are bound to find unexpected interesting objects and demand systematic searches of such outliers. As QFA summarizes the ensemble of observed quasars into a probabilistic distribution, it provides a robust way to perform outlier detection. More specifically, given the QFA model, with the optimal model parameters $\mathcal{M}^*$, the likelihood $\mathcal{L}\left(\mathrm{S}|\lambda, \mathrm{z}, \sigma_\epsilon,\mathcal{M}^*\right)$ of individual spectrum $\left(\mathrm{\lambda}, \mathrm{S}, \mathrm{z}, \sigma_\epsilon\right)$ can we evaluated according to Equation~(\ref{eq:single_ll}). The likelihood value indicates the probability of occurrence for each spectrum. A smaller likelihood value implies that the spectrum in question deviates from the majority, hence an outlier.

As a proof of concept, we evaluate the likelihood (Equation~(\ref{eq:single_ll})) for $37,548$ quasar spectra without masked pixels from the SDSS DR16 dataset. Although QFA can deal with missing pixels (see Section~\ref{subsec: pred}), we do not consider quasar spectra in the current outlier search because the marginalized likelihood has a different absolute scale than the likelihood with the full spectrum. We will defer the detailed investigations of outliers from the full SDSS catalog to future studies. 

Figure~\ref{fig:outlier} shows the density contours of the likelihood at different SNRs. Note that quasar spectra with high SNR tend to have more concentrated probability density functions and hence a higher likelihood value. In comparison, quasar spectra with low SNR tend to have more dispersed probability density distribution functions and, therefore, a lower likelihood value. Therefore, a robust outlier search must consider the SNR difference between different quasar spectra. We apply the K-Nearest Neighbor (KNN) outlier detection algorithm in \texttt{PyOD} \cite[][]{PYOD2019}. The algorithm identifies outliers by sorting all data points according to the mean distance between each data point and its nearest $k$ neighbors. As such, each spectrum is only compared with its nearest $k$ neighbors with similar SNR and likelihood. We investigate the top one percentile outliers under this metric, or $179$ outliers from $37,548$ quasar spectra.

As shown in the inset plots in Figure~\ref{fig:outlier}, further visual inspection confirms that the $179$ outlier spectra show some unexpected spectral features. These outliers include (a) undetected damped Ly$\alpha$ absorbers (DLAs); (b) associated damped Ly$\alpha$ absorbers (associated DLAs); (c) broad absorption lines (BALs); (d) Type II quasar feature -- overly strong Ly$\alpha$ emission but weak continuum; (e) erroneous redshift estimation; (f) misclassified non-quasar spectra. We provide more details of these outliers in Appendix~\ref{appendix:outlier}.

\subsection{The Evolution of the Quasar Population}\label{subsec:evolution}

Recall that QFA decomposes quasar continua into a lower dimensional embedding with a finite number of ``factors" (see Section~\ref{sec:method}, we assume $8$ factors in this study). Importantly, unlike previous studies of PCA \cite[][]{PCA2005, PARIS2011, PCA2018}, which basis is derived from a limited number of high-quality quasar spectra, QFA can make use of all SDSS DR16 quasar spectra. This allows us to study, in detail, the evolution of the quasar population as a function of their luminosity and redshift, which we will explore in this section.

The latent factors in QFA are defined up to a rotation (see Section~\ref{subsec: fac}). To ensure that the basis is physically motivated, we adopt the varimax rotation \cite[][]{VARIMAX1958}\footnote{Varimax can also be applied to PCA components, but technically, it would no longer be PCA since the basis is not orthogonal.}. Varimax determines the rotation by maximizing the sum of variances of the squared loadings\footnote{Squared loading denotes the element-wise product of the factor loading matrix $F$ in Equation~(\ref{eq: cont_model})}. In other words, varimax seeks a decomposition that decorrelates the emission features and flat continuum. Compared to the orthogonal basis enforced by PCA-based methods, the non-orthogonal basis of QFA offers more flexibility, allowing for a more physically meaningful decomposition of quasar continua. As a result, varimax enhances model interpretability compared to figure~7 of \citet[][]{PARIS2011}. As shown in the middle panel, the first component reflects the strength of the Ly$\alpha$ emission, the second panel recovers a power-law feature of quasar continua, and the third panel demonstrates the contribution from the CIV emission. We focus only on the three most notable factors and leave the other components to Appendix~\ref{appendix:com}. We note that our learned decomposition is consistent to those derived from modern-PCA methods based on larger training samples. Both methods give a power-law component in addition to components with strong features corresponding to correlations between the strengths and shifts of a wide variety of broad emission lines \cite[e.g., ][]{PCA2018}. However, QFA differs from PCA in two aspects: (a) QFA learns its components directly from millions of quasar spectra, wheras PCA components are derived from ad hoc quasar continua fitted by other automated algorithms ; (b) the probabilistic description of QFA (Equation~\ref{eq: cont_model}) enables more flexibility, such as non-orthogonality, of the learned components.

\begin{figure*}[t]
    \epsscale{1.23}
    \plotone{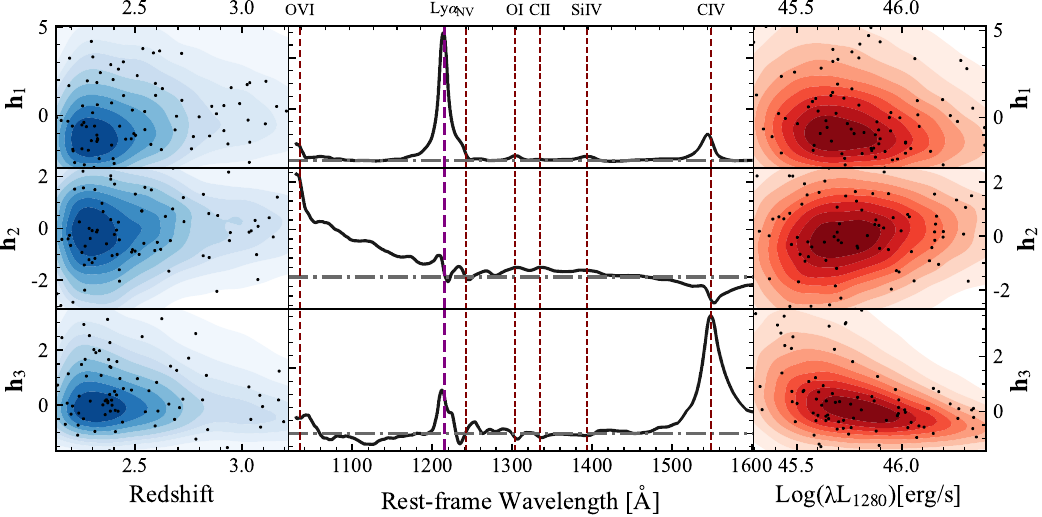}
    \caption{Evolution of the quasar population. We study the evolution of the most prominent factors in QFA. The middle panels show the corresponding factor loading as a function of rest-frame wavelength. These three notable factors signify the mechanisms that cause the Ly$\alpha$ emission, the power-law shape of the quasar spectrum, and the CIV emission, respectively. The left and right panels illustrate how the factors vary as a function of redshift and luminosity. The quasar population does not exhibit any detectable evolution from $\mathrm{z}=2$ to $\mathrm{z}=3.5$. The bottom panels show that fainter quasars have a stronger CIV emission, which is consistent with the Baldwin effect. The Pearson correlation coefficient $\mathrm{r}$ in each panel is determined with less than $1\%$ uncertainty through Monte-Carlo sampling.\label{fig:com}}
\end{figure*}

Figure~\ref{fig:com} further demonstrates the evolution of the latent factors ($\mathbf{h}$ in Equation~(\ref{eq: cont_model})) as a function of redshift (left panels) and luminosity (right panels). We evaluate the correlation with Pearson correlation coefficient $\mathrm{r}$. The uncertainty of each correlation coefficient is estimated through Monte-Carlo sampling, and we find $\lesssim 1\%$ uncertainties of the correlation coefficients for all $8$ factors. The left panels show that these three factors (as well as the other factors in Appendix~\ref{appendix:com}) do not exhibit any visible correlation with the redshift. The lack of dependency with redshift demonstrates that the quasar population has not evolved much from $1.83\,\mathrm{Gyr}$ ($\mathrm{z}=3.5$) to $3.33\,\mathrm{Gyr}$ ($\mathrm{z}=2$). Corollary, even trained on moderate-redshift ($2<\mathrm{z}<3.5$) quasar spectra, QFA might be able to infer high-redshift (e.g., $\mathrm{z}>5$) quasar continua robustly, which we will detail in Section~\ref{subsec:highz}. 

The right panels illustrate that the factors contributing to the Ly$\alpha$ emission and the power-law factor do not correlate with the monochromatic luminosity of the quasar. Interestingly, the CIV emission factor is the only exception -- it exhibits an unmistakable negative correlation ($\mathrm{r}\sim -0.44\pm 0.001$), as in fainter quasars tend to have a stronger CIV emission line and vice versa. This correlation is not unexpected and is consistent with what is known as the Baldwin effect \cite[][]{BaldwinEffect1977, JENSEN2016}. We note that, however, previous measurements \cite[e.g.,][]{JENSEN2016} focused only on the equivalent width of the CIV emission, and in our case, we study the factor embedding which contains the CIV emission. Our result also demonstrates a slightly stronger negative correlation than the literature values (e.g., $\sim -0.35\pm 0.004$ in \citet[][]{JENSEN2016}), indicating that the latent embedding learned by QFA may better reflect the mechanism that produces the Baldwin effect.

\section{Discussion}\label{sec:discussion}

We proposed in this study an unsupervised statistical algorithm, \textit{Quasar Factor Analysis} (QFA), for quasar continuum prediction. We demonstrated QFA reaches state-of-the-art continuum inference performance for $2<\mathrm{z}<3.5$ quasar spectra, regardless of SNR, subsequently reducing the systematics in the Ly$\alpha$ power spectrum measurements by $2-4\%$ compared to the existing PCA methods. We also explored various downstream tasks with QFA, including outlier selection and the evolution of the quasar population. Below, we will further discuss some other prospects of QFA, putting it in the context of other existing methods. We will also dissect some of its current limitations as well as future prospects.

\subsection{The Evolution of The Quasar Population and Its Implication to High-Redshift Quasars}\label{subsec:highz}
\begin{figure*}
    \epsscale{1.2}
    \plotone{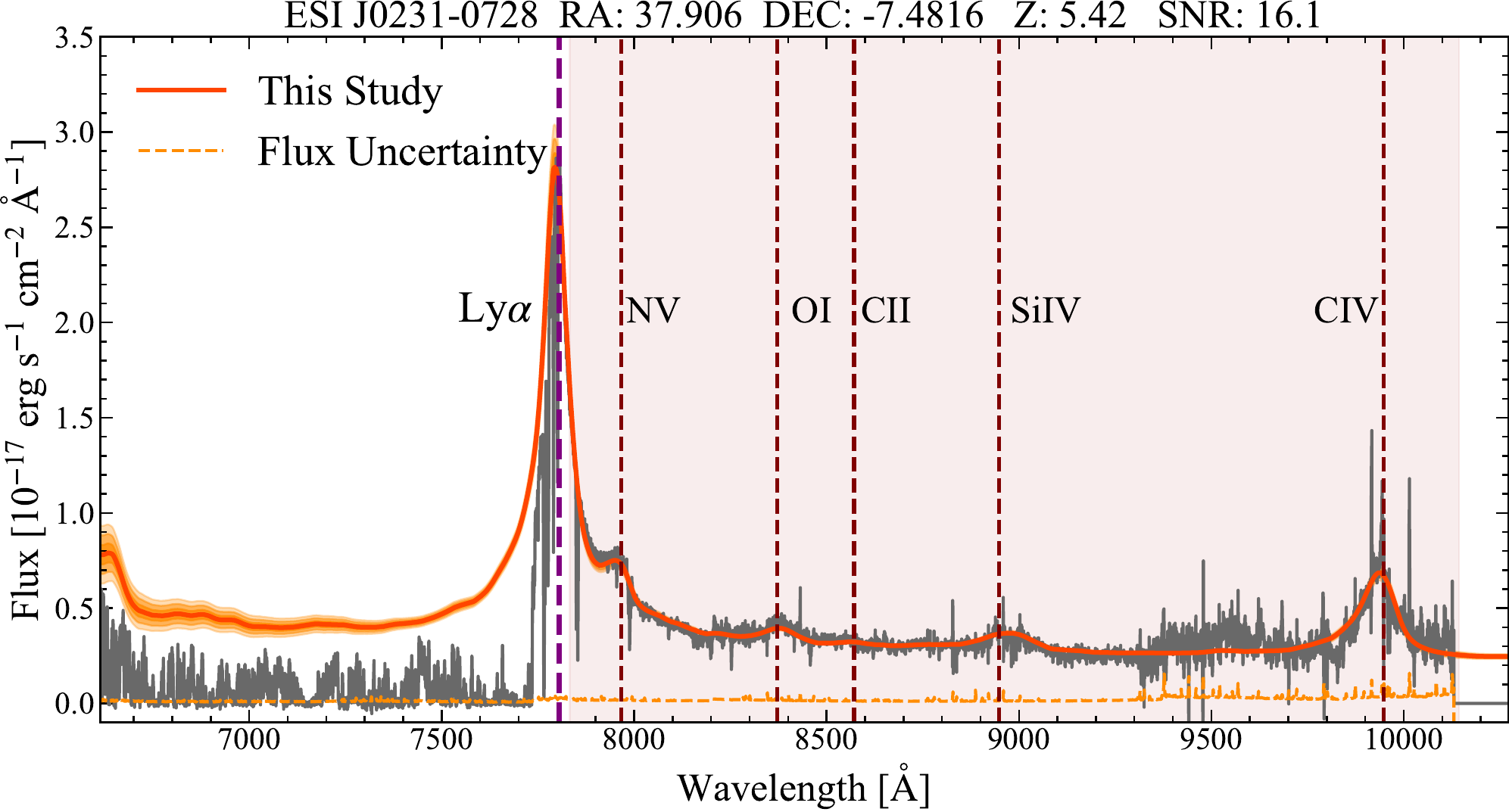}
    \caption{Continuum inference for a high-redshift $\mathrm{z}=5.42$ quasar spectrum observed by the Keck Observatory \cite[][]{RAFELSKI2012, RAFELSKI2014}. The shaded region denotes the $1\sigma,2\sigma,3\sigma$ posterior confidence intervals. As the intensive Ly$\alpha$ absorptions violate the transmission field assumption of QFA, we only adopt information in the red side (shown in the red-shaded region) as the input. QFA assigns a larger uncertain estimation toward the blue end because the red-blue correlation decreases toward the bluer wavelength.}
    \label{fig:highz}
\end{figure*}

Although rare, high-redshift quasars remain uncontested probes to the study of the IGM, and the extended Ly$\alpha$ damping wing in high-redshift quasar spectra is still perhaps the best tell-tale sign of the neural hydrogen fraction during the epoch of reionization (hereafter, EoR). Despite their importance, studying high-redshift quasars also comes with unique challenges. As the IGM becomes primarily neutral at $\mathrm{z}\gtrsim 6$, as shown in Figure~\ref{fig:highz}, the Ly$\alpha$ forest obliterates the bluer flux in moderate-redshift quasar spectra, typically known as the Gunn-Peterson Trough \cite[e.g.,][]{GP1965, FAN2006}. A key assumption for the study of high-redshift quasars \cite[e.g.,][]{PCA2018, QSMOOTH2020, NF2020} thus relies on the fact that we could extrapolate the quasar's properties at lower redshift to their higher redshift counterparts and determining the continuum at wavelengths bluer from the red-side information. 

In this study, by analyzing the entire SDSS DR16 dataset (see Section~\ref{subsec:sdss}), we did not find any statistically significant evolution of the quasar evolution from $\mathrm{z} = 2$ to $\mathrm{z} = 3.5$ (Section~\ref{subsec:evolution}), consistent with previous studies \cite[e.g.,][]{JENSEN2016}. The lack of quasar evolution might lend credence to training on moderate-redshift quasars and applying them to high-redshift quasars. Complementary to this study, \citet{JINYI2021} also reported no significant evolution from $z=6.3$ to $z=7.64$, except for a blueshift of the CIV emission line.

Assuming that we can extrapolate our inference to high redshift, as a precursor study, we have tentatively applied our QFA, trained on the $37,548$ moderate-redshift SDSS quasars, and applied that to a $\mathrm{z}=5.42$ quasar J0231-0728 observed by the Keck Observatory \cite[][]{RAFELSKI2012, RAFELSKI2014}. Since the saturated transmission fields violate our assumption on the Ly$\alpha$ forest (see Section~\ref{subsec: model}), we consider all these wavelength pixels bluer than the Ly$\alpha$ emission to be masked during the inference. To adjust for the difference in resolution between the spectra obtained from Keck and SDSS, we down-sample the high-resolution Keck spectra \cite[$R\approx 7,000 - 10,000$,][]{RAFELSKI2012, RAFELSKI2014} to match the SDSS resolution \cite[$R\approx 2,000$, ][]{SDSS2020}. We then carry out the same preprocessing procedures as elaborated in Section~\ref{subsec:sdss}. As shown in Figure~\ref{fig:highz}, QFA yields a visually plausible quasar continuum. Interestingly, unlike the other inference shown at the low-redshift (Figure~\ref{fig:compare}), as we deprive the blue information from the QFA, the inference uncertainty grows significantly toward the bluer end. This is not unexpected because there is a decline in correlation between the blue-side quasar continuum and the red-side quasar continuum. And hence when inferring only the blue side from the red side, like previously done in the PCA-based method, the inference uncertainty increases. We note that, as a preliminary experiment, the training sample and inference process are not optimized for high-redshift quasar spectra; compared to previous high-redshift quasar continuum prediction works \cite[e.g., ][]{PCA2018}, QFA mainly potential at enlarging the diversity of the training samples and being probabilistic ; more detailed considerations will be addressed in future work.

\subsection{Probabilistic Inferences and Impact for Cosmological Measurements}\label{subsec:cosmo}

In this study, we showed that the accuracy of continuum fitting of QFA can further suppress to the systematic bias of the Ly$\alpha$ power spectrum measurements to $1-3\%$, depending on the redshift (Section~\ref{subsec:pk1d}). But besides the accurate recovery, perhaps an even more critical innovation of QFA is that it also provides the posterior of the continuum. Thus far, most existing continuum-fitting methods only provide a deterministic measurement of the continuum  \cite[e.g.,][]{PCA2005, PARIS2011, IQNET2021}. The classical approach is that, when inferring the cosmology, the uncertainty introduced from the continuum inference is calibrated through synthetic data \cite[e.g.,][]{1DPOWERSPECTRUM2019}. However, this approach comes with the danger of biased estimates, especially at the percent level. For example, as we have also shown in this study (see Section~\ref{subsec: mock} and Section~\ref{subsec:testOnMock}), any PCA-generated mocks might not capture the full diversity of quasars and may lead to a biased calibration. We note that modern PCA methods \cite[e.g.,][]{PCA2018, COMPARE2021} estimate the continuum fitting uncertainty directly from the training and test spectra.

QFA does not rely on such post-hoc calibration and it provides also the posterior of the continuum prediction. The sampling of the posterior continuum is analytic and straightforward. In practice, once the posterior distribution of the latent factor ($\mathbf{h}$) is computed (Equation~(\ref{eq:pos_h})), we can then sample the posterior distribution of the latent factor and subsequently the corresponding posterior quasar spectra (Equation~(\ref{eq: cont_model})). The probabilistic nature of QFA might prove important for future missions because the posterior can be directly integrated into cosmological measurement pipelines \cite[e.g.,][]{PICCA2020}, leading to a more ab-initio Bayesian uncertainty quantification for cosmological parameters \cite[e.g.,][]{EILERS2017, GERARDI2022, SIMON2022}.

\subsection{Dissecting the Physics Behind the Quasar Continua}\label{subsec:eaa}

Since QFA assumes a latent factor decomposition of the quasar continua, it projects the high-dimensional quasar continua into low-dimensional latent embedding (in our case, eight latent factors, see Figure~\ref{fig:com} and Figure~\ref{fig:com-full}). Compared to PCA methods, latent factor analysis is not confined to an orthogonal basis. As shown in Section~\ref{subsec:evolution}, this flexibility in choosing the basis has led to a somewhat more sensible decomposition of the quasar continua. In particular, as shown in Figure~\ref{fig:com} and Appendix~\ref{appendix:com}, most of the components consist of a handful of prominent broad features.

It has been long postulated that the broad emission lines in quasar spectra are produced by the line-emitting gas in the broad line regions and are closely associated with the accretion process of the supermassive black holes \cite[e.g.,][]{SHEN2011, JINYI2021}. Recall that the properties of the black holes (primarily, the mass) determine the temperature and pressure profile of the accretion disk. As such, the various latent components learned by QFA, e.g., the component representing the CIV emission line, might thus reveal the physics of the accretion disk, including the virial motion of the line-emitting gas in the broad line region \cite[e.g.,][]{CZERNY2011}, and the radiation-driven outflows \cite[e.g., ][]{MEYER2019}.

Since the spectral embeddings from QFA likely relate to the supermassive black hole properties, an intriguing possibility would be to find the mapping between the supermassive black hole properties (including their masses and Eddington ratios) and these spectral embedding \citep[see][]{EILERS2022}. Along the same vein, a possible way to further improve on QFA is to harness the subset of quasar spectra of which we know the more ``ground truth" properties of the supermassive black holes (e.g., through reverberation mapping \cite[e.g.,][]{MARIANNE2006}). Guided by this small set of ``labels," an even better latent factor decomposition might be possible through a mixture of supervised and unsupervised training.  

Finally, QFA might also help study changing look AGNs - AGNs that show substantial time variations due to the changes in the accretion process of the supermassive black holes, outflows, and clouds \cite[e.g.,][]{RICCI2022}. While the signatures in the spectral space might be subtle, the variation in the lower dimensional embedding learned by QFA should be more prominent.

\subsection{The Correlation and the Information Content in Quasar Continua}\label{subsec:corr}

\begin{figure}
    \centering
    \epsscale{1.2}
    \plotone{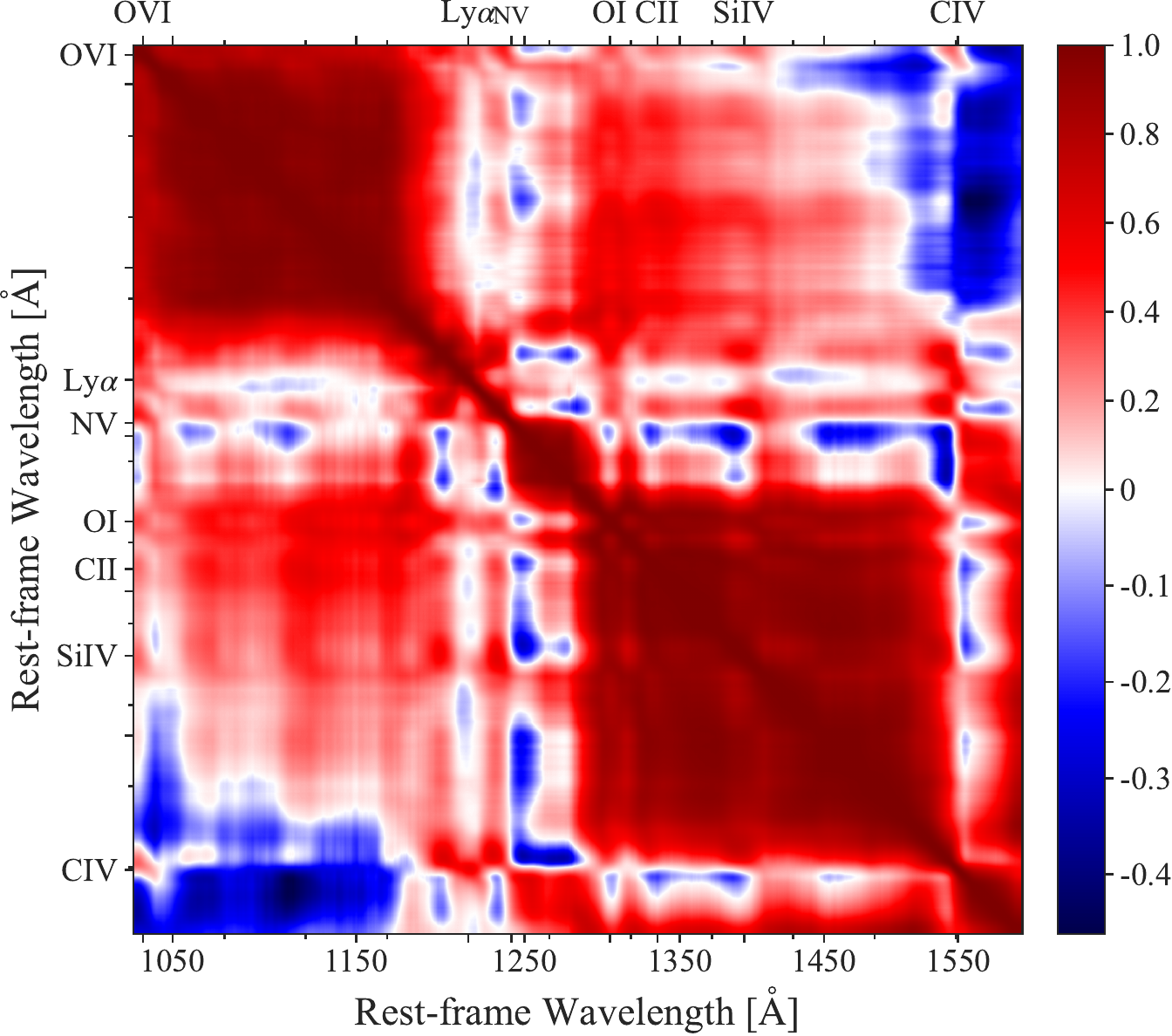}
    \caption{Correlation matrix of quasar continuum learned by QFA from the whole SDSS DR16 quasar spectra. The stronger inter-correlation (the diagonal entry) compared to the intra-correlation (blue-red correlation) inspires this study which aims to harness information in the blue while making the inference on the continuum. The correlation matrix found by QFA generally agrees with previous finding \cite[e.g.,][]{PCA2005, PARIS2011}, albeit with a weaker intra-correlation (see text for details).}
    \label{fig:cmatrix}
\end{figure}

The generative nature of QFA also allows us to sample continua, evaluate pixel-wise correlations, as illustrated in Figure~\ref{fig:cmatrix}. The correlation matrix was estimated based  on quasar continua sampled from a well-trained QFA model. By doing so, we reveal the information being leveraged. As a whole, the correlation matrix QFA recovers are reminiscent of the one found in other studies \cite[e.g.,][]{PCA2005, PARIS2011}. The blue-red inter-correlation has a typical value of $0.1-0.5$, weaker than the intra-correlation ($0.2-0.9$). The weaker blue-red inter-correlation than the intra-correlation has motivated this study. Recall that most existing continuum fitting methods \cite[,e.g.,][]{PCA2005, PCA2018, QSMOOTH2020, NF2020, IQNET2021} rely on relating the blue side with the red side. And QFA goes beyond this and also considers the entire quasar spectrum when making the continuum inference. The fact that QFA also harnesses the information in the blue might explain the superior performance of QFA.

Comparing Figure~\ref{fig:compare} and Figure~\ref{fig:highz} further supports this idea. As shown in Figure~\ref{fig:compare}, the continuum posterior remains tightly constrained for the moderate redshift spectra, where we can model the transmission field on the blue side and condition on them. However, the continuum uncertainty inflates at the bluer end for the high-redshift quasar when we can only harness the information from the red (Section~\ref{subsec:highz}). The significant deviation between QFA and PCA methods when applied to the SDSS spectra also resides in the blue, further suggesting that the PCA method fails to capture the blue continuum accurately because the blue-red correlation subsides.

Interestingly, compared to the literature studies (e.g., figure~2 in \citet[][]{PARIS2011}), QFA favors a weaker blue-red inter-correlation. While QFA has a correlation of $0.1-0.5$ between the the Ly$\alpha$ forest region ($1030-1216\,\mathrm{\AA}$) and the Ly$\alpha$-CIV region ($1216-1550\,\mathrm{\AA}$), previous studies attain a correlation value of $0.3-0.5$. Similarly, between the blue side and the region redder than the CIV emission line ($\lambda_\mathrm{RF}>1550\,\mathrm{\AA}$), QFA suggests a correlation value of $-0.4$ to $-0.1$, whereas the literature values cluster around $-0.6$ to $-0.4$. 

The difference might suggest that when training only on the handful of ($\sim 100$) high SNR sample, previous methods might have overestimated the correlation, leading to poorer generalization ability (also borne out with the SDSS test in this study). When QFA is trained on $\sim 200$ SNR$\geq 7$ SDSS DR16 quasar spectra, we recover correlation values closer to these literature values ($\sim 0.3-0.9$ intra-correlation between the Ly$\alpha$ forest regions and the regions between the Ly$\alpha$ and the CIV emission; $\sim -0.8$ to $-0.4$ intra-correlation between the Ly$\alpha$ forest regions and the regions with wavelengths longer than the CIV emission line). However, when we expand the sample size to learn the quasar properties from the diverse range of quasars in SDSS, the intra-correlation diminishes. We conducted experiments to validate whether the decrease in correlation was caused by the low signal-to-noise ratio (SNR) in our sample. However, we observed a similar decrease in correlation even when using only high-SNR (e.g., SNR$>7$) spectra for training, suggesting that including low-resolution SNR sample in the training has little effect on the estimated correlation. Thus, our results suggest that the previously observed intra-correlation might be overly optimistic, which further underlines the importance of harnessing the information from the blue side when reconstructing the quasar continua.

\subsection{The Power of Unsupervised Learning}\label{subsec:svu}

Besides harnessing the information in the blue, a key advantage of QFA is that it directly learns the distribution of quasar continua and the transmission fields by modeling the entire set of observed spectra through their combined effects. Most existing methods thus far rely on supervised learning \cite[e.g.,][]{PCA2005, PARIS2011, PCA2018, QSMOOTH2020, NF2020, IQNET2021}, learning the mapping between red-side continua to blue-side continua. The key innovation of our study is that QFA can learn the continua directly from the observed spectra, alleviating the need to have a training set with ground truth continua. Unsupervised learning aims to learn the distribution of quasar spectra through appropriate prior knowledge of the underlying data structure (Section~\ref{sec:method}). In the case of QFA, the prior knowledge we impose comes from the underlying structure of the continua and the transmission fields. 

While we made a clear distinction between supervised learning and unsupervised learning to highlight the fundamentally different concept that underscores QFA, in many ways the two methods are related. Supervised and unsupervised learning reflect our different prior beliefs of the system. In the case of supervised learning, the model emphasizes the validity of the training quasar continua. However, the lack of high-quality ground-truth continua can often lead to a biased model. In the unsupervised learning of QFA, we relay to a different form of prior knowledge, focusing only on our understanding of how the quasar continua and transmission fields operate by assigning a specific functional form for these individual components. As we have seen in this study, this weaker physical prior, bolstered by the massive data sets we have garnered, can lead to much superior performance in continuum inference.

Finally, while we focus on unsupervised learning in this study, a hybrid form of weak unsupervised learning, fine-tuned with a subset of ``supervised labels," has led to many new ideas in the machine learning community. These methods have coined the term semi-supervision \cite[e.g.,][]{DAVID2019} or self-supervision \cite[e.g.,][]{SIMCLR2020, XIE2021}. The same concepts have also seen some successes in their application in astronomy, including galaxy morphology classification \citep[][]{WALMSLEY2022}, modeling stellar spectroscopy \citep[][]{BRIAIN2021}, and transient identifications \cite[e.g.,][]{VILLAR2020, MARIANER2021, INIGO2021}. The future of quasar continuum inference might therefore lie in such a hybrid mode, comprehensively using all available ground-truth labels when they are available and at the same time, harnessing our insight into the underlying physical process of quasars, as we did in this study.

\subsection{Other Existing Methods}\label{subsec:other}

In this study, we compare QFA with only two methods, the PCA algorithm proposed by \cite[][]{PARIS2011} and \texttt{PICCA}. We focus on only these two methods because they remain some of the most adopted methods in cosmological measurements \cite[e.g.,][]{BAO2014, 1DPOWERSPECTRUM2019, PICCA2020, COMPARE2021}. But we note that there are many other more advanced PCA-related methods \cite[e.g.,][]{MFPCA2012, PCA2018, QSMOOTH2020} being proposed since the seminal paper of \citet[][]{PARIS2011}.

While these variations have undoubtedly enriched the possibilities of the PCA-based methods, \citet[][]{COMPARE2021} performed a comprehensive study of the differences between these methods and concluded that the lack of ground truth training set bottlenecks the PCA-based methods, rendering them to offer similar performance (within $\sim 10\%$) for continuum prediction. As different PCA-based methods demonstrate similar performance, we contend that adopting these different variances is less likely to alter the qualitative conclusions of this study. However, the limitations of these PCA-based methods, as demonstrated in Sections~\ref{subsec:testOnMock} and \ref{subsec:visual}, are that they rely primarily on pre-defined continua which invariably lead to biased training sets, as discussed in Section~\ref{subsec:sdss}, and only utilize information from the red side. These shortcomings have not been addressed in updated methods. Nonetheless, we note that modern PCA-based methods \cite[e.g., ][]{PCA2018} were primarily developed for high-redshift quasar spectra in which little blue-side information is available. As such, as discussed in Section~\ref{subsec:highz}, the advantages of QFA compared to PCA methods are likely to be less prominent for high-redshift quasar spectra, apart from enlarging the training dataset to include fainter and lower SNR quasars and being probabilistic.

Finally, in recent years, the study of quasars has also seen the rise of deep learning methods \cite[e.g.,][]{QSMOOTH2020, NF2020, IQNET2021}. Comparing all these methods is clearly beyond the scope of this study. But we note that most of these deep-learning methods still focus on supervised learning and thus inherit the same problem as the PCA-based approaches. 

\subsection{Caveats and Limitations}\label{subsec:cal}
We demonstrate that the unsupervised nature of QFA has led to superior performance in continuum inference. However, the model assumptions also currently limit the performance of QFA. In particular, we assume a predefined (not trainable) mean optical depth because the mean absorption degenerates with the quasar continua. While prior studies \cite[e.g.,][]{FG2008, BECKER2013} have largely agreed on mean optical depth measurements, at least at the redshifts of interest in this study, the mean optical depth is not well constrained at higher redshifts. As we expand beyond the current redshift range, the data may necessitate a model in which the mean optical depth is trainable. We also discuss the effects of different mean optical depth functions in Appendix~\ref{ablation:opt}. We found that the variations in continuum predictions are commensurate with mean optical depth function measurements. This suggests the importance of evaluating alternative mean optical depth functions in practice.

Perhaps the more important limitation of QFA is the assumption that the Ly$\alpha$ forest constitutes an independent Gaussian distribution. In QFA, we assume such a distribution because the independent Ly$\alpha$ forest assumption is essential for the analytic derivation. While this assumption is adequate for this study because, for SDSS quasar spectra with typical resolution $\lambda/\Delta \lambda \approx 2000$, the correlations between adjacent pixels are generally weak; the assumption is clearly false in detail. For instance, significant absorbers in the IGM, such as DLA, have demonstrated the absorption in the adjacent wavelength pixel is anything but uncorrelated. On top of that, \citet[][]{FARR2020} has shown that the Ly$\alpha$ forest also contains higher-order moment information beyond the Gaussian assumption. 

Furthermore, the stochastic ``error" term $\boldsymbol{\Psi}$, which stands for the differences between the quasar continuum and its dimension-reduced form $\mu + F\mathbf{h}$, is assumed to be independent over wavelength, which is undoubtedly an oversimplified assumption since coherent structures can contribute to the continuum fitting error. However, in order for the stochastic optimization process to converge, we deem this simplification necessary and justified as $\boldsymbol{\Psi}$ is typically small compared to $\mu+F\mathbf{h}$, i.e., $\sim 3\%$ from our experiments.

Due to these limitations, this is why, when deriving the Ly$\alpha$ power spectrum in this study, we only use QFA to make the inference on the continuum instead of directly using the inferred transmission field. A better QFA thus requires us to develop a more generalized formalism that can take into account the inter-pixels correlation and the higher-order moment simultaneously while ensuring that the models are still analytic or easily optimizable. This is undoubtedly a tall order that we will leave to future studies.

\section{Conclusion}\label{sec:conclusion}

In this study, we propose an unsupervised learning method, \textit{Quasar Factor Analysis}, to infer quasar continua. QFA learns the distribution of quasar continua and the transmission field directly from the ensemble of observed spectra, regardless of their SNR. QFA does not depend on any pre-defined continua as a training set and can provide uncertainty quantification of the continua. The probabilistic nature of QFA allows the method to deal with missing pixels, capture a more physically motivated lower dimensional embedding, and find spectra outliers. Our main findings are summarized as follows:

\begin{itemize}
\item Testing on mock datasets, we demonstrate that QFA reaches state-of-the-art performance, $\sim 2\%$ absolute fractional flux error at wavelength bluer than the Ly$\alpha$ emission and $\lesssim 1\%$ at wavelength redder the Ly$\alpha$ emission, as opposed to the $\gtrsim 3\%$ error from the PCA-based method and \texttt{PICCA}. 

\item Besides a better mean recovery, QFA also incurs the least case-by-case scatter. The absolute fractional flux error from QFA in continuum recovery ranges from $\sim 1\%$ in the best cases to $\sim 5\%$ in the worst cases. In contrast, PCA has an error of $1\% - 10\%$, and \texttt{PICCA} $1\% - 7\%$.

\item QFA generalizes better. When introducing a $10\%$ linear perturbation beyond the training set, the errors from the PCA-based method double, while QFA remains adaptable and achieves the same performance.

\item We further applied the method to the SDSS DR16 dataset and showed that QFA yields more robust continua. The PCA-based method can deviate $\gtrsim 6\%$ from the QFA's inferences for $50\%$ of the SDSS spectra, and $\gtrsim 10\%$ for $25\%$ of the SDSS quasar spectra, while \texttt{PICCA} gives $\gtrsim 3\%$ differences for $50\%$ SDSS quasar spectra from QFA's inferences and $\gtrsim 5\%$ for $50\%$ SDSS quasar spectra.

\item QFA's superior performance in extracting the continuum reduces the bias when estimating the Ly$\alpha$ forest estimation to $3-4\%$ at $\mathrm{z} = 2.4 -2.6$ and to $\lesssim 1\%$ at $\mathrm{z} = 2.9-3.1$, which is a slight improvement to \texttt{PICCA}. In contrast, PCA induces a systematic bias of $5\%-7\%$ for both redshift ranges.

\item Through the latent embedding extracted through QFA, we illustrate that the quasar population does not manifest any detectable evolution from $\mathrm{z}=2$ to $3.5$. 
\end{itemize}

In a nutshell, QFA provides a new framework to model the quasar populations. We showed that, even with a minimal prior assumption on the quasar continua and the transmission, modern-day optimization frameworks allow us to model the quasar continua and transmission field from a vast set of quasar spectra simultaneously, without ad hoc human intervention. As large-scale spectroscopic surveys such as DESI and 4MOST will continue to revolutionalize the field, QFA provides a path to maximally extract every bit of information from all these heterogeneous quasar datasets, allowing the quasars to transmit to us the information about their own formation as well as the history throughout the cosmic evolution.

\section{Code and Data Availability}

To ensure reproducibility, all source codes of \textit{Quasar Factor Analysis} 
are made publicly available on Zenodo
(\citet{sun_zechang_2023_8025967}, \dataset[doi:10.5281/zenodo.8025967]{https://doi.org/10.5281/zenodo.8025967}). 
The data used in this study can be found on Zenodo:
\dataset[doi:10.5281/zenodo.8050660]{https://doi.org/10.5281/zenodo.8050660}.

\section{Ackowledgement}
The authors thank Paulo Montero-Camacho and Pablo Renard for the detailed comments on the manuscript. The authors also appreciate useful comments and discussions with Song Huang, Bradley Greig, Siwei Zou, Yunjing Wu, Xiaojing Lin, Ben Wang, Jiaqi Zou, Mingyu Li, Youquan Fu, Yongming Liang, Rongmon Bordoloi, Bin Liu, and Jiashu Pan. Z.S. specially thanks for Prof. Mingsheng Long in Tsinghua University for the invaluable machine learning course.  The authors acknowledge \texttt{numpy} \cite[][]{NUMPY2020}, \texttt{astropy} \cite[][]{ASTROPY2022}, \texttt{PyTorch} \cite[][]{PYTORCH2019}, and \texttt{pandas} \cite[][]{PANDAS2023} python packages used in this research. The authors also gratefully acknowledge the Sloan Digital Sky Survey IV for providing the data used in this work. Y.S.T. acknowledges financial support from the Australian Research Council through DECRA Fellowship DE220101520. Z.S. and Z.C. are supported by the National Key R\&D Program of China (grant no.\ 2018YFA0404503), the National Science Foundation of China (grant no.\ 12073014), and the science research grants from the China Manned Space Project with No. CMS-CSST2021-A05.

\appendix
\section{Complexity Analysis}\label{appendix:complexity}

The training and inference process of QFA involves manipulations of large matrices, which can be computationally expensive. Furthermore, direct large matrix inversion can lead to catastrophic divergence problems in practice. A crucial part of QFA is thus to design robust computation strategies, which we will detail in this appendix.

The computational complexity for matrix inversion and log determinant are both $\mathcal{O}\left(\mathrm{N}^3_{\mathrm{pix}}\right)$.
The dimension of quasar spectrum $(\mathrm{N}_{\mathrm{pix}} \gg 1)$ and the number of quasar spectra is large $(\mathrm{N}_{\mathrm{spec}} \gg1)$. As such, naive implementations of evaluating the loss function in Equation~\ref{eq:single_ll} and inferring the posterior distribution of quasar continua in Equation~\ref{eq:pos_h} can lead to a prohibitively large computational complexity. However, we can make use of the special structure of the covariance matrices of QFA to reduce the computation \cite[][]{GP2017, GP2020}.

In particular, the covariance matrix $\mathrm{\Sigma}$ given by QFA has the form as $\mathrm{M}\mathrm{M}^\mathrm{T}+\mathrm{D}$, in which $\mathrm{\Sigma}$ is a $\mathrm{N}_\mathrm{pix}\times\mathrm{N}_\mathrm{pix}$ matrix, $\mathrm{M}$ is a $\mathrm{N}_\mathrm{pix}\times \mathrm{N}_\mathrm{h}$ matrix and $\mathrm{D}$ is a $\mathrm{N}_\mathrm{pix}\times \mathrm{N}_\mathrm{pix}$ diagonal matrix. Inverting this large matrix is computationally expensive, even when it is possible. Fortunately, we can reduce the complexity of matrix inversion by applying the Woodbury identity:
\begin{equation}
    \mathrm{(MM^\mathrm{T}+D)^{-1}=D^{-1}-D^{-1}M(I+M^\mathrm{T}D^{-1}M)^{-1}M^\mathrm{T}D^{-1}},
\end{equation}
where $\mathrm{I}$ is a $\mathrm{N}_{\mathrm{h}}\times\mathrm{N}_{\mathrm{h}}$ identity matrix. Thus, we can compute the inverse matrix of $\mathrm{\Sigma}$ through calculating the inverse of a much smaller $\mathrm{N}_{\mathrm{h}}\times\mathrm{N}_{\mathrm{h}}$ matrix $\mathrm{I+M^\mathrm{T}D^{-1}M}$, which is significantly more numerically stable. This simplification leads to a total computational complexity of only $\mathcal{O}(\mathrm{N}_{\mathrm{pix}}^2\mathrm{N}_{\mathrm{h}})$. 

Similarly, for the determinant, we apply the Sylvester determinant theorem,
\begin{equation}
    \mathrm{\log\det(MM^\mathrm{T}+D)=\log\det D + \log\det (I+M^\mathrm{T}D^{-1}M)},
\end{equation}
\noindent
reducing the complexity of computing the log determinant from $\mathcal{O}\left(\mathrm{N}^3_{\mathrm{pix}}\right)$ to $\mathcal{O}(\mathrm{N}_{\mathrm{pix}}+\mathrm{N}^3_{\mathrm{h}})$.

\section{Derivative of the Loss function}\label{appendix:derivative}

When optimizing the model parameters to maximize the likelihood (Equation~(\ref{eq:total_ll})) of the data, derivatives of the loss function (here the likelihood function) with respect to the model parameters are needed to update the model parameters during the optimization process. Although the derivatives (or ``gradients") can be calculated via auto-diff algorithms in deep learning frameworks, implementing the analytical expressions of the derivatives when it is possible can speed up the GPU calculations.

Recall that, for a single spectrum $\mathrm{(\lambda, S, z, \sigma_{\epsilon})}$, its likelihood function $\mathcal{L}(\mathrm{\lambda, S, z, \sigma_{\epsilon}}|\mathcal{M})$ is shown in Equation~(\ref{eq:single_ll}). For ease of derivation, we further rewrite $\Sigma$ and $\Delta$ as
\begin{equation}
\begin{split}
    &\Sigma = \mathrm{A}FF^\mathrm{T}\mathrm{A}+\mathrm{A}\Sigma_{\mathrm{\Psi}}\mathrm{A} + \Sigma_{\mathrm{\omega}} + \Sigma_{\mathrm{\epsilon}}\\
    &\Delta = \mathrm{S} - \mathrm{A}\mu\\
    &\tilde{\mathrm{A}} = \mathrm{diag}^{-1}(\mathrm{A})\\
    &\tilde{\mathrm{\omega}} = \mathrm{diag}^{-1}(\mathrm{\Sigma_{\omega}})\\
    &h(\mathrm{z}_{\mathrm{abs}}) = (1-\exp(-\tau_0(1+\mathrm{z}_\mathrm{abs})^\beta)+c_0)\\
    &f(\mathrm{z}_\mathrm{abs}) = h(\mathrm{z}_{\mathrm{abs}})^2\\
\end{split}
\end{equation}
Here we define ``$\mathrm{diag}$" as the operation of returning a matrix whose diagonal elements are the input vector, and ``$\mathrm{diag}^{-1}$" denotes the operation of returning the diagonal elements of the input matrix. ``$\mathrm{Sum}$" denotes the summing up of all elements within a vector or a matrix. It is easy to check that the derivatives of Equation~(\ref{eq:single_ll}) for all the model parameters are as follow:
\begin{equation}
    \begin{split}
    &\frac{\partial\mathcal{L}}{\partial \Sigma}(\mathrm{\lambda, S, z, \sigma_{\epsilon}}|\mathcal{M}) = -\frac{1}{2}(\Sigma^{-1}-\Sigma^{-1}\Delta\Delta^\mathrm{T}\Sigma^{-1})\\
    &\frac{\partial\mathcal{L}}{\partial F}(\mathrm{\lambda, S, z, \sigma_{\epsilon}}|\mathcal{M}) = \mathrm{A}(\Sigma^{-1}-\Sigma^{-1}\Delta\Delta^\mathrm{T}\Sigma^{-1})\mathrm{A}\\
    &\frac{\partial\mathcal{L}}{\partial \sigma^2_{\mathrm{\Psi}}}(\mathrm{\lambda, S, z, \sigma_{\epsilon}}|\mathcal{M}) = \frac{1}{2}\tilde{\mathrm{A}}\circ\mathrm{diag}^{-1}\left(\Sigma^{-1}-\Sigma^{-1}\Delta\Delta^\mathrm{T}\Sigma^{-1}\right)\circ\tilde{\mathrm{A}}\\
    &\frac{\partial\mathcal{L}}{\partial\omega_0}(\mathrm{\lambda, S, z, \sigma_{\epsilon}}|\mathcal{M}) = \frac{1}{2}\mathrm{diag}^{-1}\left(\Sigma^{-1}-\Sigma^{-1}\Delta\Delta^\mathrm{T}\Sigma^{-1}\right)\circ f(\mathrm{z}_\mathrm{abs})\\
    &\frac{\partial\mathcal{L}}{\partial\tau_0}(\mathrm{\lambda, S, z, \sigma_{\epsilon}}|\mathcal{M}) = -\mathrm{Sum}\left\{\mathrm{diag}^{-1}\left(\Sigma^{-1}-\Sigma^{-1}\Delta\Delta^\mathrm{T}\Sigma^{-1}\right)\circ\tilde{\mathrm{\omega}}\circ f(\mathrm{z}_\mathrm{abs})\circ h(\mathrm{z}_\mathrm{abs})\circ(1+\mathrm{z}_\mathrm{abs})^\beta\right\}\\
    &\frac{\partial\mathcal{L}}{\partial \beta}(\mathrm{\lambda, S, z, \sigma_{\epsilon}}|\mathcal{M}) = -\mathrm{Sum}\left\{\mathrm{diag}^{-1}\left(\Sigma^{-1}-\Sigma^{-1}\Delta\Delta^\mathrm{T}\Sigma^{-1}\right)\circ\tilde{\mathrm{\omega}}\circ f(\mathrm{z}_\mathrm{abs})\circ h(\mathrm{z}_\mathrm{abs}) \circ \tau_0(1+\mathrm{z}_{\mathrm{abs}})\circ \ln\left(1+\mathrm{z}_\mathrm{abs}\right)
    \right\}\\
    &\frac{\partial\mathcal{L}}{\partial c_0}(\mathrm{\lambda, S, z, \sigma_{\epsilon}}|\mathcal{M})=-\mathrm{Sum}\left\{\mathrm{diag}^{-1}\left(\Sigma^{-1}-\Sigma^{-1}\Delta\Delta^\mathrm{T}\Sigma^{-1}\right)\circ\tilde{\mathrm{\omega}}\circ f(\mathrm{z}_\mathrm{abs})\circ h(\mathrm{z}_\mathrm{abs})\right\}\\
    \end{split}
    \label{eq:grad}
\end{equation}

\noindent
These analytic expressions are explicitly calculated and included in the gradient descent process to optimize for the model parameters.

\section{Examples of continuum reconstruction on the mock dataset}\label{appendix:dataset}

\begin{figure}
    \gridline{
    \fig{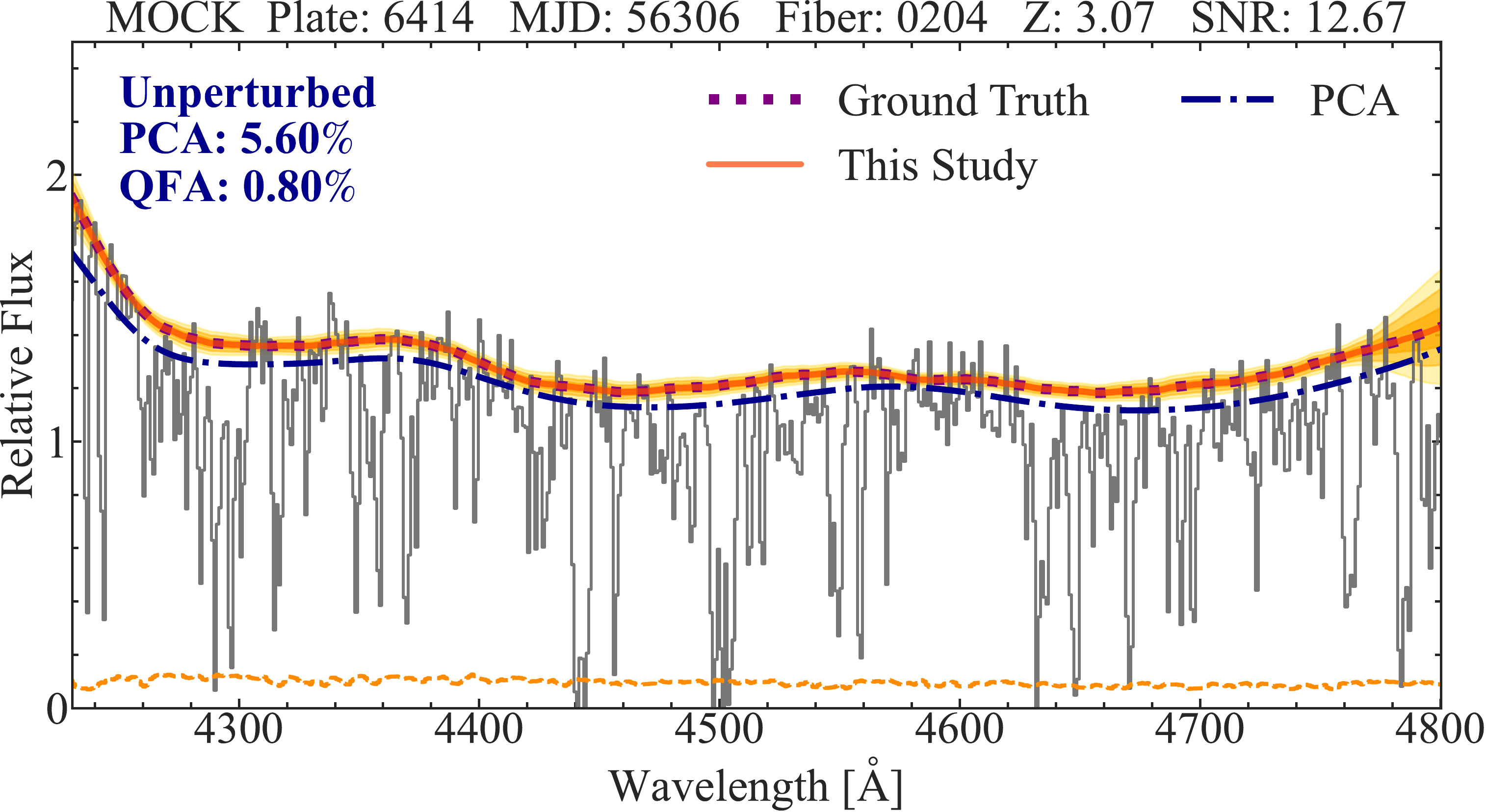}{0.5\textwidth}{(a)}
    \fig{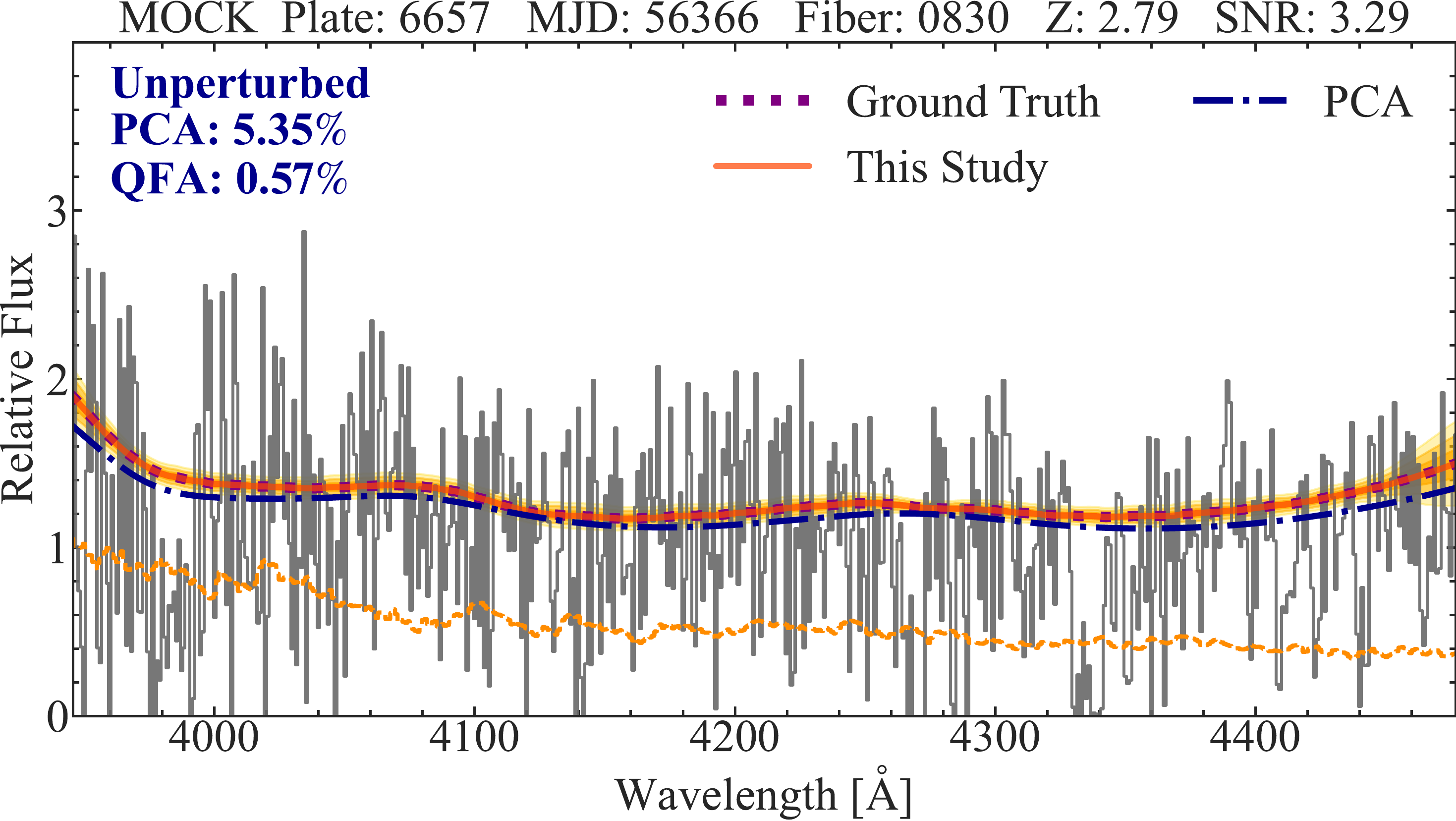}{0.5\textwidth}{(b)}
    }
    \gridline{
    \fig{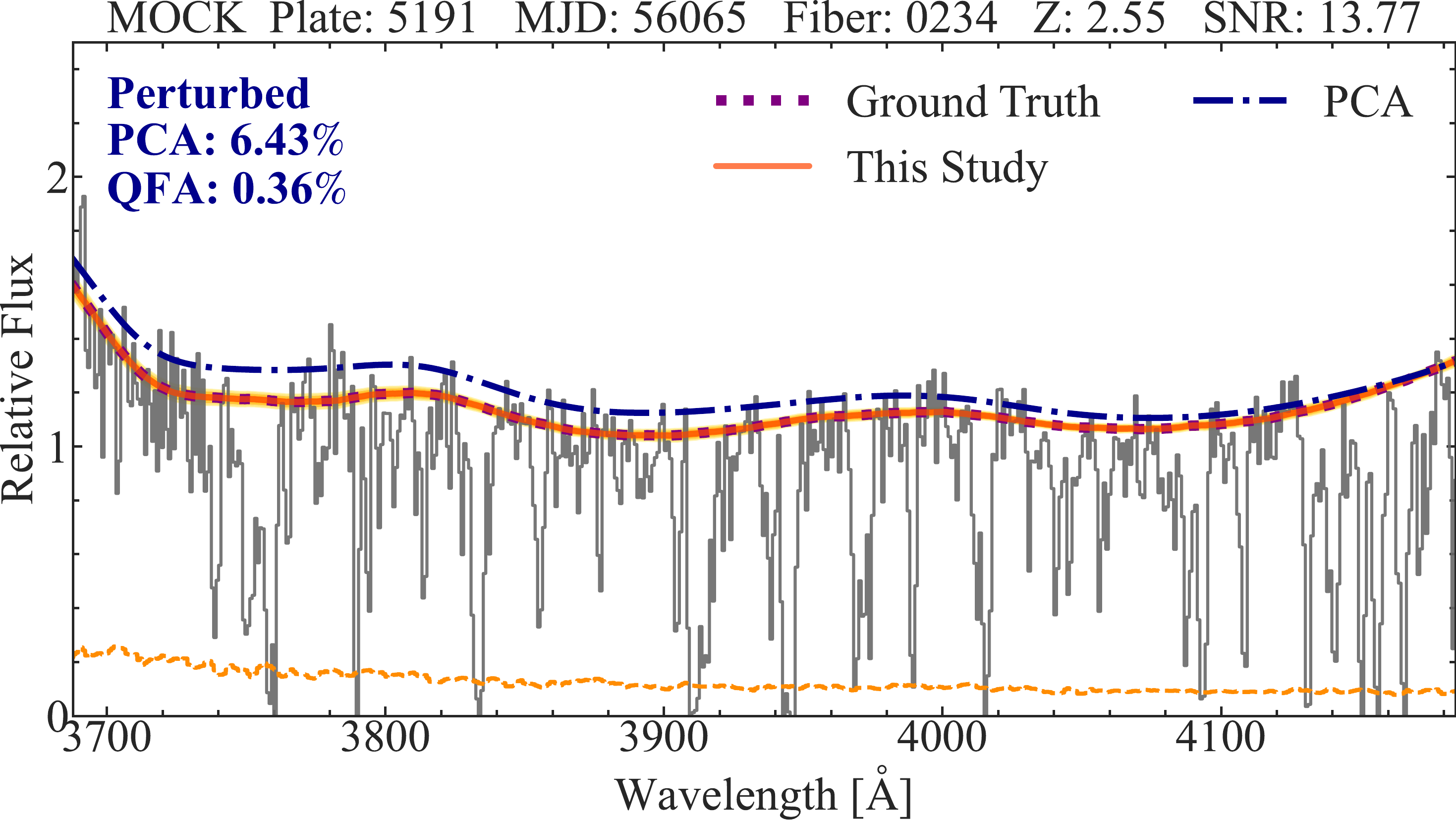}{0.5\textwidth}{(c)}
    \fig{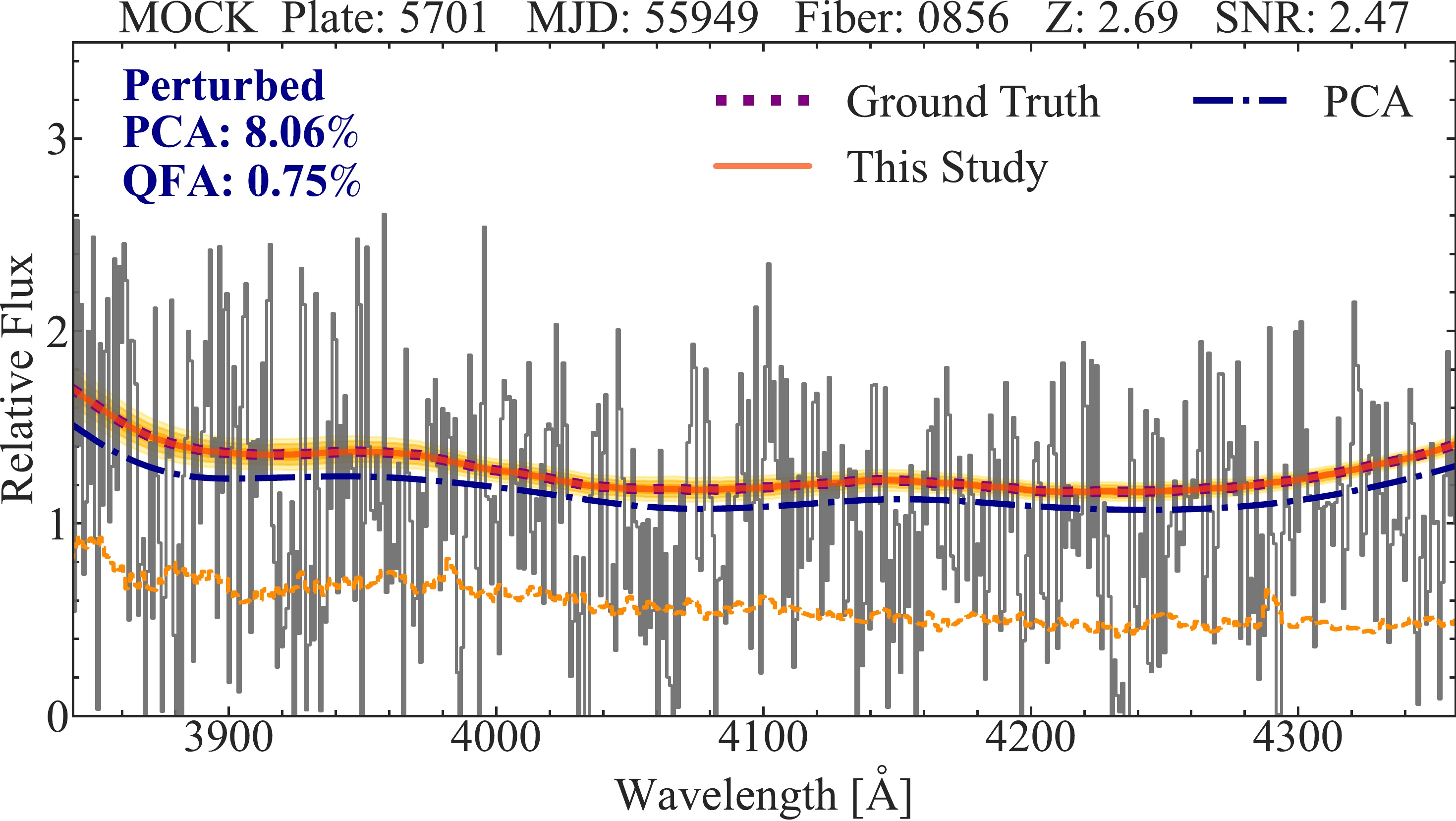}{0.5\textwidth}{(d)}
    }
    \caption{Similar to Figure~\ref{fig:compare}. We show examples on the mock dataset in which PCA fails but QFA performs well.}
    \label{fig:example-mock-pca}
\end{figure}

\begin{figure}
    \gridline{
    \fig{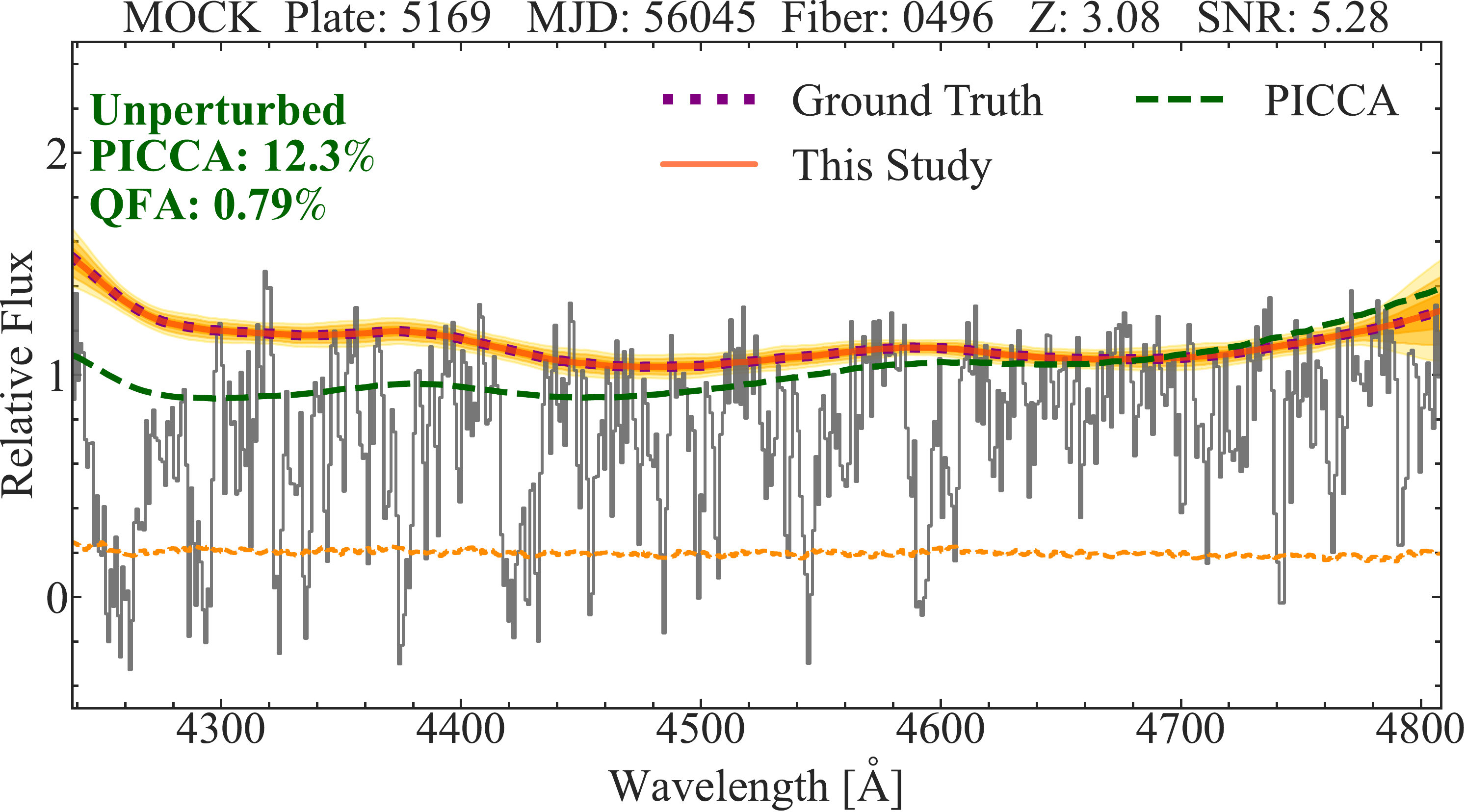}{0.5\textwidth}{(a)}
    \fig{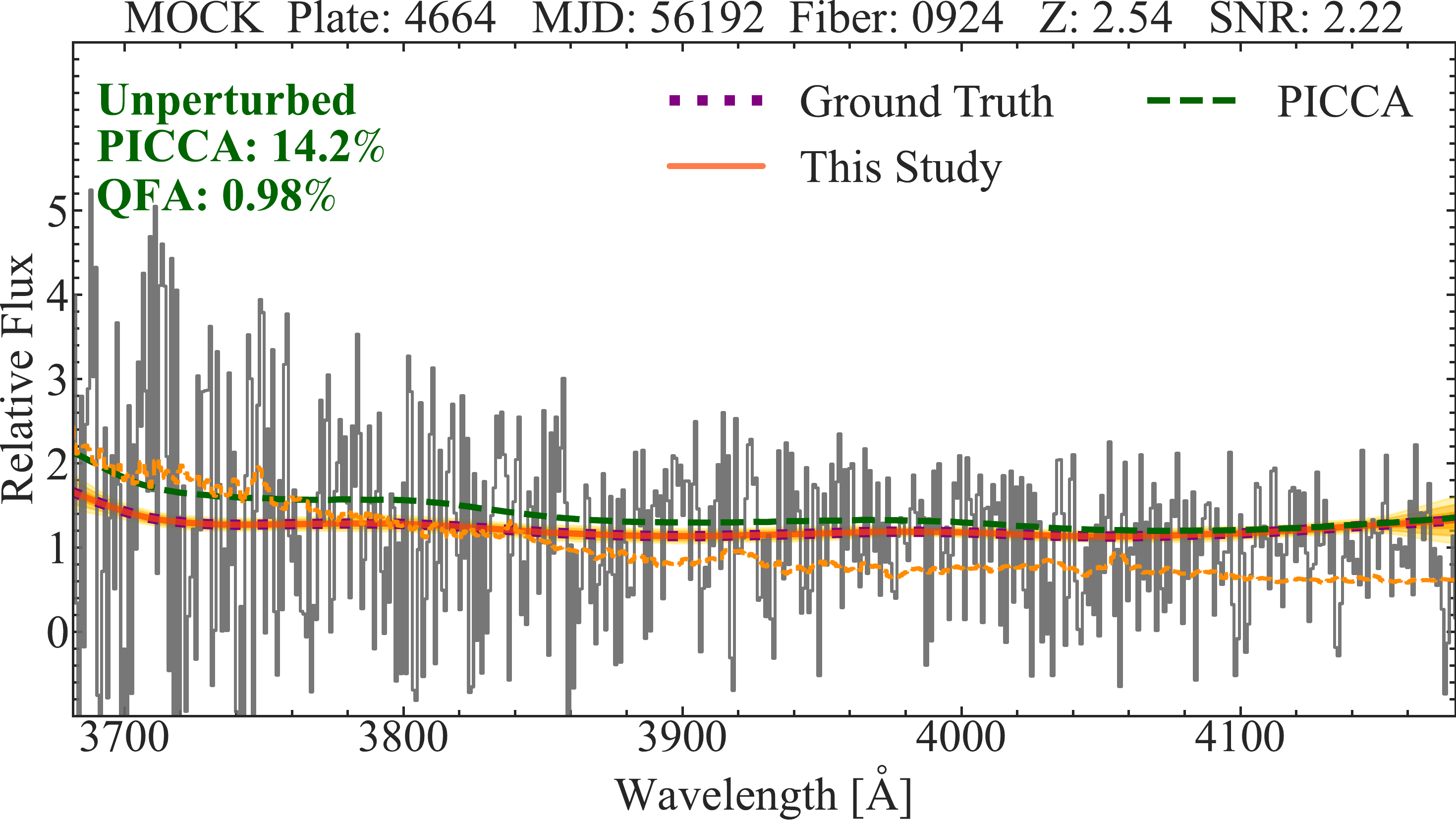}{0.5\textwidth}{(b)}
    }
    \gridline{
    \fig{MOCKPICCA-WITH-PERTURB-HIGHSNR.pdf}{0.5\textwidth}{(c)}
    \fig{MOCKPICCA-WITH-PERTURB-LOWSNR.pdf}{0.5\textwidth}{(d)}
    }
    \caption{Similar to Figure~\ref{fig:compare}. We show examples on the mock dataset in which \texttt{PICCA} fails but QFA performs well.}
    \label{fig:example-mock-picca}
\end{figure}

To help readers intuitively evaluate the results of continuum fitting presented in Section~\ref{subsec:visual}, we provide eight examples in which PCA/\texttt{PICCA} fails but QFA performs well. These examples are taken from the mock datasets discussed in Section~\ref{subsec: mock}, and therefore, the ground truth continua are shown. From Figure~\ref{fig:example-mock-pca} and Figure~\ref{fig:example-mock-picca}, it is evident that inferring ground truth continua from moderate-resolution quasar spectra is a challenging task, even for high-SNR quasar spectra.

\section{Continuum Reconstruction on wavelengths longer than Ly$\alpha$ emission}\label{appendix:red}

\begin{figure*}
    \epsscale{1.3}
    \centering
    \gridline{
    \fig{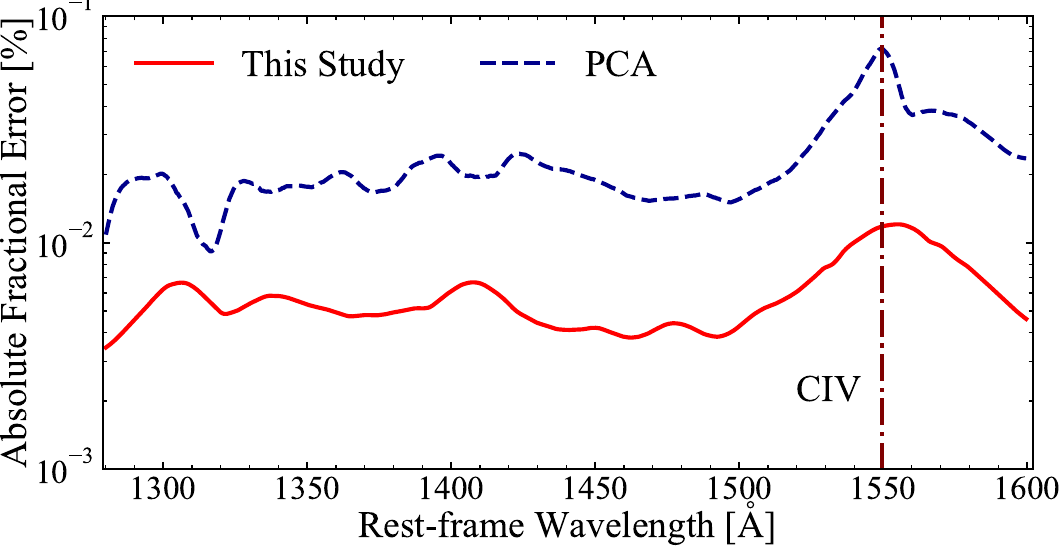}{0.5\textwidth}{(a) Continuum prediction error on the unperturbed dataset}
    \fig{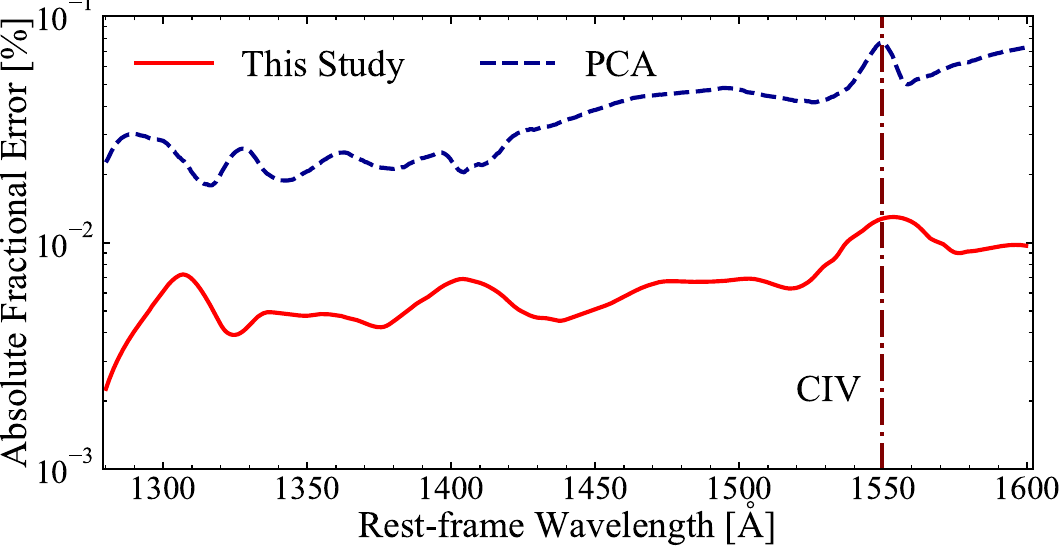}{0.5\textwidth}{(b) Continuum prediction error on the perturbed dataset}
    }
    \caption{
    Similar to Figure~\ref{fig:compare}, but here we evaluate the median continuum prediction error as a function of wavelength for wavelengths longer than the Ly$\alpha$ emission. The left figure shows the model performance on the unperturbed dataset, and on the right, the perturbed dataset. QFA achieves $\lesssim 1\%$ absolute fractional error, as opposed to $1-4\%$ from the PCA-based method. \label{fig:red_performance}}
\end{figure*}
    
\begin{figure*}
    \epsscale{1.3}
    \centering
    \gridline{
    \fig{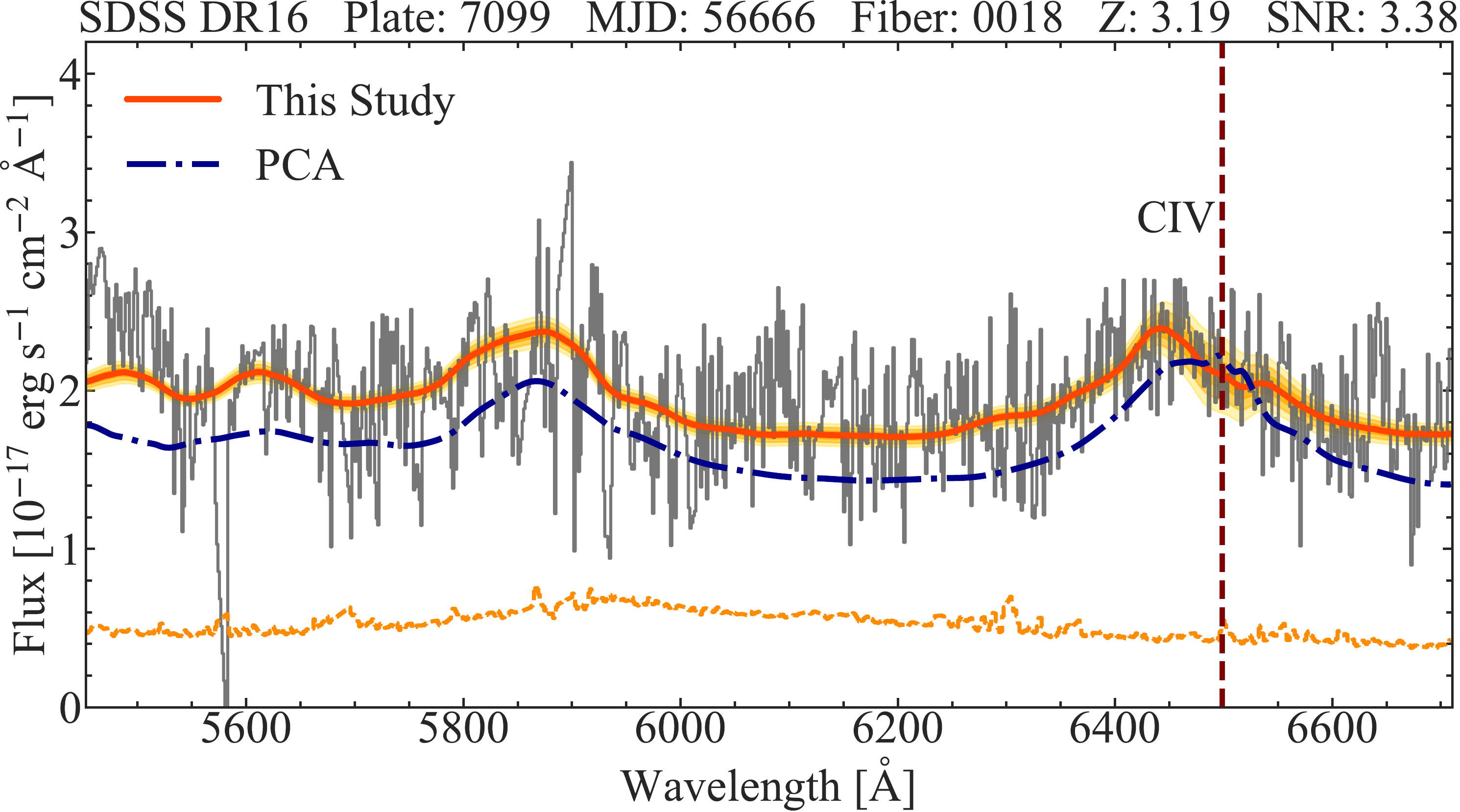}{0.5\textwidth}{(a)\label{fig:reda}}
    \fig{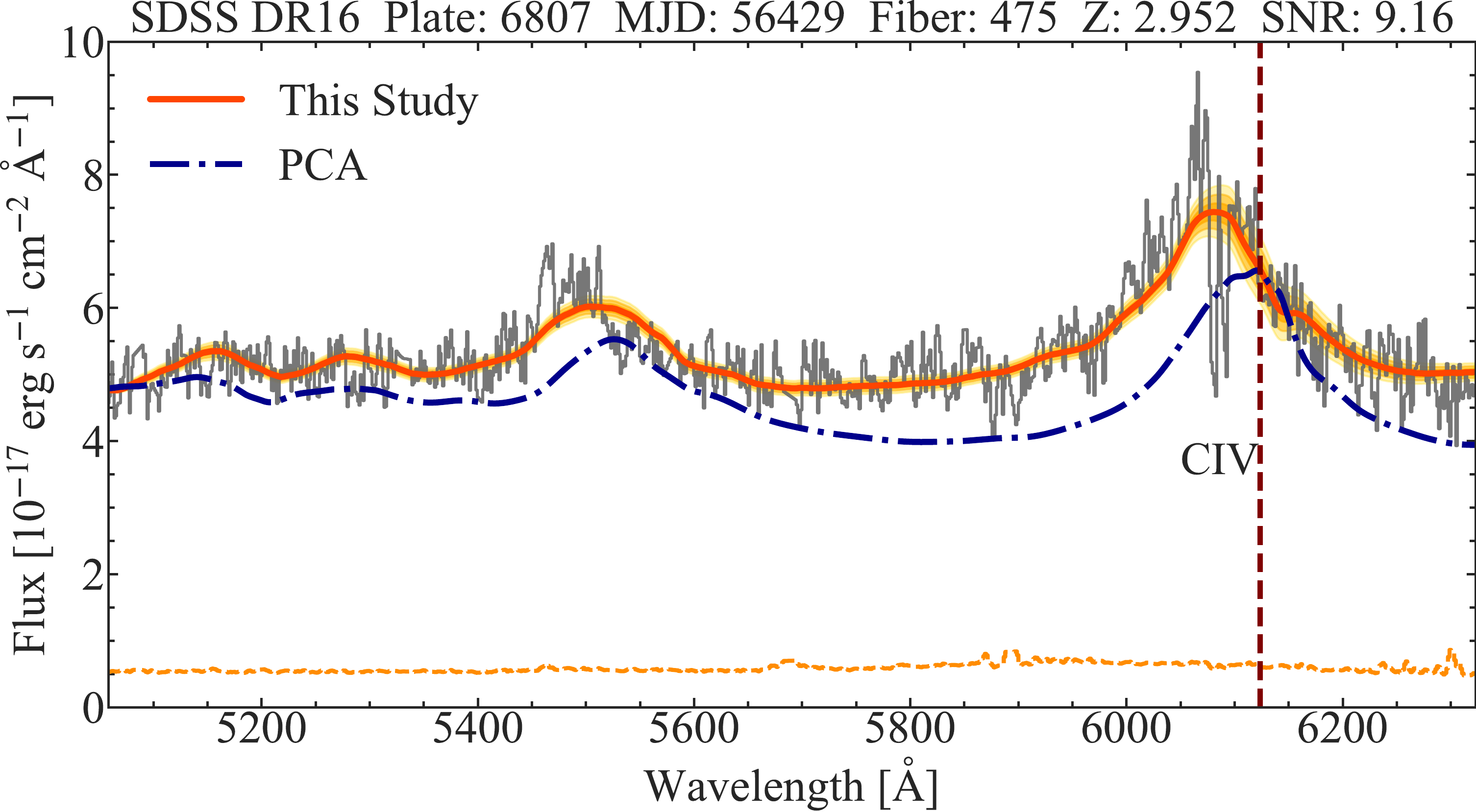}{0.5\textwidth}{(b)\label{fig:redb}}
    }
    \caption{Two examples of quasar continuum inferences on the red side with SDSS DR16 spectra. QFA performs well in both cases. The predicted continua are not affected by the absorption features on the red side and are consistent with the mean flux of the observed SDSS spectra. Conversely, PCA tends to underestimate the red-side continua in both cases due to the inconsistency between the PCA template and the SDSS DR16 spectra.\label{fig:red_example}}
\end{figure*}

In Section~\ref{sec: result}, we focused on the model performance at the Ly$\alpha$ forest regions because the Ly$\alpha$ forest regions carry valuable information on the density fluctuations. For completeness, in this appendix, we also evaluate the model performance on wavelengths longer than the Ly$\alpha$ emission (rest-frame wavelength from $1280\,\mathrm{\AA}$ to $1600\,\mathrm{\AA}$). The results are summarized in Table~\ref{tab:model_performance_red}.

QFA yields, on average, more accurate continuum predictions on the red side ($\lesssim 1\%$ AFFE) than the blue side ($\sim 2\%$ AFFE). This is to be expected because the model can focus on only capturing the continuum without being set back by degeneracy between the quasar continua and the Ly$\alpha$ forest. As before, the unsupervised nature of QFA enables it to work robustly for both the perturbed and unperturbed mock quasar spectra. QFA reaches $\lesssim 1\%$ AFFE on the red side for both mock datasets. As shown in Figure~\ref{fig:red_performance}, QFA performs well across the entire wavelength range, achieving an accuracy of $\lesssim 1\%$ over the full wavelength coverage. In contrast, similar to its performance in the Ly$\alpha$ forest regions (see Section~\ref{subsec:testOnMock}), PCA is less robust compared to QFA due to its supervised nature. It yields $\sim 1\%$ AFFE in the unperturbed dataset but $\sim 4\%$ AFFE in the perturbed dataset. Also, PCA's performance shows a larger scatter than QFA in both datasets.

We also apply the methods to the SDSS DR16 data set and illustrate two randomly selected examples in Figure~\ref{fig:red_example}. The figure demonstrates QFA also works well on the observed spectra. QFA gives reasonable continuum predictions in both examples; the QFA predictions generally pass through the mean flux and are not affected by the absorptions on the red side. PCA, in contrast, yields subpar continua, generally overshooting or undershooting the observed spectra.

\begin{table}[htbp!]
\centering
\begin{tabular}{lrrr|rrr}
\hline\hline
&\multicolumn{3}{c}{Unperturbed}&\multicolumn{3}{c}{Perturbed}\\
 \hline
 AFFE [\%]& 5th& 50th & 95th  & 5th &  50th & 95th\\
 \hline
QFA {\it vs.} Truth &0.56&\textbf{1.08}&\textbf{2.11}&\textbf{0.35}&\textbf{0.77}&\textbf{1.71}\\
PCA {\it vs.} Truth&\textbf{0.29}&1.33&4.29&1.71&3.87&6.95\\
\hline
\end{tabular}
\caption{Similar to Table~\ref{tab:model_performance}, but here we evaluate the model performance on wavelengths longer than the Ly$\alpha$ emission.
\label{tab:model_performance_red}}
\end{table}

\section{Examples of Quasar Spectra Outliers in SDSS DR16}\label{appendix:outlier}

\begin{figure}
    \centering
    \epsscale{1.2}
    \plotone{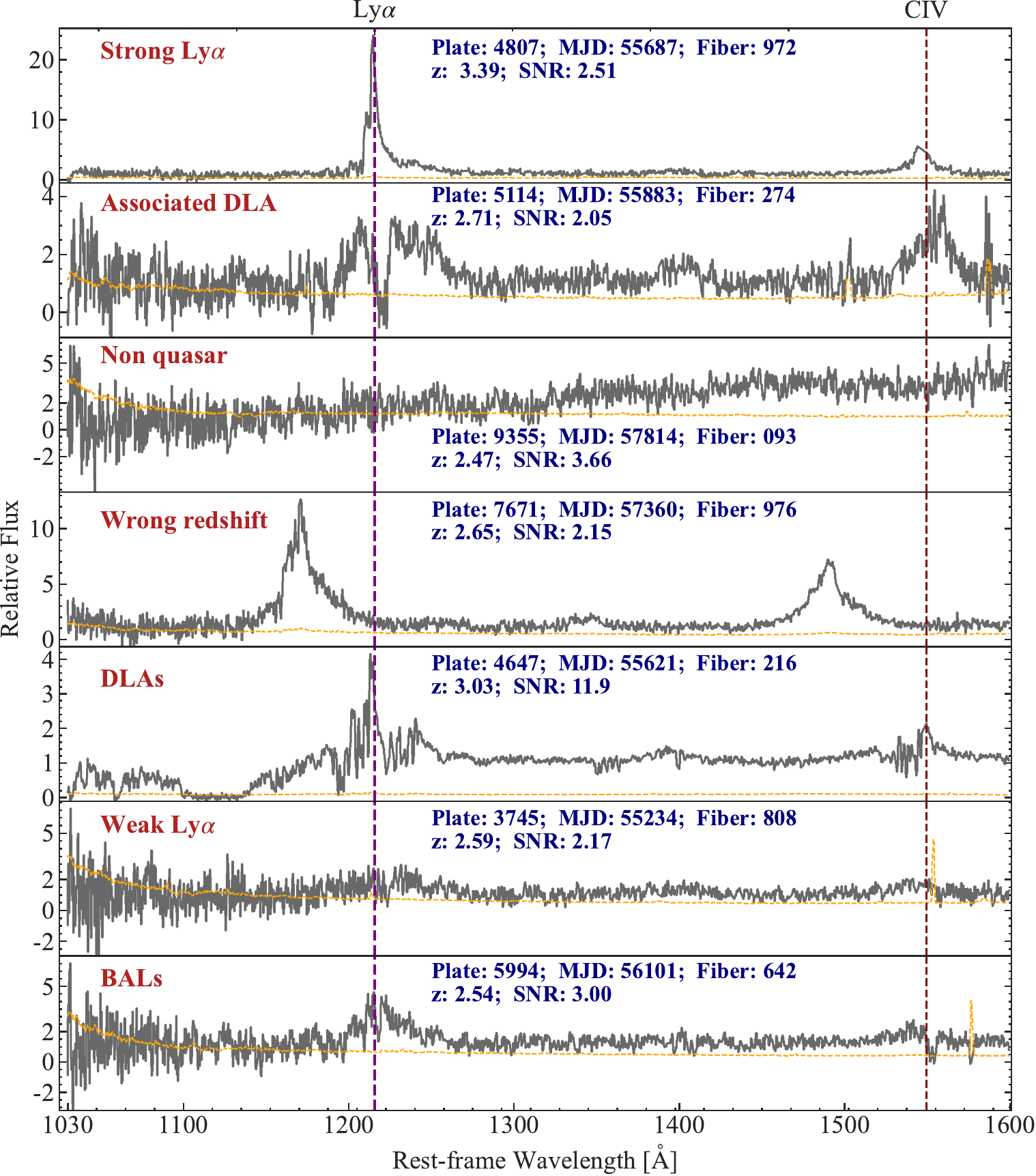}
    \caption{Examples of the quasar spectra outliers selected by QFA as described in Section~\ref{subsec:outlier}. The solid grey lines denote quasar spectra, and the orange dashed lines show the flux uncertainties. We visually classified these outliers into seven outlier classes (see Section~\ref{subsec:outlier} and Appendix~\ref{appendix:outlier}).
    \label{fig:outlier-example}}
\end{figure}

In Section~\ref{subsec:outlier}, we have discussed how QFA can effectively find spectra outliers. From the $37,548$ spectra we studied in SDSS, we determine $179$ outliers (see Section~\ref{subsec:outlier} for details). We visually inspected these outliers and classified them based on their features. Figure~\ref{fig:outlier-example} shows a representative spectrum from each of these visual classes.

These classes are defined, from top to bottom, primarily based on the following features. (1) Quasar spectra with unusually strong Ly$\alpha$ emission. The continuum is almost $20$ times weaker than the Ly$\alpha$ emission in the example shown. (2) Quasar spectra with associated DLA \cite[e.g.,][]{ADLA2013}. Although we have attempted to mask out DLA through the DLA catalog provided by the SDSS DR16 data pipeline as in Section~\ref{subsec:sdss}, some of these spectra passed through because the damped Ly$\alpha$ system resides in the Ly$\alpha$ emission. (3) Misclassified quasar spectra by the SDSS DR16 data pipeline. (4) Quasar spectra with erroneous redshift determination. (5) Quasar spectra with damped Ly$\alpha$ absorptions but not flagged by the SDSS DR16 data pipeline. (6) Quasar spectra with an unusually weak Ly$\alpha$ emission. The continuum is almost the same level as the Ly$\alpha$ emission in the example shown. (7) Quasar spectra with BAL around the CIV emission line.

\section{1D Ly$\alpha$ forest power spectrum measurement on the unperturbed dataset}\label{appendix:p1d}

\begin{figure}
    \plotone{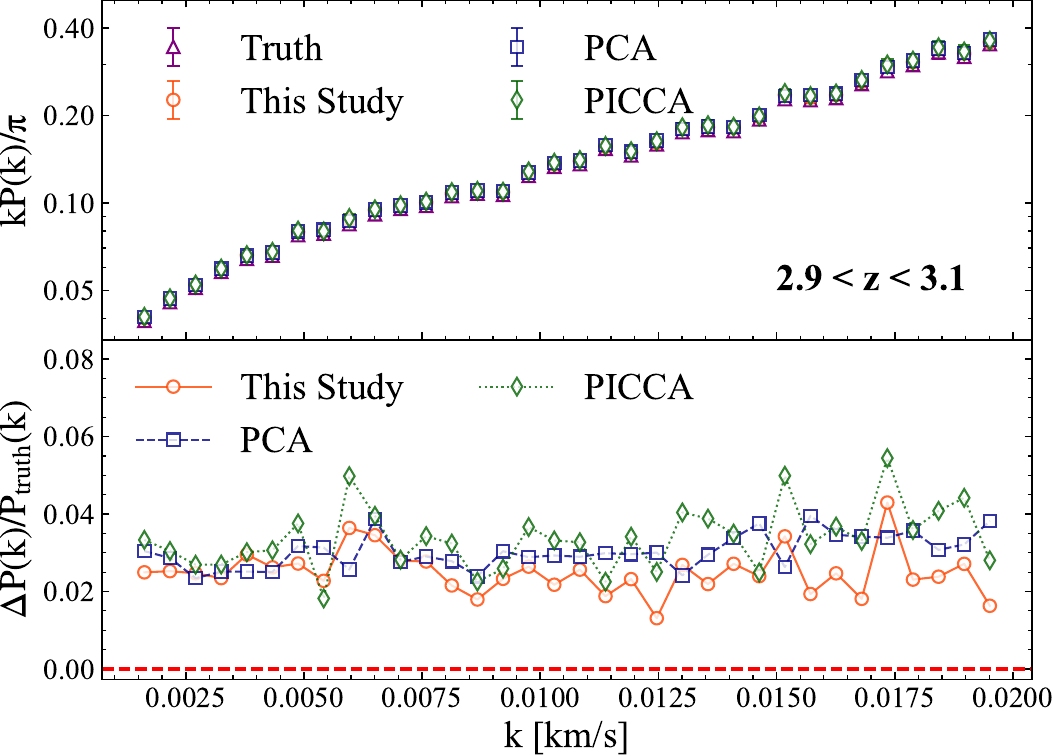}
    \caption{Similar to Figure~\ref{fig:pk1d}. We show the 1D Ly$\alpha$ forest measurements on the unperturbed dataset ($2.9<\mathrm{z}<3.1$) here. Three methods give similar uncertainty.}
    \label{fig:pk1d-unperturb}
\end{figure}

In Figure~\ref{fig:pk1d}, we show the 1D Ly$\alpha$ forest power spectrum measurements from the perturbed dataset. For completeness， we show the 1D Ly$\alpha$ forest power spectrum measurements in Figure~\ref{fig:pk1d-unperturb}. As in the unperturbed dataset, three methods give only $\sim 1\%$ difference for continuum prediction, no significant difference is found for the 1D Ly$\alpha$ forest power spectrum measurements.

\section{Latent Embedding Learned by QFA}\label{appendix:com}

\begin{figure}
    \centering
    \epsscale{1.2}
    \plotone{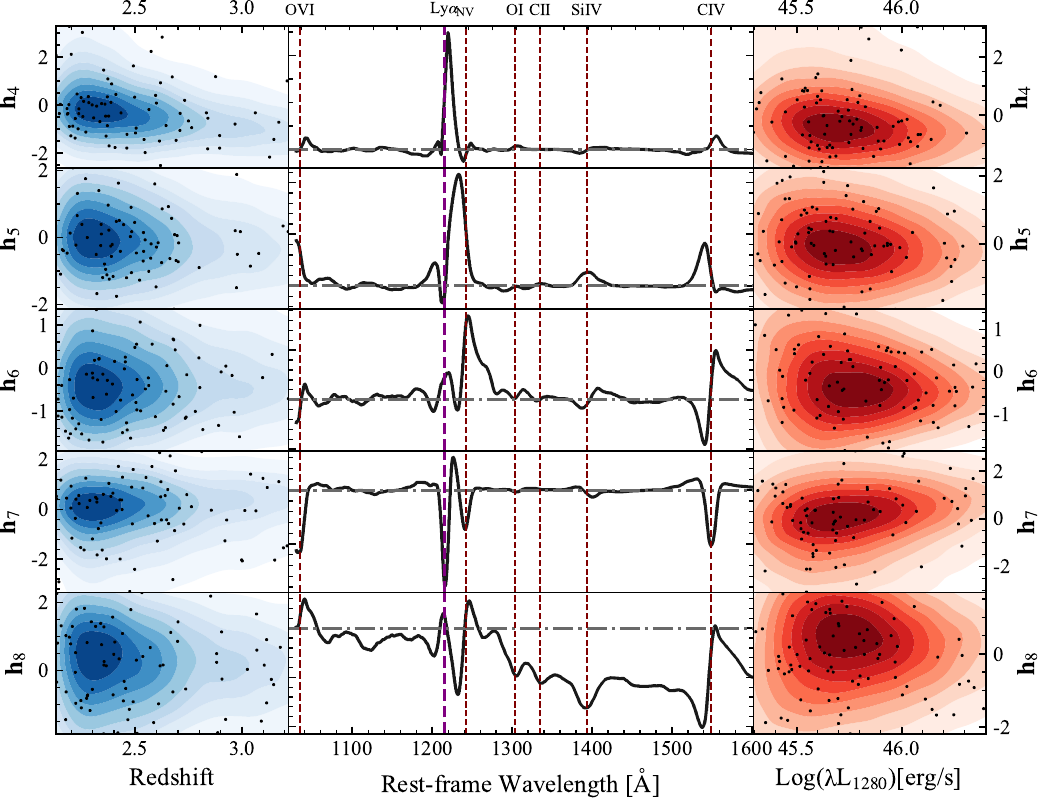}
    \caption{Similar to Figure~\ref{fig:com}, here we show the other five components learned by QFA from the SDSS DR16 dataset. Like the other three components in the main text, all components do not display any redshift evolution. Furthermore, all components (apart from the CIV component in the main text) have no visible dependency on luminosity either.\label{fig:com-full}}
\end{figure}

In Section~\ref{subsec:evolution}, we focused on the three of the most prominent factor components and their dependency on redshift and luminosity. For completeness, Figure~\ref{fig:com-full} shows the other five components learned by QFA. As before (Section~\ref{subsec:evolution}), the components shown are subjected to a varimax rotation.

Unlike PCA analysis (e.g., figure~2 in \citet{PARIS2011}), all eight components (five here and three in the main text) learned by QFA demonstrate clear and prominent features or even singular features. These features include Ly$\alpha$ emission line, power-law slope, and CIV emission line (see Figure~\ref{fig:com}). Most components are also dominated by visible positive or negative correlation features from two-to-three features, including  Ly$\alpha$, CIV, NV, OI, CII, SiIV and CIV emission lines. All components do not show any clear evolution with redshift, demonstrating that the quasar population has not evolved from $\mathrm{z}=3.5$ to $\mathrm{z=2}$. Also, no evident luminosity dependency is found within these factors except the negative correlation from the particular CIV component, as discussed in the main text.

\section{Ablation Study}\label{appendix:ablation}

\subsection{Model Dependency on Hidden Dimensions}\label{ablation:Nh}

The number of hidden dimensions -- $\mathrm{N}_\mathrm{h}$, is a critical hyperparameter in our method. As have described in Section~\ref{sec:method}, $\mathrm{N_h}$ is much smaller than the dimension of the spectra ($\mathrm{N_{pix}}$), and is set to $8$ in the main text. Here we perform ablation study to investigate the effect of different $\mathrm{N_h}$ on the model performance. As shown in Table~\ref{tab:Nh}, the model remains good performance when $\mathrm{N_h}\lesssim 10$. Too large $\mathrm{N_h}$, e.g., $\mathrm{N_h}=15$, will cause the model to overfit some absorption features and degrade the model performance, while too small $\mathrm{N_h}$, e.g., $\mathrm{N_h}=3$, may lead to a underfitted model , and also degrade the model performance. From our ablation study, $\mathrm{N_h}$ is appropriate at the range of about $5\sim 10$.

\begin{deluxetable}{lccccccc}[htbp!]
\tablecaption{Ablation Study on $\mathrm{N_h}$\label{tab:Nh}}
\tablehead{\colhead{$\mathrm{N_h}$}&\colhead{5}&\colhead{6}&\colhead{7}&\colhead{8}&\colhead{9}&\colhead{10}&\colhead{15}}
\startdata
AFFE [\%] （50th）&2.43&2.44&2.47&2.47&2.48&2.65&4.41\\
\enddata
\tablenotetext{\dagger}{We increase $\mathrm{N_h}$ from 5 to 10, and display the corresponding 50 percentile absolute fractional flux error (AFFE, see Section~\ref{subsec:testOnMock} for details) below. Generally, for $\mathrm{N_h}\lesssim 10$, no obvious performance different is found. For too small $\mathrm{N_h}$, the model lacks of complexity to describe the data, which lead to a slightly worse performance, while for too large $\mathrm{N_h}$, e.g., $\mathrm{N_h}=15$, the model tends to over fit some absorption features, and degrade its performance.}
\end{deluxetable}

\subsection{Model Dependency on Pre-defined Mean Optical Depth Function}\label{ablation:opt}

The choice of pre-defined mean optical depth function is an important prior for our method. As the quasar continuum is intrinsically degenerate with the Ly$\alpha$ forest, it's unavoidable to assign some prior information to separate the quasar continua from the Ly$\alpha$ forest reasonably. Although different mean optical depth measurements are generally at the same level, it may be problematic for quasar continuum fitting with a precision requirement of few percent. Here we investigate the effect of different mean optical depth priors. We will compare the variation of model performance with respect of three literature mean optical depth functions from \citet{FG2008}, \citet{BECKER2013}, and \citet{KAMBLE2020}. 

\begin{figure}
    \centering
    \plotone{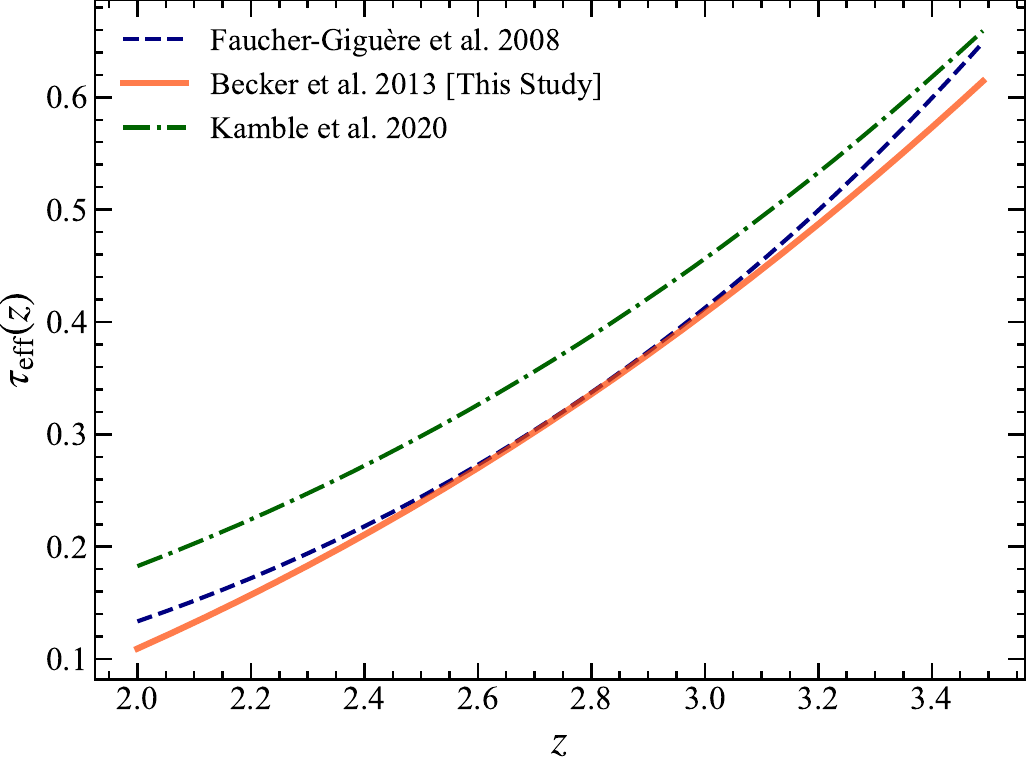}
    \caption{Mean optical depth measurements from \citet{FG2008}, \citet{BECKER2013}, and \citet{KAMBLE2020}. We have adopted the result from \citet{BECKER2013} (see Equation~\ref{eq:optical_depth}) in the main text.}
    \label{fig:mean-optical-depth}
\end{figure}

The mean optical depth functions determined by \citet{FG2008}, \citet{BECKER2013}, and \citet{KAMBLE2020} are shown in Figure~\ref{fig:mean-optical-depth}. While the two measurements, \citet{FG2008} and \citet{BECKER2013}, from high-resolution and high-SNR quasar spectra show almost consistent result, the measurement \cite[][]{KAMBLE2020} from a large number of moderate-resolution and SNR quasar spectra, however, is slightly different from the other two measurements. We have adopted the measurement from \citet{BECKER2013} in the main text as described in Section~\ref{subsec: model}.

\begin{deluxetable}{lccc}[htbp!]
\tablecaption{Ablation Study on $\tau_\mathrm{eff}$\label{tab:meanopt}}
\tablehead{\colhead{$\tau_\mathrm{eff}$}&\colhead{\citet{FG2008}}&\colhead{\citet{BECKER2013}}&\colhead{\citet{KAMBLE2020}}}
\startdata
AFFE [\%] （50th）&2.15&2.48&5.23\\
\enddata
\tablenotetext{\dagger}{Variations in continuum modeling resulting from different choices of mean optical depth are on par with the differences between the mean optical depths themselves. In application, adopting alternative mean optical depth measurements may prove necessary.}
\end{deluxetable}

The results in Table~\ref{tab:meanopt} indicate that the discrepancies amongst continuum predictions are commensurate with variations in the mean optical depth function. This suggests that precise measurements of the mean optical depth function are imperative to disambiguate the quasar continuum and Ly$\alpha$ forest absorption. At present, the mean optical depth function is only employed as an a priori assumption. Therefore, in practice, evaluating alternative mean optical depth functions may be critical.  

\subsection{Model Dependency on Regularization Strength}\label{ablation:alpha}

To impose a prior that our model parameters should be close to zero, we add a regularization term to the log-likelihood function in Section~\ref{subsec:regularization}. The strength of the regularization term is controlled by a hyperparameter $\alpha$ in Equation~\ref{eq: loss}. Here, we investigate the dependency of model performance on $\alpha$. As shown in Table~\ref{tab:alpha}, too small $\alpha$ won't provide enough constraints and will lead to poor performance, while too large $\alpha$ will make the variations of model parameters too small, thereby weakening the model's performance. $\alpha$ is appropriate at the level of $10^{-1}$.

\begin{deluxetable}{lccccccc}[htbp!]
\tablecaption{Ablation Study on $\alpha$\label{tab:alpha}}
\tablehead{\colhead{$\alpha$}&\colhead{0.001}&\colhead{0.005}&\colhead{0.05}&\colhead{0.1}&\colhead{0.5}&\colhead{1}&\colhead{2}}
\startdata
AFFE [\%] （50th）&3.07&3.14&2.42&2.65&2.31&2.37&3.26\\
\enddata
\tablenotetext{\dagger}{We increase $\alpha$ from left to right, and display the corresponding 50 percentile absolute fractional flux error (AFFE, see Section~\ref{subsec:testOnMock} for details) below. Too small $\alpha$ will not give enough constraints, too large $\alpha$ will lead to too small parameters, both cases will degrade the model performance. $\alpha$ is appropriate in the level of $\sim 10^{-1}$.}
\end{deluxetable}

\section{Underestimation of continuum uncertainty and possible calibrations}\label{appendix:uncertainty}

\begin{figure*}
    \plotone{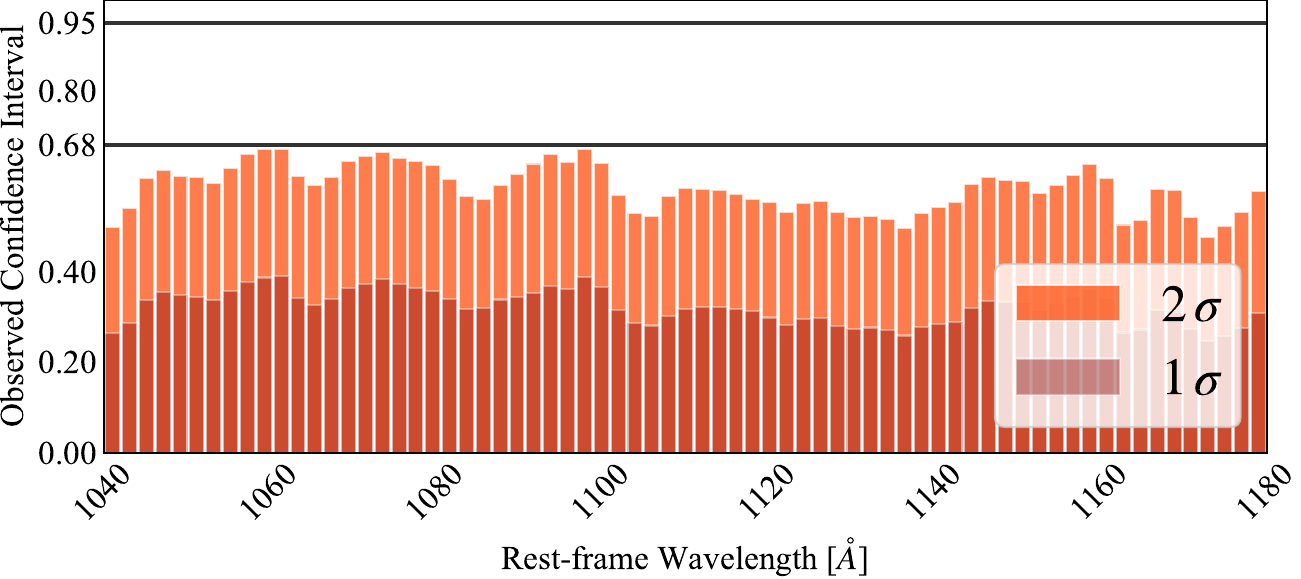}
    \caption{Uncalibrated observed confidence intervals for QFA as a function of rest-Frame wavelength (similar to figure~11 in \citet{NF2020}).
If the uncertainty given by QFA were calibrated accurately, $\mathbf{P}\%$ of the observed absolute errors would fall within the $\mathrm{P}\%$ confidence interval. Under the Gaussian assumption of QFA, we show here the observed confidence intervals for $1\,\sigma$ and $2\,\sigma$ (e.g.， $\mathbf{P} = 68\%$ and $\mathbf{P} = 95\%$). The solid horizontal lines correspond to the expected confidence intervals. QFA produces overly optimistic confidence intervals because of the omission of the stochastic error term $\boldsymbol{\Psi}$.}
    \label{fig:uncal}
\end{figure*}

\begin{figure*}
    \plotone{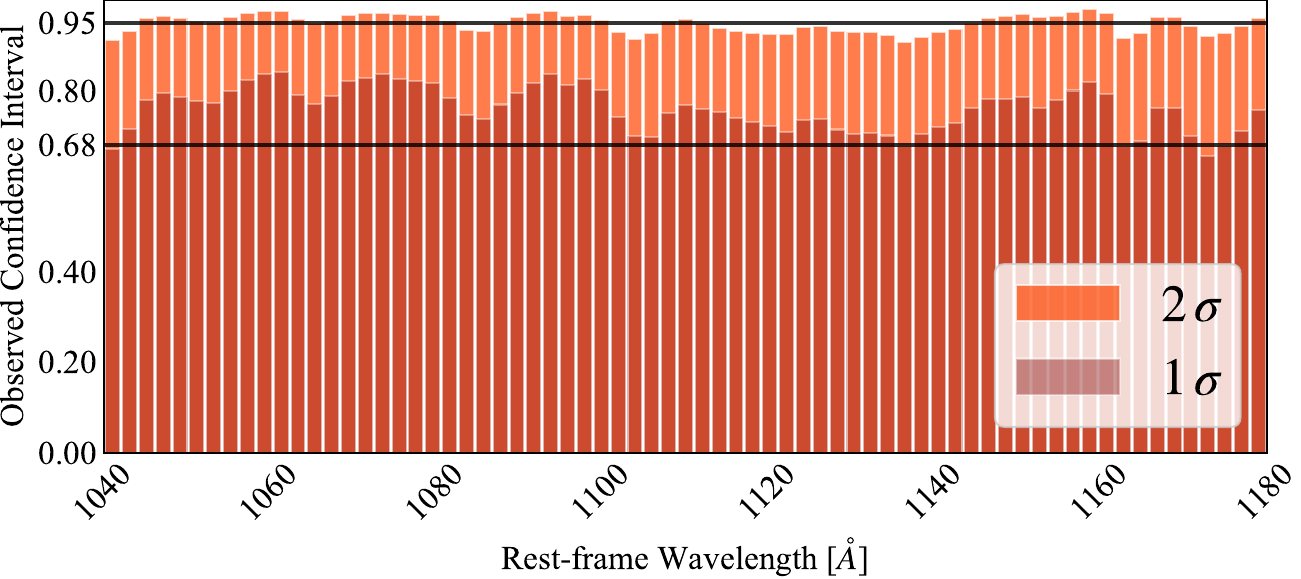}
    \caption{Similar as Figure~\ref{fig:uncal}, but scale the inferred uncertainty to a factor of three. In that case, the observed and inferred uncertainty values exhibited close correspondence.}
    \label{fig:cal}
\end{figure*}

As discussed in Section~\ref{subsec: pred} and Section~\ref{subsec:cosmo}, omitting the stochastic error term $\boldsymbol{\Psi}$ during inference is necessary to optimize the model's performance in estimating the posterior mean of the continua, as certain erratic features tend to be subsumed into this term. However, such an approach will significantly underestimate the uncertainty in QFA. As illustrated in Figure~\ref{fig:uncal}, by performing an analogous experiment as \citet{NF2020}, which assesses the performance of the observed confidence intervals by calculating the fraction of observed absolute errors that fall within the $1\,\sigma$ (68\%) and $2\,\sigma$ (95\%) confidence intervals under the Gaussian assumption, we concluded that the empirical error observed is appreciably greater than the inferred uncertainty. However, if we scale the inferred uncertainty by a factor of 3, we find that the inferred uncertainty closely matches the empirical uncertainty, as demonstrated in Figure~\ref{fig:cal}.

\bibliography{sample631}{}
\bibliographystyle{aasjournal}
\end{CJK*}
\end{document}